\documentclass[a4paper,11pt]{article}

\usepackage{graphicx}
\usepackage{subfig}
\usepackage[normalem]{ulem}
\usepackage{amsmath}

\pdfoutput=1
\usepackage{jcappub}

\newcommand{\be}{\begin{eqnarray}}
\newcommand{\ee}{\end{eqnarray}}

\def\vn{\mathbf{n}}

\def\ensemble{\textit{ensemble}}

\title{Is the lack of power anomaly in the CMB correlated with the orientation of the Galactic plane?}

\author[a,b]{U.Natale,}
\author[c,d]{A.Gruppuso,}
\author[a,b,c]{D.Molinari,}
\author[a,b,c]{and P.Natoli}

\affiliation[a]{Dipartimento di Fisica e Scienze della Terra, Universit\`a degli Studi di Ferrara, via Saragat 1, I-44122 Ferrara, Italy}
\affiliation[b]{INFN, Sezione di Ferrara,via Saragat 1, I-44122 Ferrara, Italy}
\affiliation[c]{INAF OAS Bologna, Osservatorio di Astrofisica e Scienza dello Spazio di Bologna,via Gobetti 101, I-40129 Bologna, Italy}
\affiliation[d]{INFN, Sezione di Bologna,Via Irnerio 46, I-40126 Bologna, Italy}

\emailAdd{umberto.natale@unife.it}
\emailAdd{alessandro.gruppuso@inaf.it}
\emailAdd{diego.molinari@unife.it}
\emailAdd{paolo.natoli@unife.it}

\abstract{The lack of power at large angular scales in the CMB temperature anisotropy pattern is a feature known to depend on the size of the Galactic mask. Not only the large scale anisotropy power in the CMB is lower than the best-fit $\Lambda$CDM model predicts, but most of the power seems to be localised close to the Galactic plane, making high-Galactic latitude regions more anomalous. We assess how likely the latter behaviour is in a $\Lambda$CDM model by extracting simulations from the {\it Planck} 2018 fiducial model. By comparing the former to {\it Planck} data in different Galactic masks, we reproduce the anomaly found in previous works, at a statistical significance of $\sim 3 \, \sigma$. This result suggests the existence of a bizzarre correlation between the particular orientation of the Galaxy and the lack of power anomaly. To test this hypothesis, we perform random rotations of the {\it Planck} 2018 data and compare these to similarly rotated $\Lambda$CDM realisations. We find that, among all possible rotations, the lower-tail probability of the observed high-Galactic latitude data variance is still low at the level of $2.8 \, \sigma$. Furthermore, the lowering trend of the variance when moving from low- to high-Galactic latitude is anomalous in the data at $\sim 3\,\sigma$ when comparing to $\Lambda$CDM rotated realisations. This shows that the lack of power at high Galactic latitude is substantially stable against the ``look elsewhere'' effect induced by random rotations of the Galaxy orientation. Moreover, this analysis turns out to be substantially stable if we employ, in place of generic $\Lambda$CDM simulations, a specific set whose variance is constrained to reproduce the observed data variance.
}


\begin{document} 
\maketitle
\flushbottom

\section{Introduction}
\label{intro}

The low-variance anomaly is a feature of the Cosmic Microwave Background (CMB) temperature anisotropy pattern present in both WMAP \cite{Monteserin:2007fv,Cruz:2010ud,Gruppuso:2013xba} and {\it Planck} data \cite{Ade:2013nlj,Ade:2015hxq,Akrami:2019bkn}. It shows up at large angular scales, where the instrumental noise is negligible, with a statistical significance around $2$-$3 \, \sigma$ C.L. depending on the estimator employed.
This effect is correlated with other CMB anomalies, see e.g.~\cite{Copi:2006tu,Copi:2008hw,Copi:2010na,Schwarz:2015cma}, which are sensitive to the lack of power with respect to expectations of the $\Lambda$CDM model, see \cite{Muir:2018hjv} for further details. 
For this reason, we will use the expressions lack-of-power and low-variance as synonyms.

A statistical fluke is of course the simplest explanation for this phenomenon. However, in this case, one has to accept to live in a rare $\Lambda$CDM realisation.
In any case, there are at least three reasons why this anomaly is worth of further investigations \cite{Gruppuso:2015zia}:
\begin{enumerate}
\item it is unlikely that the effect is due to an unaccounted instrumental systematics: both WMAP and {\it Planck} observe it with similar significance despite being two separate experiments with different data gathering schemes and scanning strategies;
\item it is not natural to attribute this effect to foreground residuals: the latter are not expected to be correlated to the CMB, so a foreground residual should increase and not lower the total anisotropy power\footnote{Note also that typically (and in particular at large scales where this work is focused) the foreground mitigation is performed at the map level (in the harmonic or pixel space) and not at the $C_{\ell}$ level.}. 
A similar argument would also apply to possible extensions of the $\Lambda$CDM as long as their source is statistically independent from the primary CMB anisotropy \cite{Gruppuso:2007ya,Bunn:2008zd}.
\item it is suspiciously dependent on the Galactic mask: its statistical significance increases when only high Galactic regions are considered, which is usually a conservative choice in CMB data analysis \cite{Gruppuso:2015xqa}.
It was also shown \cite{Gruppuso:2017nap} that this effect was dominated by odd over even multipoles, see e.g.~\cite{Kim:2010gf,Kim:2010gd,Gruppuso:2010nd}.
\end{enumerate}

This paper wants to focus on the last item by estimating, from a statistical point of view, how likely is to find a CMB map of the $\Lambda$CDM model with such a behaviour between low- and high-Galactic latitudes. 
To perform this analysis we will use random rotations (see Appendix \ref{sec:rotations}) of simulated CMB maps in order to evaluate among all the possible orientations what is the probability of having most of the power at low-Galactic latitudes.
The adopted estimator is the variance, $V$, of the temperature anisotropies, $\delta T (\hat{\vn})$, 
\begin{equation}
V \equiv \langle {\left( \delta T (\hat{\vn})\right)}^2 \rangle \, ,
\end{equation}
where $\hat{\vn}$ is the unit-vector pointing a given direction of observation. 
$V$ is built through the angular power spectrum (APS), $C_{\ell}$: 
\begin{equation}
V = \sum_{\ell=2}^{\ell_{max}}\frac{2\ell+1}{4\pi}C_\ell \, ,
\label{variancett}
\end{equation}
where the maximum multipole, $\ell_{max}$, is set to 29 in the following since we want to be consistent with the maximum multipole considered in the \textit{Planck} pixel-based low-$\ell$ Likelihood functions \cite{Aghanim:2015xee}. However, the dependence of $V$ upon $\ell_{max}$ is very weak for $\ell_{max} \gtrsim 10$ and therefore such a choice does not impact significantly on our results.

The paper is organised as follows: in Section \ref{dataset} we describe the dataset we consider, and  how we generate Monte Carlo simulations; in Section \ref{preliminary} we perform the analysis of the {\it Planck} 2018 dataset comparing the results with $\Lambda$CDM simulations. After recovering results in agreement with previous works, we consider random rotations of the data and the simulations, to assess the a posteriori choice of assuming a particular orientation for the Galactic plane; in Section \ref{analysis} we repeat the same analyses focusing on a specific set of $\Lambda$CDM simulations that show the same low-variance of the observed map. Conclusions are drawn in Section \ref{conclusions}. 

\section{Data set and simulations}
\label{dataset}

\subsection{CMB maps and masks}

\begin{figure}[t]
	\centering
	\includegraphics[width=.45\textwidth]{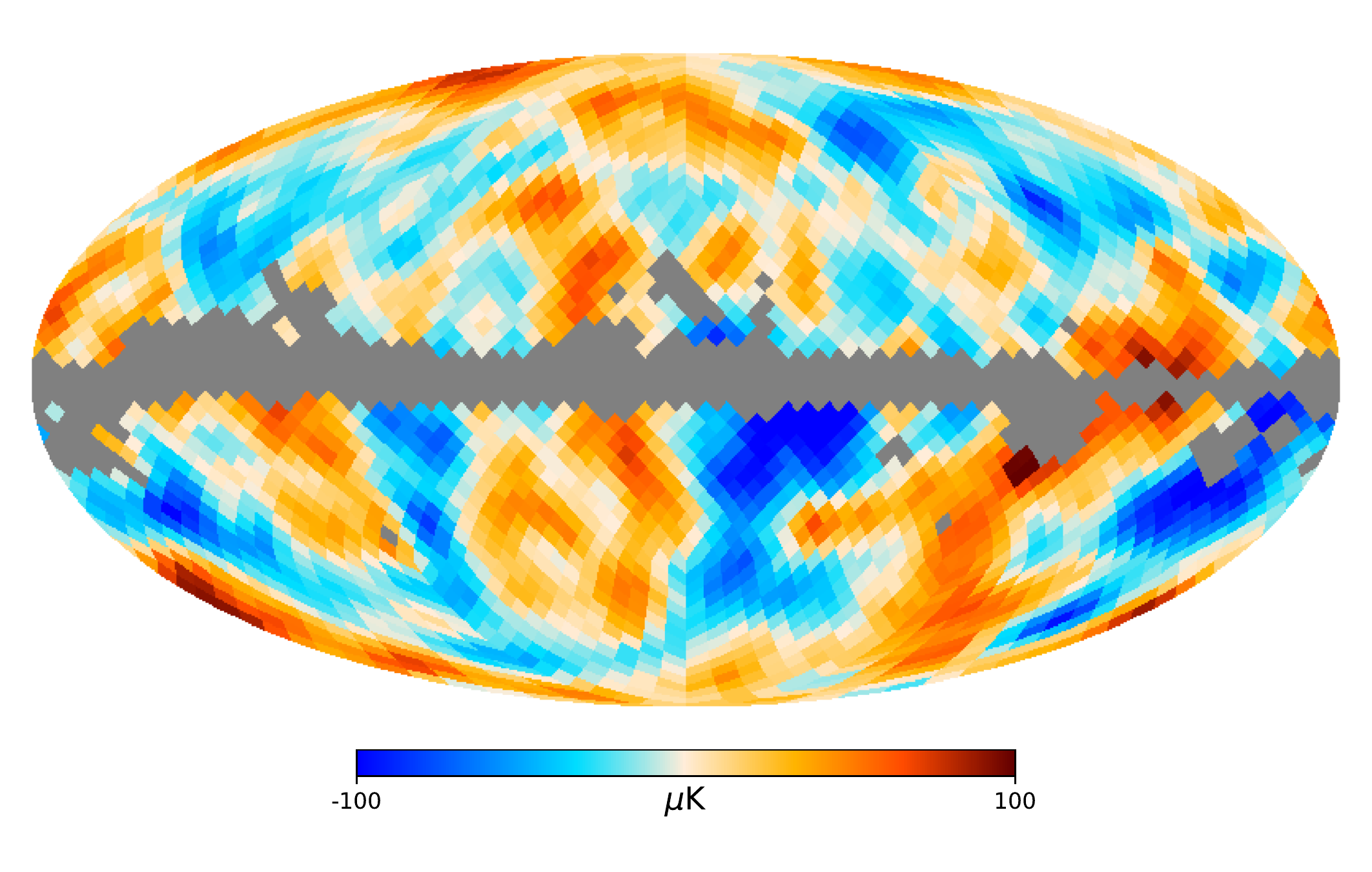}
	\includegraphics[width=.45\textwidth]{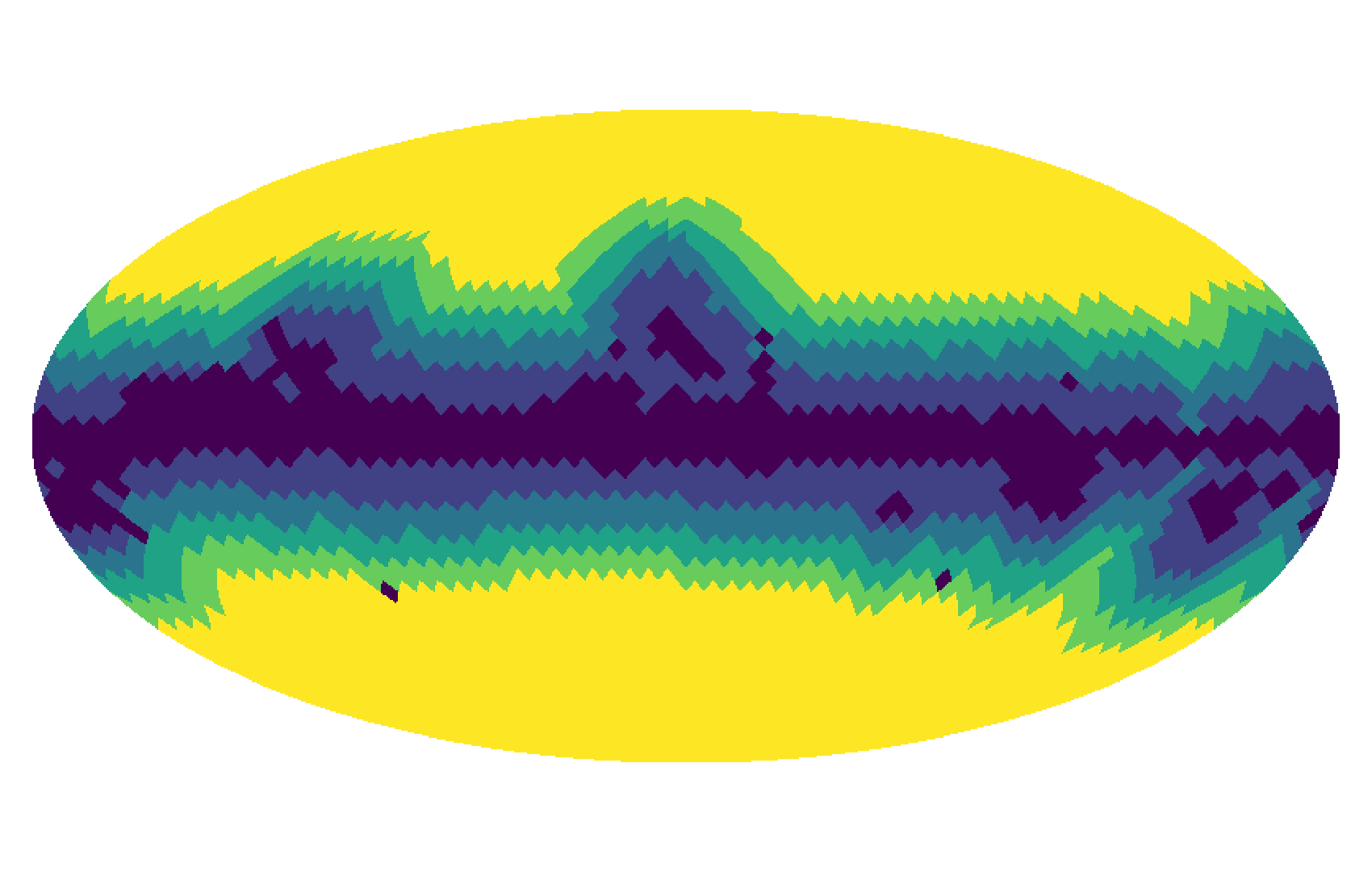}
	\caption{Left panel: \texttt{Commander} 2018 map smoothed at $440$ arcmin, where the Std 2018 mask is applied. Right panel: Galactic temperature masks considered in this paper. The dark blue region is for the 2018 standard case. The blue region is for the Ext$_{12}$ case. The light blue region is for the Ext$_{18}$ case. The dark green region is for the Ext$_{24}$ case. The green region is for the Ext$_{30}$ case.	
} \label{fig:masks}
\end{figure}

We use data products from the \textit{Planck} 2018 data release, available in the {\it Planck} Legacy Archive\footnote{https://www.cosmos.esa.int/web/planck/pla}. 
In particular we employ the temperature \texttt{Commander} 2018 map \cite{Akrami:2018mcd} downgraded to \texttt{HEALPix}\footnote{http://healpix.sourceforge.net} \cite{Gorski:2004by} resolution $N_{side}=16$ with a Gaussian beam with full width half maximum, FWHM, of 440 arcmin. The map is shown in the left panel of Fig.~\ref{fig:masks}. As a consistency check we also employ the \texttt{SMICA} temperature map \cite{Akrami:2018mcd}, also downgraded from high resolution to $N_{side}=16$. These CMB maps have been delivered already with a constrained CMB realisation along the Galactic plane. We have added to those maps a regularisation noise realisation with 2 $\mu$K rms, consistently considered in the extraction of the APS. This choice is consistent with the procedure adopted in \cite{Aghanim:2019ame}. We checked that such a noise has a negligible impact on our results. The maps have been masked with several Galactic masks, shown in the right panel of Fig.~\ref{fig:masks} and whose sky fractions are listed in Table~\ref{tab:masks}. More specifically, the considered masks are the $N_{side}=16$ confidence mask provided with the 2018 \texttt{Commander} solution \cite{Akrami:2018mcd}, named Std 2018, and other four masks built extending the edges of the Likelihood 2015 standard mask \cite{Aghanim:2015xee} by 12, 18, 24 and 30 degrees, called respectively Ext$_{12}$, Ext$_{18}$, Ext$_{24}$ and Ext$_{30}$. This choice is done in order to make contact with previous works, i.e. \cite{Gruppuso:2015xqa,Gruppuso:2017nap}, and to compare the impact of the most recent {\it Planck} 2018 data with respect to that of the 2015 release, see Appendix \ref{comparison_2018_2015}.

\begin{table}[!h]
	\begin{center}
		\begin{tabular}{ l | c   }
			\hline
			\hline
			Mask & Sky Fraction [\%]\\
			\hline
			Std 2018& 85.6 \\

			Ext$_{12}$ & 70.8\\

			Ext$_{18}$ & 59.1\\
	
			Ext$_{24}$ & 48.7\\
	
			Ext$_{30}$ & 39.4 \\
			\hline 
			\hline
		\end{tabular}
		\caption{Observed sky fractions for the masks shown in Figure \ref{fig:masks}.}\label{tab:masks}
	\end{center}
\end{table}

\subsection{Sets of simulations}

We generate $10^5$ CMB temperature maps at \texttt{HEALPix} resolution $N_{side}=16$  randomly extracted from the \textit{Planck} 2018 best-fit model through the \texttt{synfast} function of \texttt{healpy} \cite{Gorski:2004by} with a Gaussian beam of 440 arcmin FWHM. To provide numerical regularisation, a different random noise realisation, with rms of 2 $\mu$K, is added to each of the CMB simulations, as done for the observed \texttt{Commander} and \texttt{SMICA} 2018 maps. This set is used to estimate the statistical significance of the low-variance in a $\Lambda$CDM framework.
A subset of $10^3$ simulations of this set of $\Lambda$CDM realisations is referred to as \textit{ensemble} 0. 
Another subset of $10^3$ simulations constrained to have variance $V$ close to the value observed by \texttt{Commander} 2018, $V_\textrm{c}=2090.02 \;  \mu$K$^2$ obtained with the Std 2018 mask, is called \textit{ensemble} 1.
More precisely a map $\textbf{m}_i$ with variance $ V_i$ belongs to \textit{ensemble} 1, if $V_\textrm{c}-20 \;\mu$K$^2$ $\le V_i \le V_\textrm{c}+20 \;\mu$K$^2$.
The analysis of the stability of our results with respect to the choice of the threshold of $20 \;\mu$K$^2$ is given in Appendix \ref{threshold}. 
Note that in the case of \texttt{SMICA} the variance is also constrained in the same range which contains the value observed in the data ($V_\textrm{s}=2085.57\,\mu$K$^2$).

\subsection{Angular power spectrum estimator}

\begin{figure}[t]
	\centering
	\includegraphics[width=.6\textwidth]{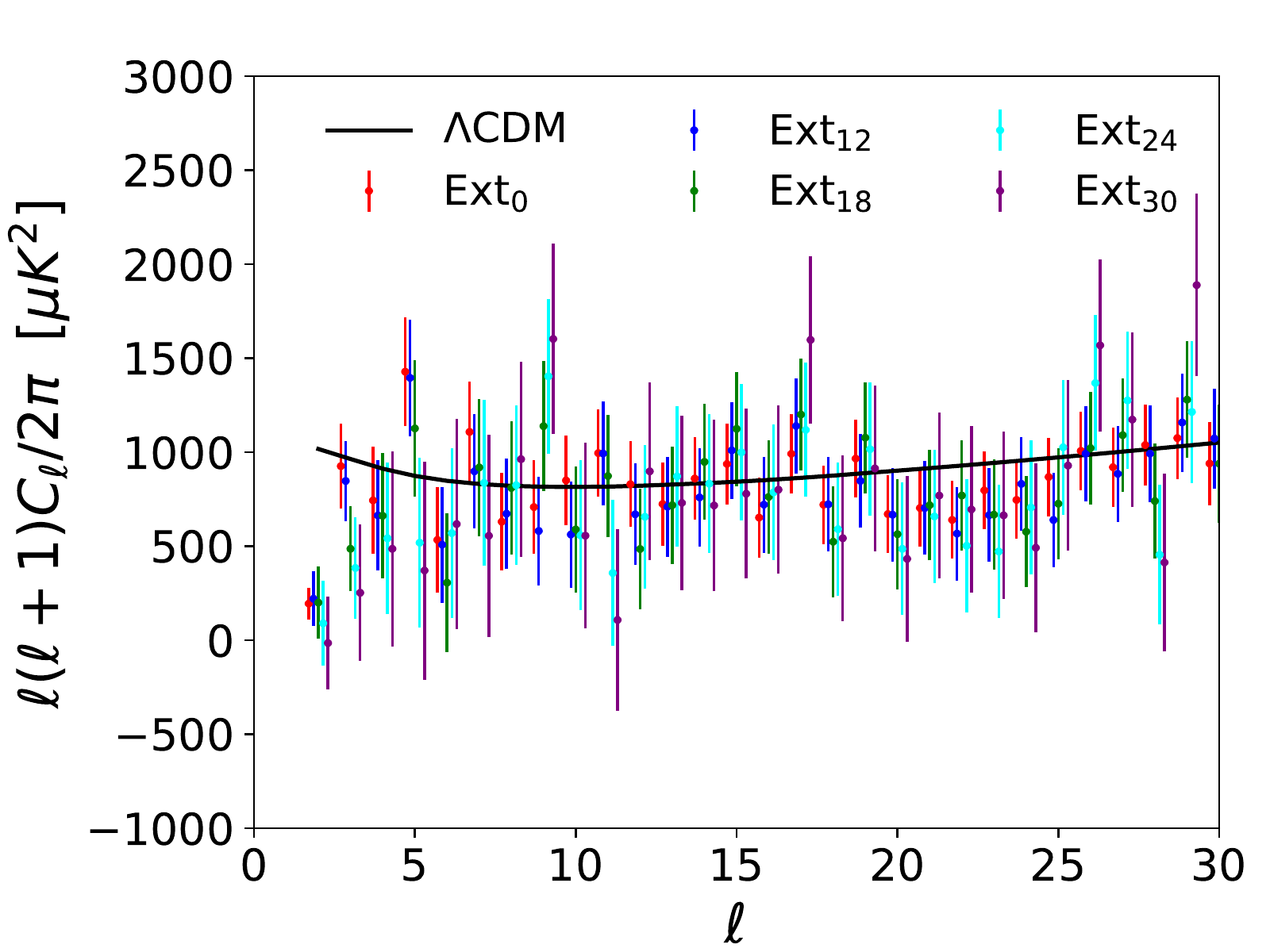}
	\caption{APS of the \textrm{Commander} 2018 temperature map estimated with the five masks shown in Fig.~\ref{fig:masks} and whose sky fraction is reported in Table~\ref{tab:masks}. The uncertainty shown for each multipole do not include cosmic variance.} \label{fig:nocosmvar}
\end{figure}

As anticipated in Section \ref{intro}, we use the variance $V$ as estimator for the lack of power, built through Eq.~(\ref{variancett}). 
The $C_{\ell}$ are obtained with an optimal angular power spectrum estimator, namely \texttt{BolPol} \cite{Gruppuso:2009ab}, an implementation of the Quadratic Maximum Likelihood (QML) method \cite{Tegmark:1996qt,Tegmark:2001zv}. The choice of the QML algorithm minimises the introduction of extra statistical uncertainty in our analysis with respect to other, suboptimal, APS estimators \cite{Molinari:2014wza}.
For each of the simulated maps and for the various masks defined above, we have used the estimates of \texttt{BolPol} to build the variance, $V$.In Fig.~\ref{fig:nocosmvar} we show the APS of the \textrm{Commander} 2018 temperature map estimated with the five masks shown in Fig.~\ref{fig:masks} and whose sky fraction is reported in Table~\ref{tab:masks}.

\section{Analysis in $\Lambda$CDM framework}
\label{preliminary}
	
\begin{figure}[t]
	\hspace{-0.85cm}
	\centering
	\subfloat{\includegraphics[width=.35\textwidth]{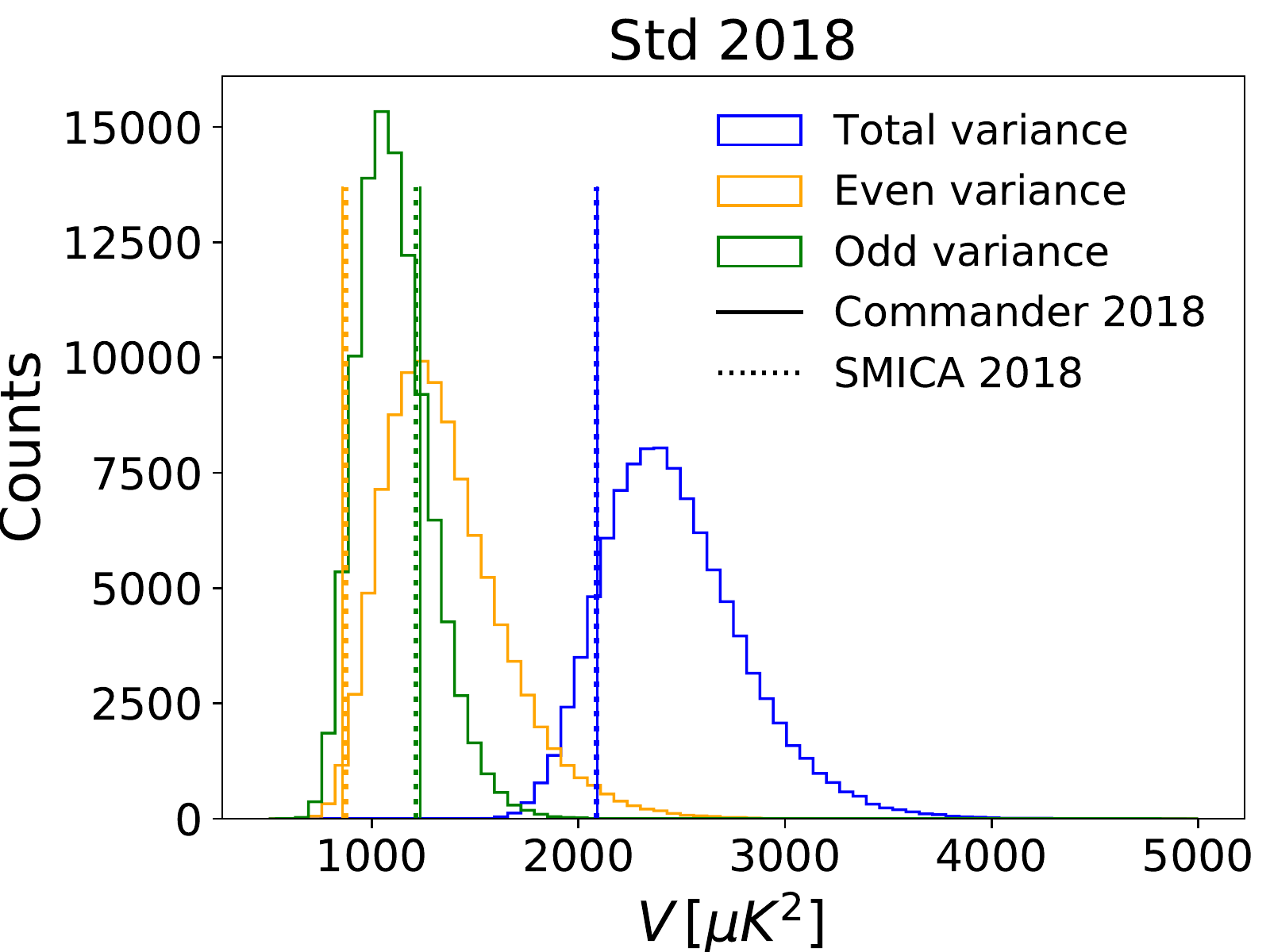}}
	\subfloat{\includegraphics[width=.35\textwidth]{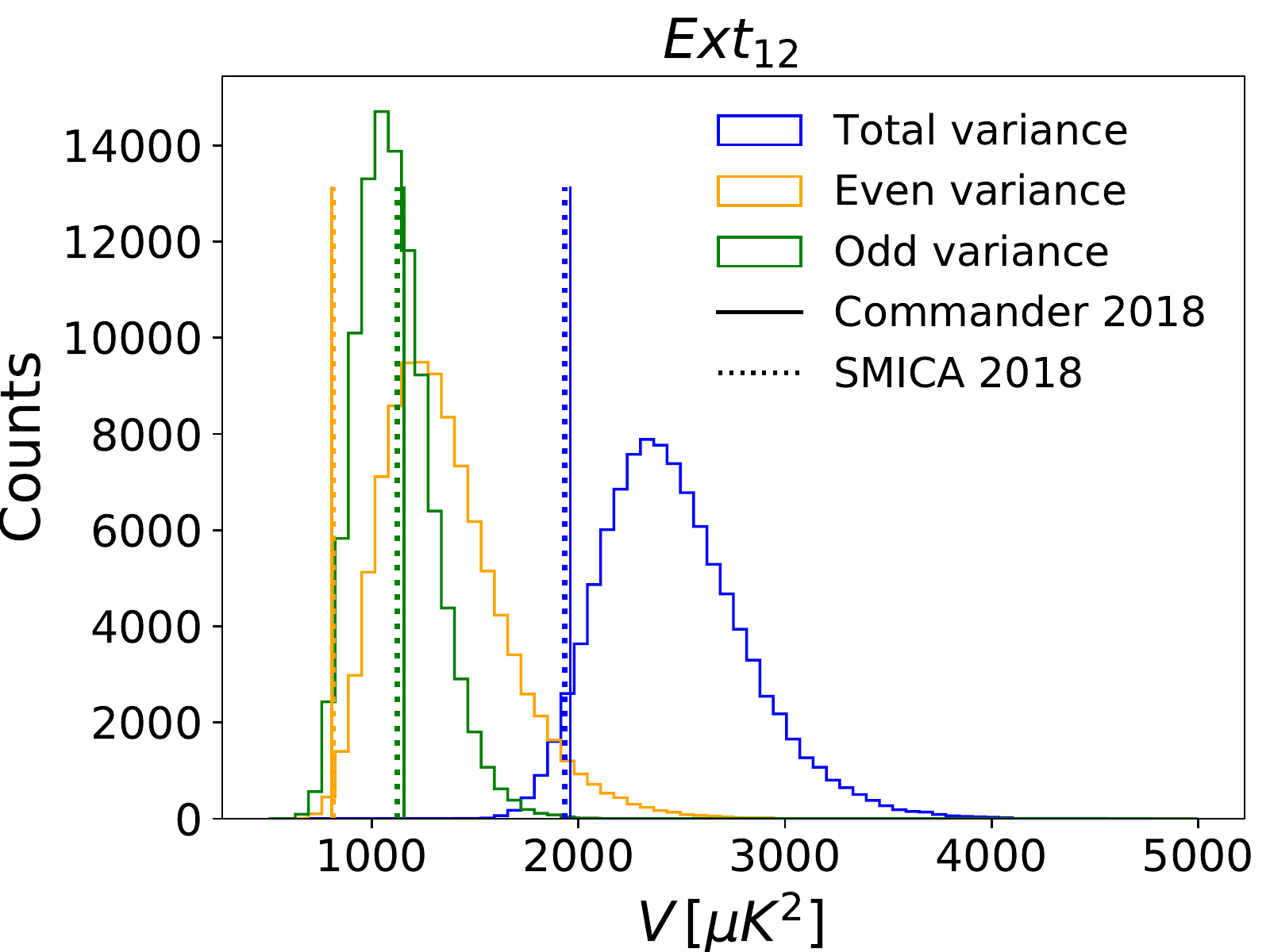}}
	\subfloat{\includegraphics[width=.35\textwidth]{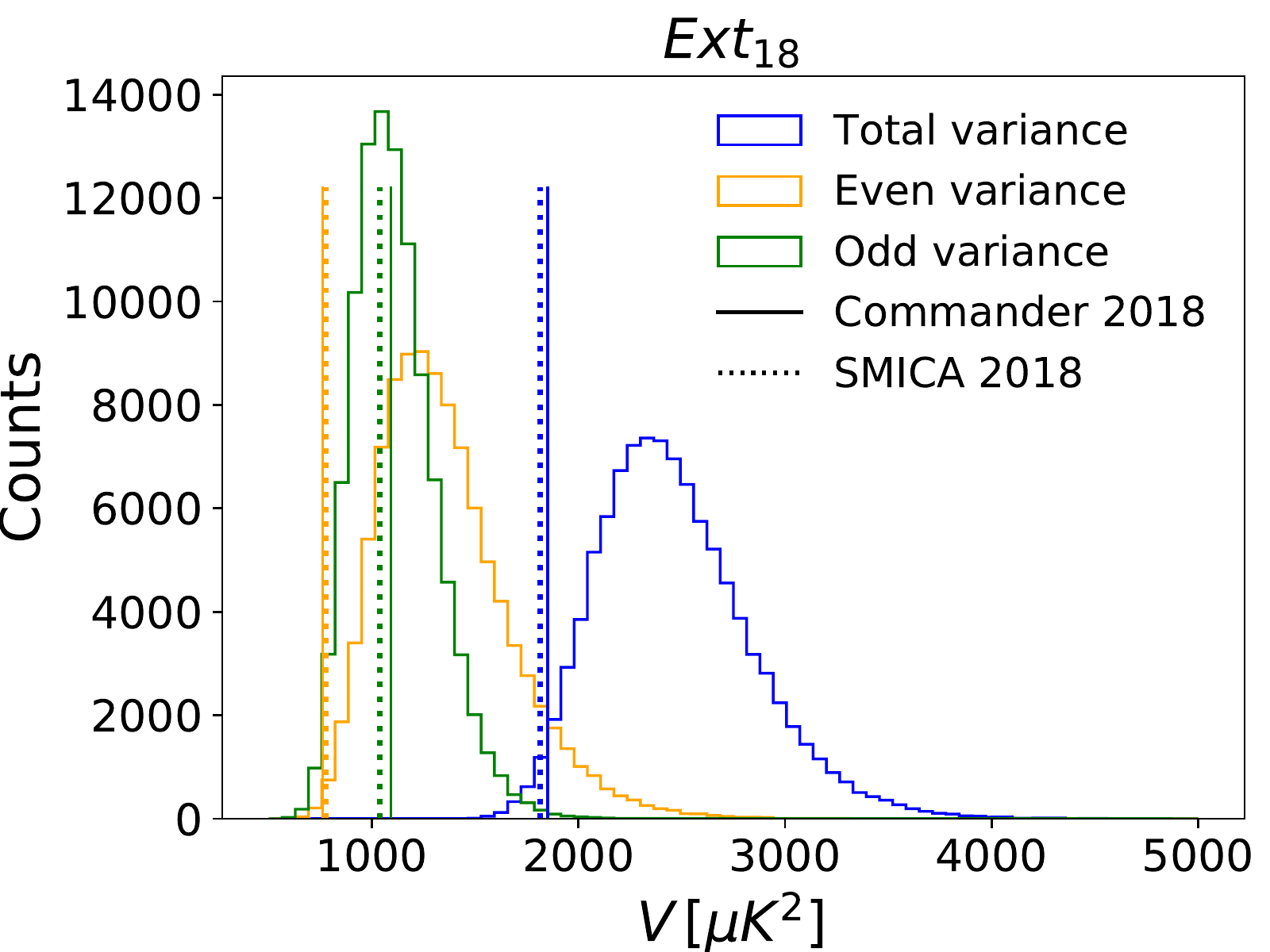}} \\
	\subfloat{\includegraphics[width=.35\textwidth]{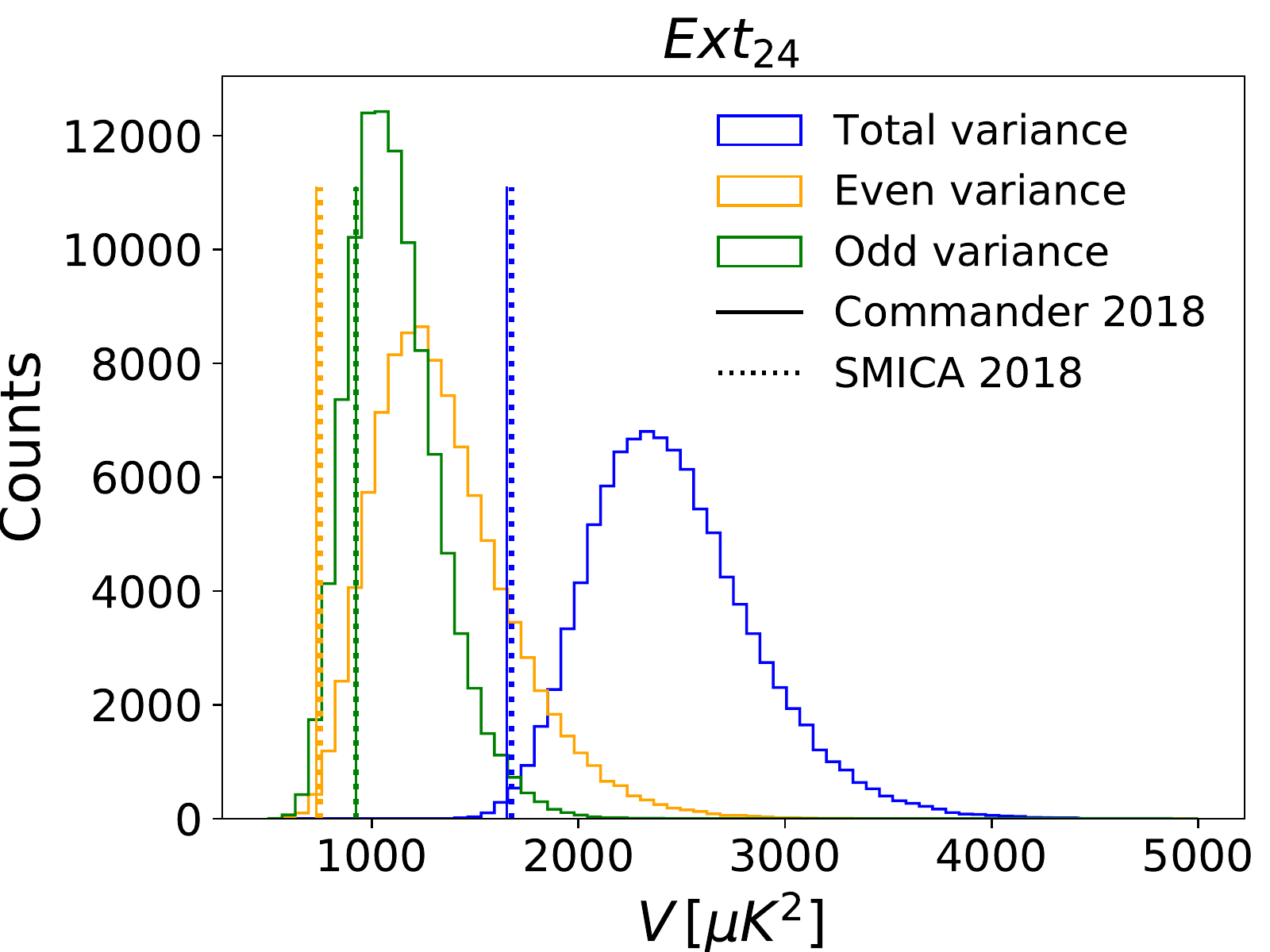}}
	\subfloat{\includegraphics[width=.35\textwidth]{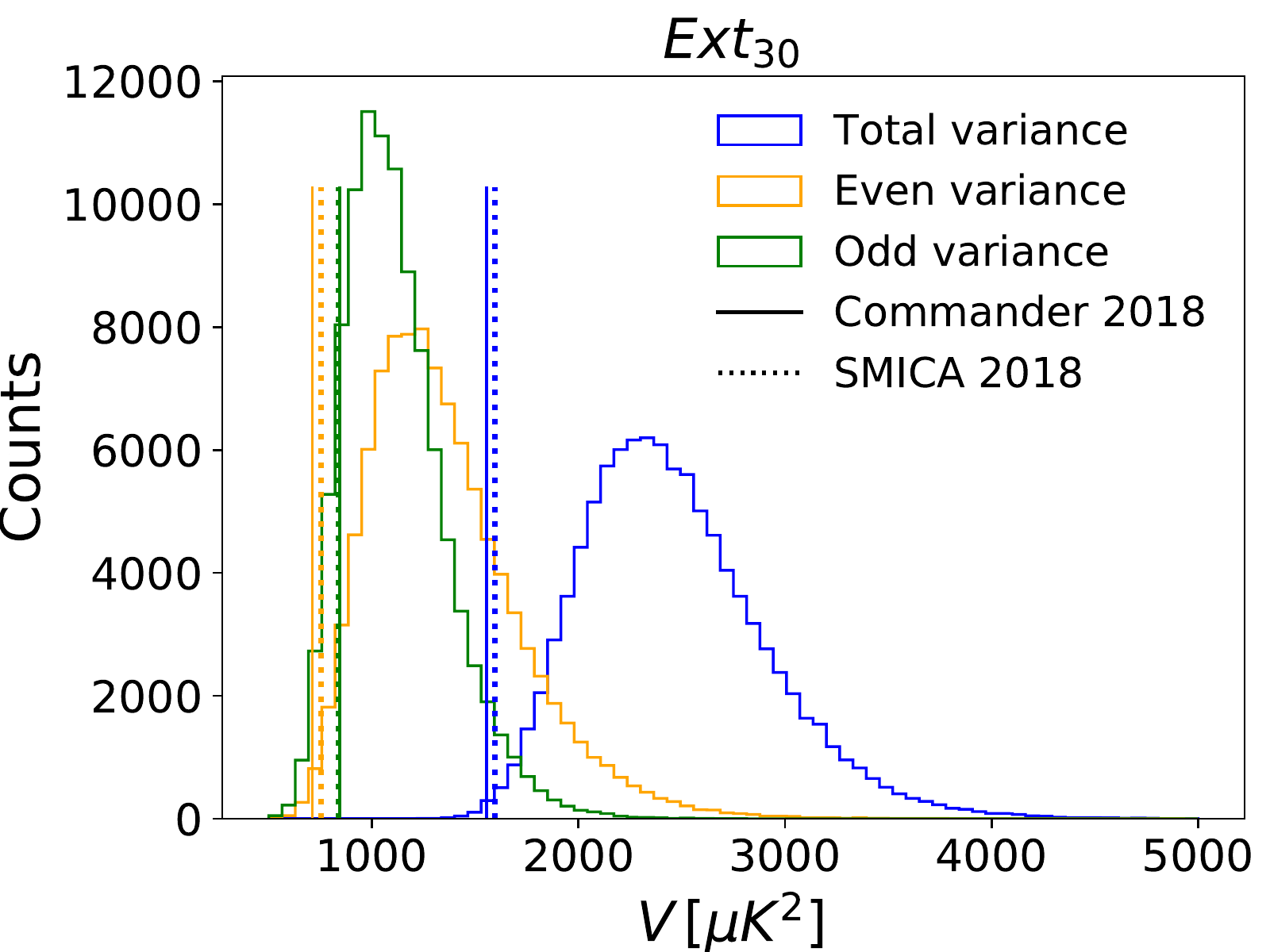}}
	\caption{Each panel shows the empirical distribution of $V$ in $\mu$K$^2$ expected in a $\Lambda$CDM model (blue) and computed throught Eq.~(\ref{variancett}) for the masks listed in Table \ref{tab:masks}. We report also the even and odd splits of the variance, through Eq. \ref{eqn:variance_EO} (orange and green, respectively). Vertical dashed and dotted bars correspond to the {\it Planck} 2018 \texttt{Commander} and \texttt{SMICA} CMB solutions, respectively. 
	}	\label{fig:MC_variance}
\end{figure}

As already known in the literature, the observed value of $V$ is low and its statistical significance increases considering regions at high Galactic latitude, 
see e.g. \cite{Monteserin:2007fv,Cruz:2010ud,Gruppuso:2013xba,Ade:2013nlj,Ade:2015hxq,Akrami:2019bkn}. Employing the \texttt{Bolpol} code to extract the TT APS for each of the $10^5$ $\Lambda$CDM simulations, we have built the probability distribution functions of $V$ for each of the five masks shown in Fig.~\ref{fig:masks}. The MC distributions are displayed in Fig.~\ref{fig:MC_variance} where they are compared to the corresponding \textit{Planck} 2018 observed values shown as vertical bars.

In the same panels we provide also $V_{+}$ ($V_{-}$), shown in orange (green), defined as $V$ but where the sum in Eq.~(\ref{variancett}) is performed only over the even (odd) multipoles, i.e.
\begin{equation}\label{eqn:variance_EO}
	V_{\pm}=\sum_{\ell=2}^{\ell_{max}}\left[\frac{1\pm(-1)^\ell}{2}\right]\frac{2\ell+1}{4\pi}C_\ell \,.
\end{equation}
In addition, we display as vertical bars, with the same color convention, the corresponding observed {\it Planck} 2018 values for $V_{\pm}$.

\begin{figure}[t]
\centering
\includegraphics[scale=0.6]{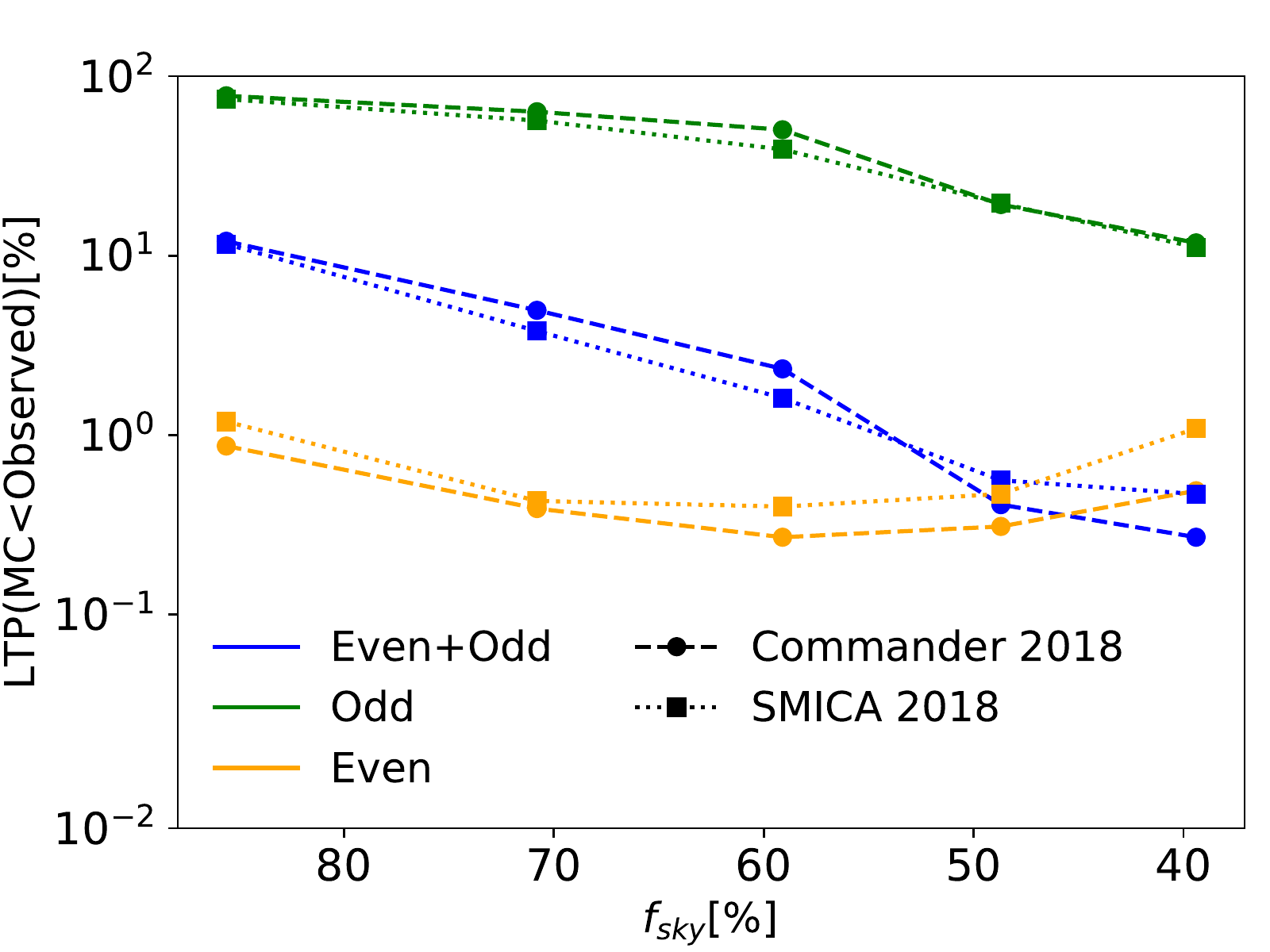}
\caption{Lower tail probability of the {\it Planck} 2018 \texttt{Commander} and \texttt{SMICA} maps with respect to the $10^5$ $\Lambda$CDM simulations as a function of the sky fraction.}
\label{fig:p_value_MC_seme}
\end{figure}

Fig.~\ref{fig:p_value_MC_seme} shows the three lower tail probabilities, henceforth LTP, for $V_+$ (orange), $V_-$ (green) and $V$ (blue) against the observed sky fraction of the five cases of Fig.~\ref{fig:MC_variance}. 
$V$ shows a monotonic behaviour: as one considers regions at higher and higher Galactic latitude the {\it Planck} observed values shift towards lower variances more rapidly than the increase of the width of the distribution due to sampling variance because of the smaller observed sky fraction considered. In other words, the observed values are more and more unlikely and for the extreme case, i.e.~Ext$_{30}$ mask, we find a compatibility with $\Lambda$CDM model only at $0.3 \%$ C.L. for the \texttt{Commander} map and $0.5 \%$ for \texttt{SMICA}. 
This is dominated by $V_{+}$ which is constantly low, independently on the considered sky fraction. Indeed, for \texttt{Commander}, its LTP varies around $0.3-0.5\%$, for all the considered sky fractions lower than the Std 2018 one. For \texttt{SMICA}, instead, its LTP varies in a slightly higher but still low range $[0.5\%\,,1.1\%]$.
On the other hand, $V_-$ is more sensitive to the sky fraction, decreasing monotonically as one takes into account regions at higher and higher Galactic latitude. However, its LTP remains inside the 1~$\sigma$ dispersion of the MC's, reaching $\sim11\%$ in the Ext$_{30}$ mask, independently from the employed CMB solution. 

The fact that the LTP of $V$ decreases when using more aggressive masks suggests that the low power of the {\it Planck} data is somehow anisotropically distributed on the map. In other words, the increasing discrepancy of the data with respect to $\Lambda$CDM when we exclude from the analysis pixels around the Galactic plane, indicates a sort of ``localisation'' of most of the power around the Galactic plane itself.

Moreover, Fig.~\ref{fig:MC_variance} and \ref{fig:p_value_MC_seme} show that, at large angular scales, such a low-Galactic-latitude power turns out to be dominated by the odd 
multipoles,
see also \cite{Gruppuso:2017nap}.

\subsection{Variance analyses including rotations}

\begin{figure}[t]
	\hspace{-0.85cm}
	\centering
	\subfloat{\includegraphics[width=.35\textwidth]{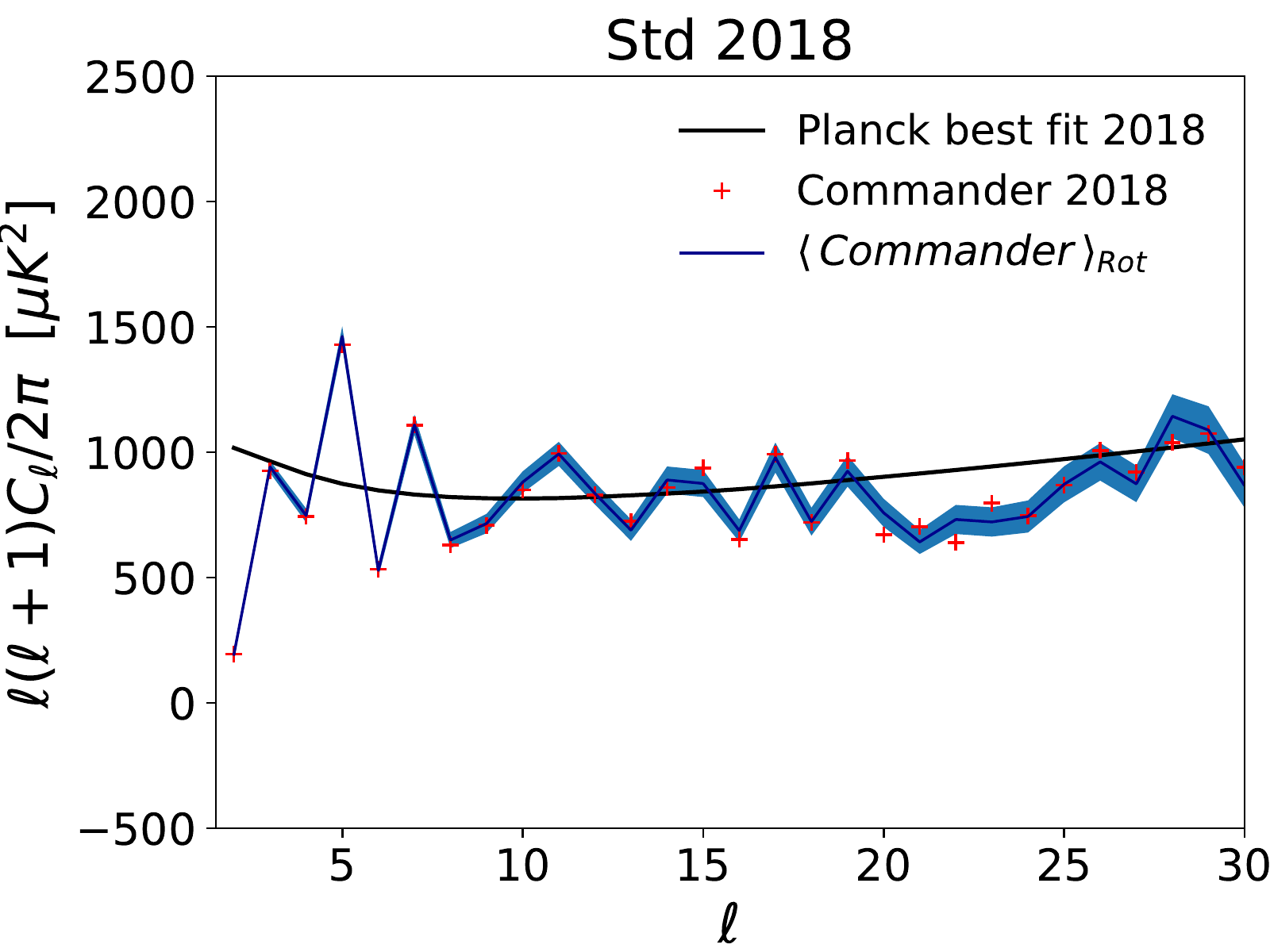}}
	\subfloat{\includegraphics[width=.35\textwidth]{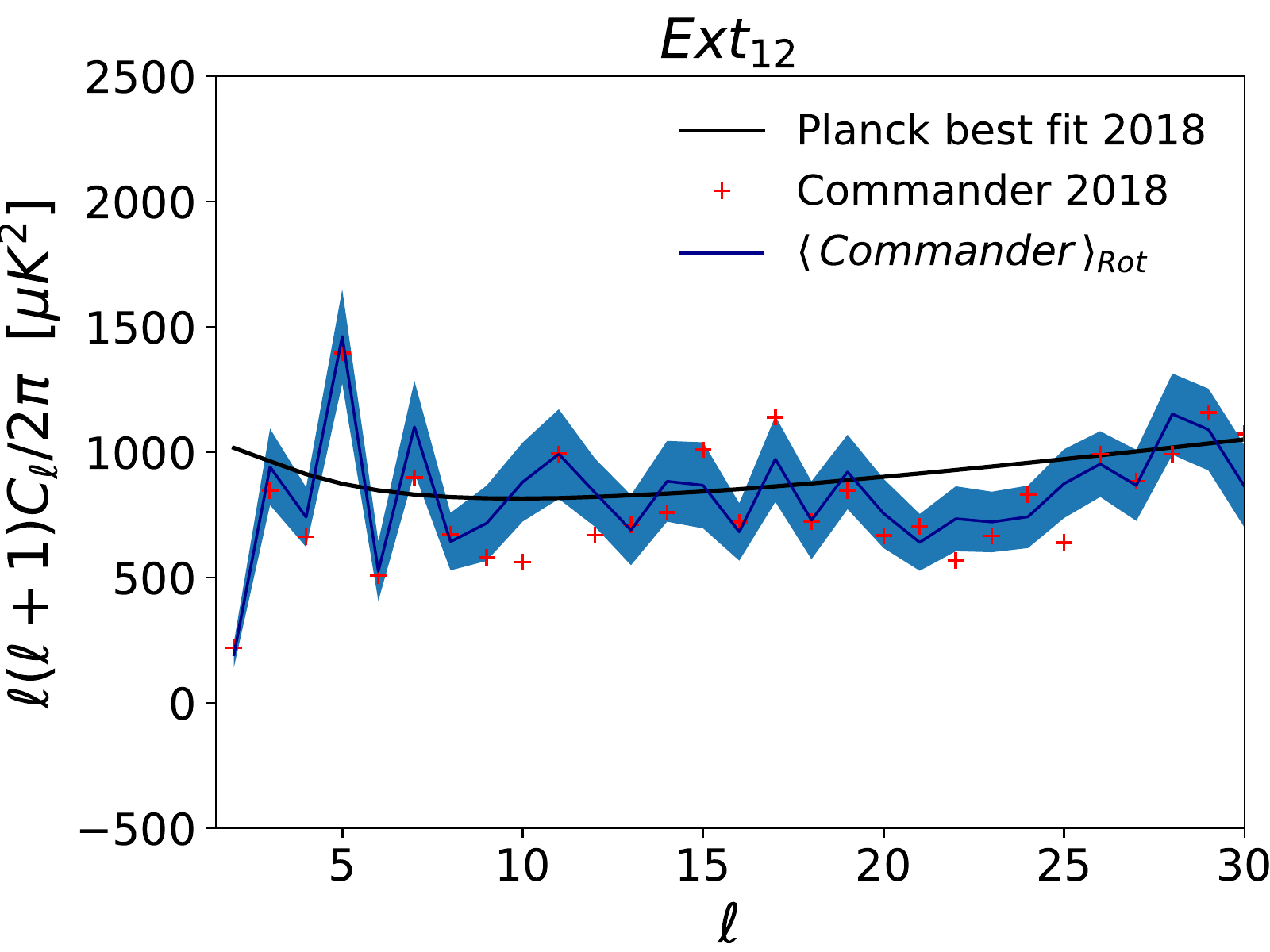}}
	\subfloat{\includegraphics[width=.35\textwidth]{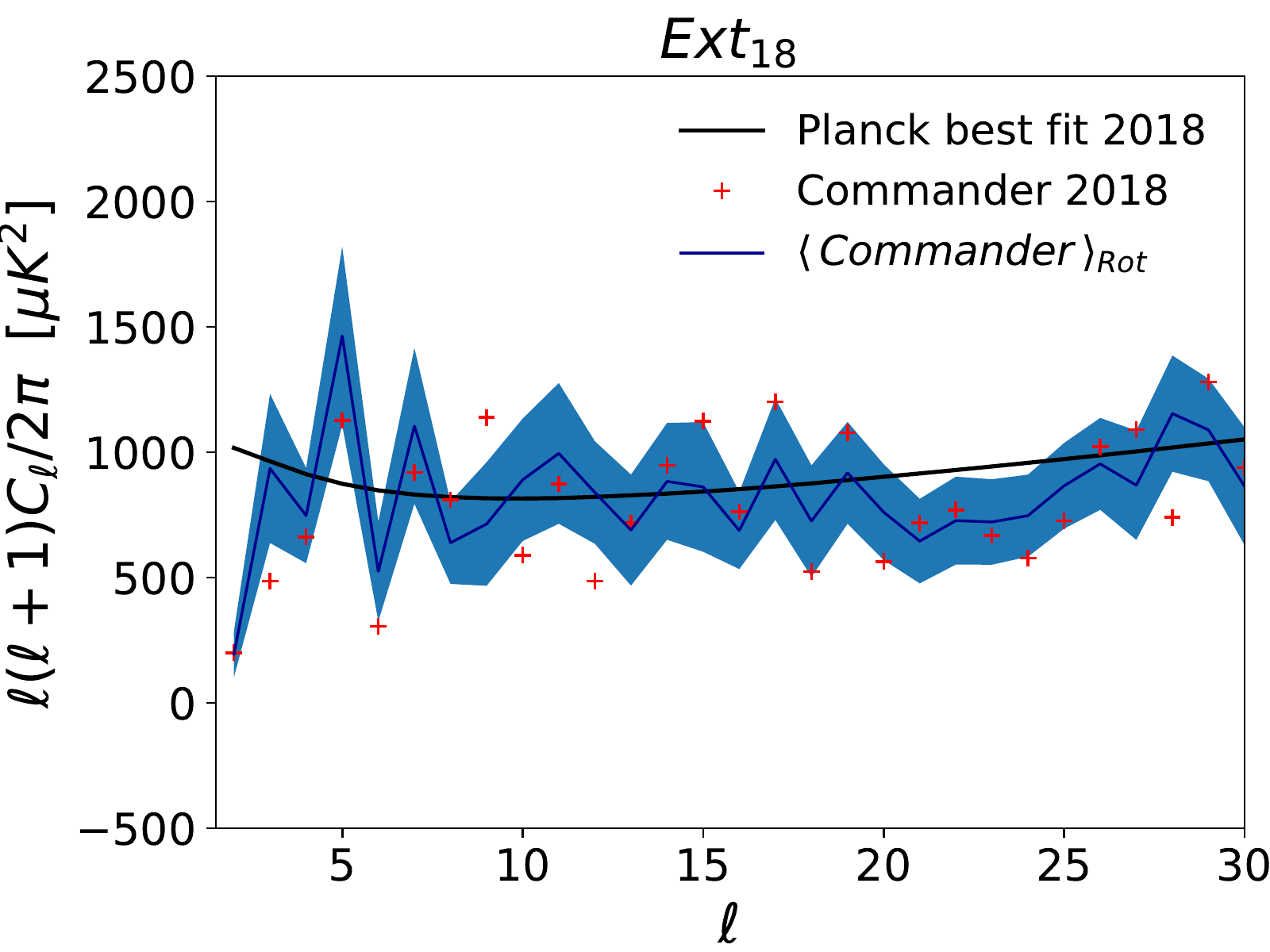}} \\
	\subfloat{\includegraphics[width=.35\textwidth]{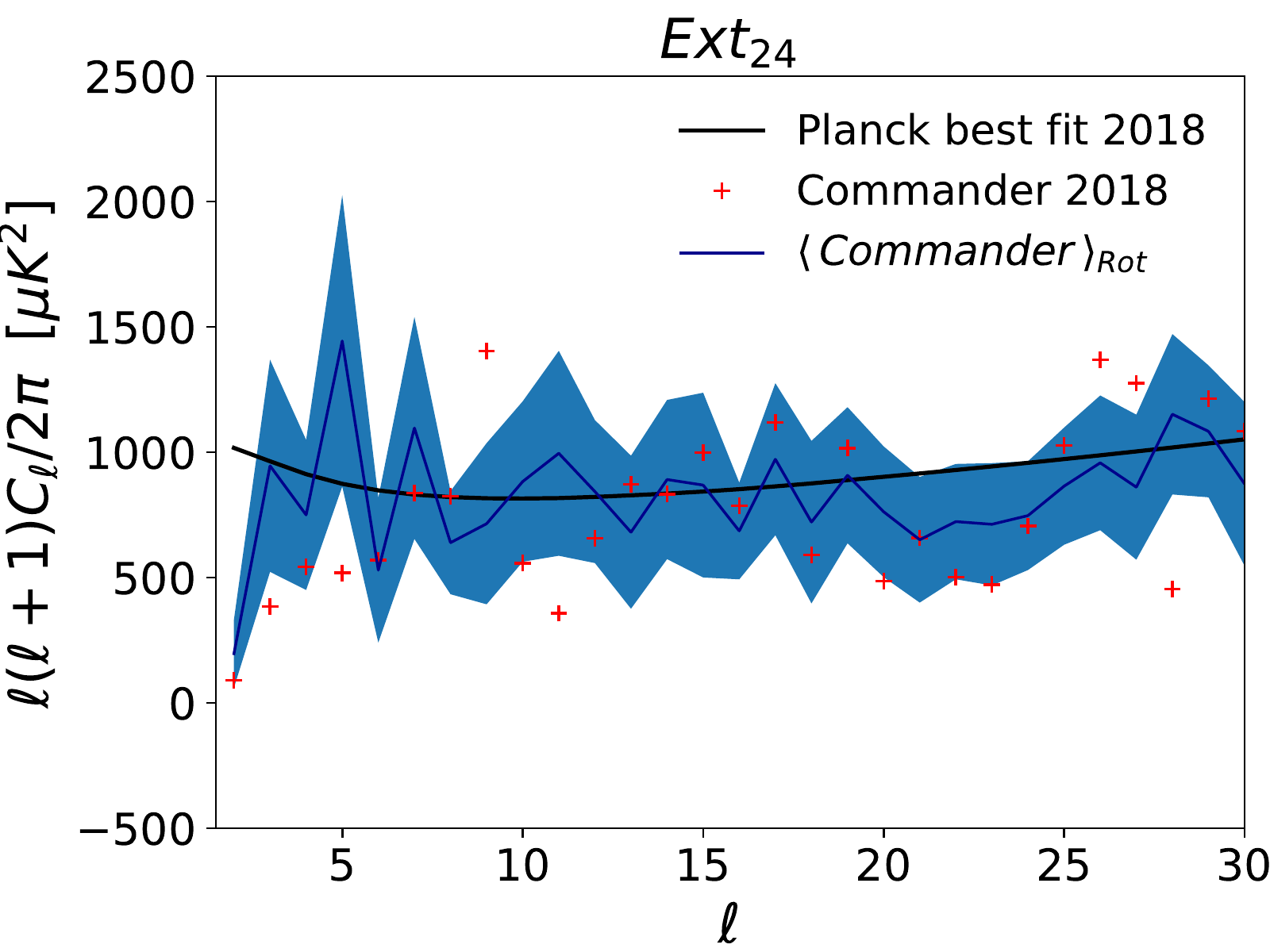}}
	\subfloat{\includegraphics[width=.35\textwidth]{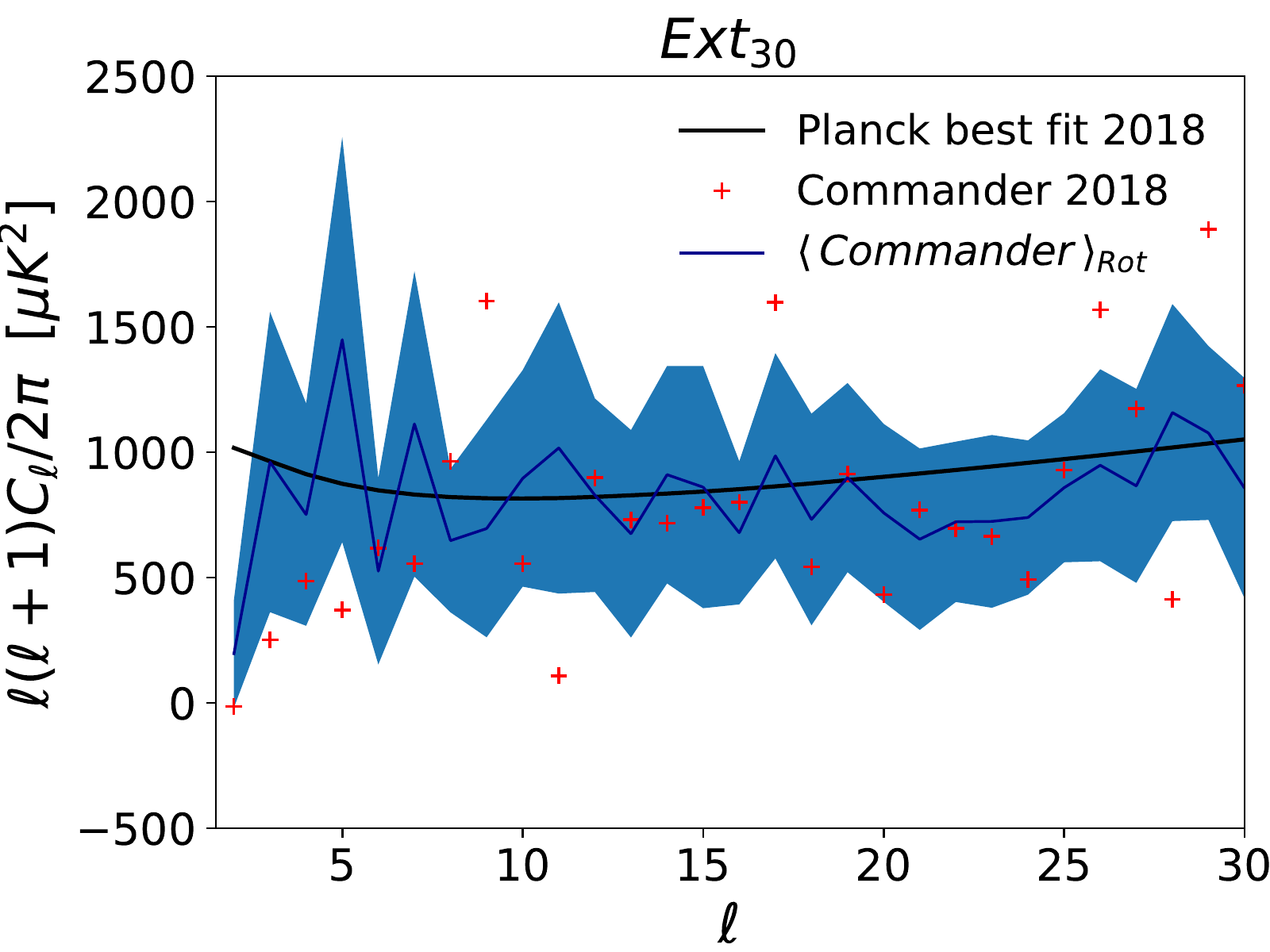}}
	\caption{Each panel shows the TT APS of the \texttt{Commander} 2018 map  estimated using different masks (red symbols). Blue line and blue region are respectively the average and the standard deviation of 10$^3$ random rotations of the \texttt{Commander} 2018 map. Note that in each mask the MC average is equal to the estimates obtained in the Std 2018 mask demonstrating that the variance is a mathematical object invariant under rotations only on average: the presence of a mask breaks the rotational symmetry for the single realization.}
	\label{fig:Commander_rotated}
\end{figure}

We now further investigate the dependency of $V$ with respect to the Galactic mask by implementing random rotations of the maps (see Appendix \ref{sec:rotations} for details which include the validation). This is performed in order to evaluate among all the possible orientations what is the probability of having most of the power at low Galactic latitude. The above procedure can be seen as a sort of look-elsewhere effect on the orientation of the mask. For computational reasons we reduce the number of MC simulations by considering the \ensemble\ 0 made of $10^3$ maps generated from the \textit{Planck} 2018 best-fit model. Note that $V$ is invariant under rotation of the input maps by construction only in the full sky case. In fact, when a mask is applied, the variance $V$ is not conserved under rotation for a single realisation but invariance is restored only on ensemble average. This effect is nicely captured already at the angular power spectrum level: in Fig.~\ref{fig:Commander_rotated} each panel shows the average and the statistical uncertainty at 1$\sigma$ of the TT spectra of 10$^3$ random rotations of the \texttt{Commander} map for the various masks\footnote{We obtain a similar behaviour for \texttt{SMICA} that is not shown here for sake of brevity.}. Notice that the APS estimates obtained with the Std 2018 mask are recovered only on average (blue lines) in the other masks. Moreover, as expected, the standard deviation (blue region) increases as the mask gets larger, allowing less observed sky for the analysis. 
In addition, still in Fig.~\ref{fig:Commander_rotated} we show the TT spectrum of the \texttt{Commander} 2018 map without any rotation (red symbols).

We analyse random rotations of the \ensemble\ 0 and corresponding observed data building two estimators, the LTP-estimator (Section \ref{sec:LTP_estimator0}) and the $r$-estimator (Section \ref{sec:r-estimator0}). With the former we investigate separately for each mask how anomalous is the particular orientation of the Galactic plane. With the latter we quantify the statistical significance of the lowering trend of $V$ with respect to its value in the Std 2018 mask with all the possible orientations.
 
\subsubsection{LTP estimator}
\label{sec:LTP_estimator0}

\begin{figure}[t]
		\hspace{-0.85cm}
	\centering
\subfloat{\includegraphics[width=.35\textwidth]{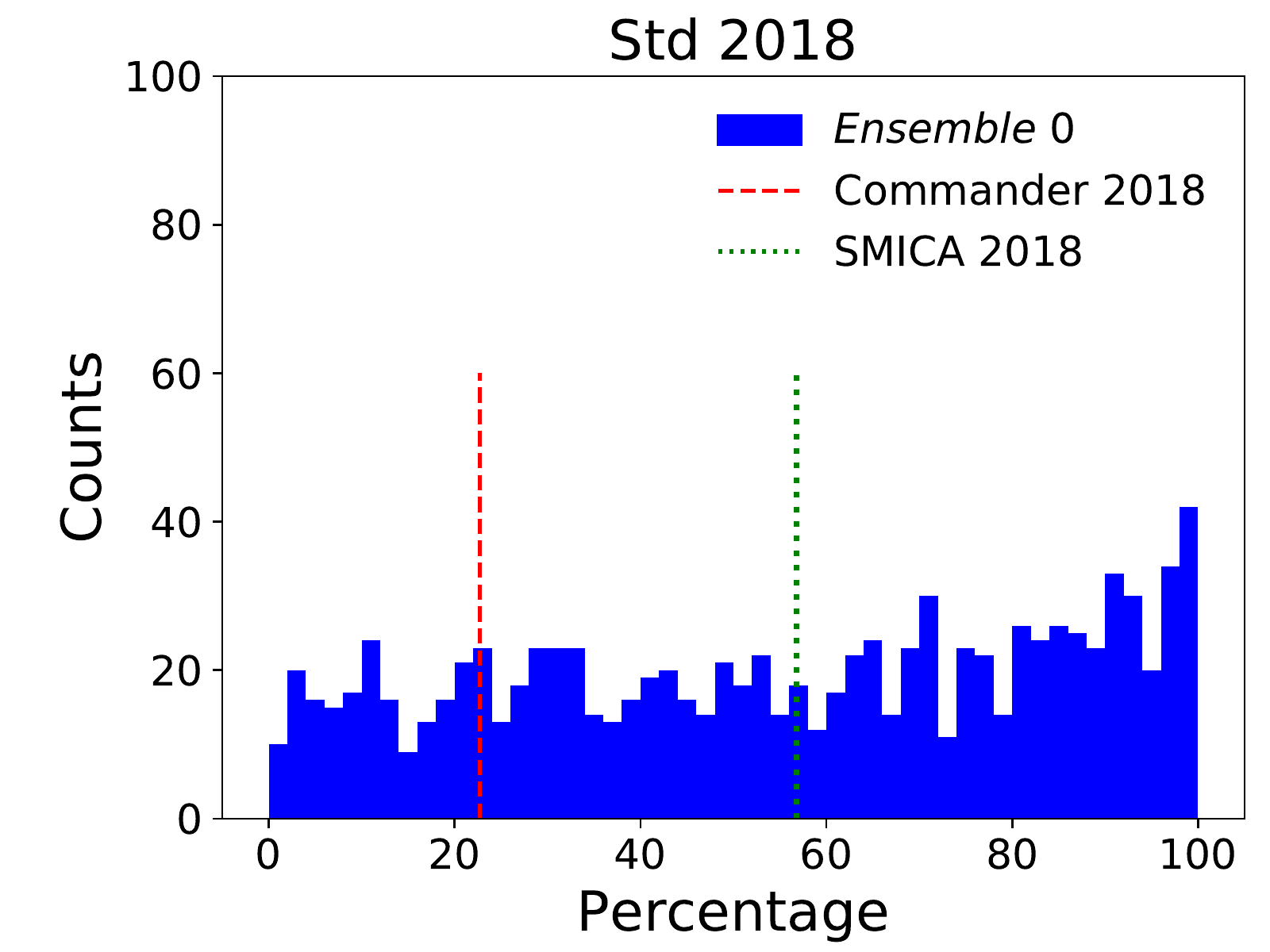}}
\subfloat{\includegraphics[width=.35\textwidth]{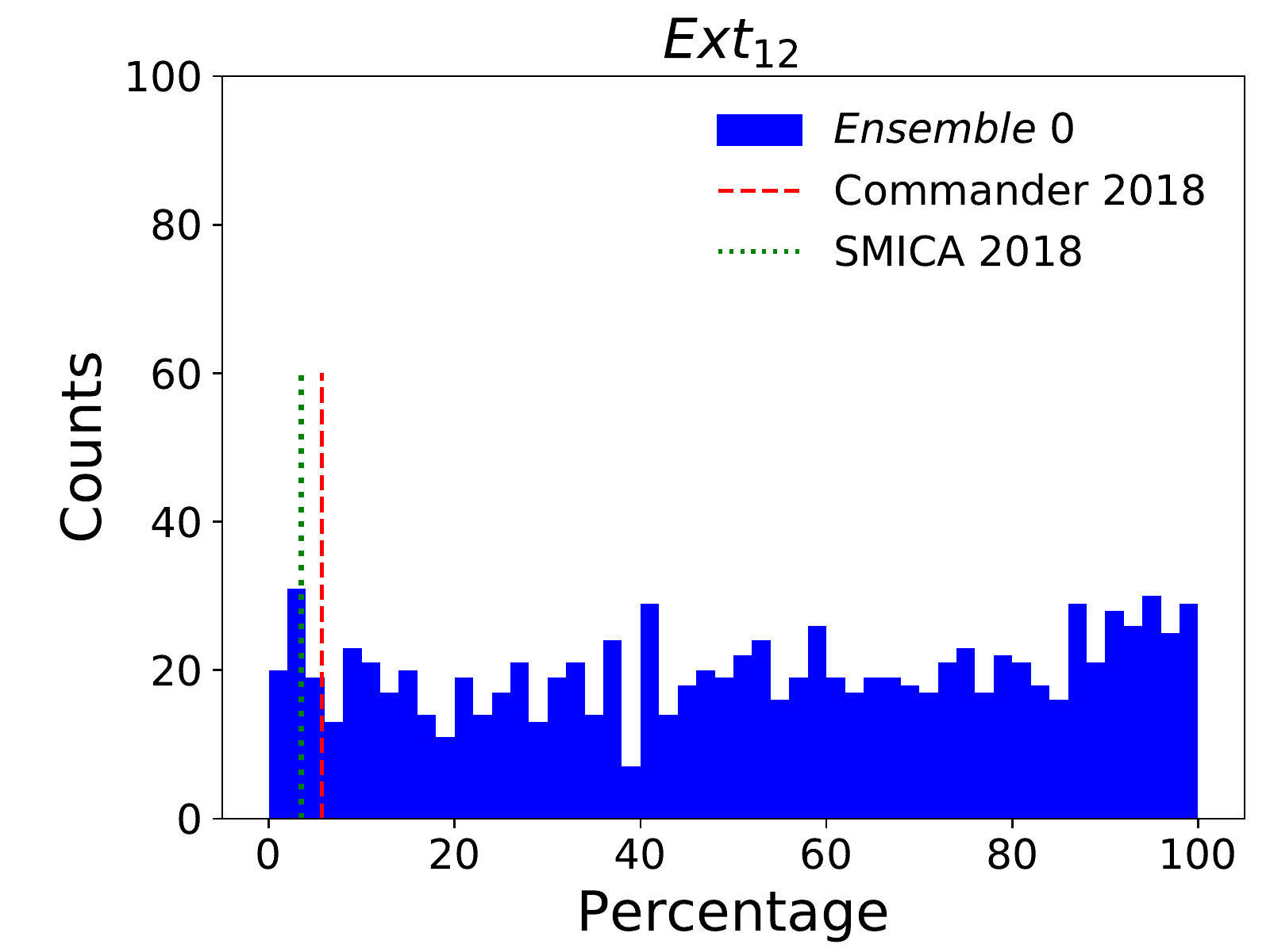}}
\subfloat{\includegraphics[width=.35\textwidth]{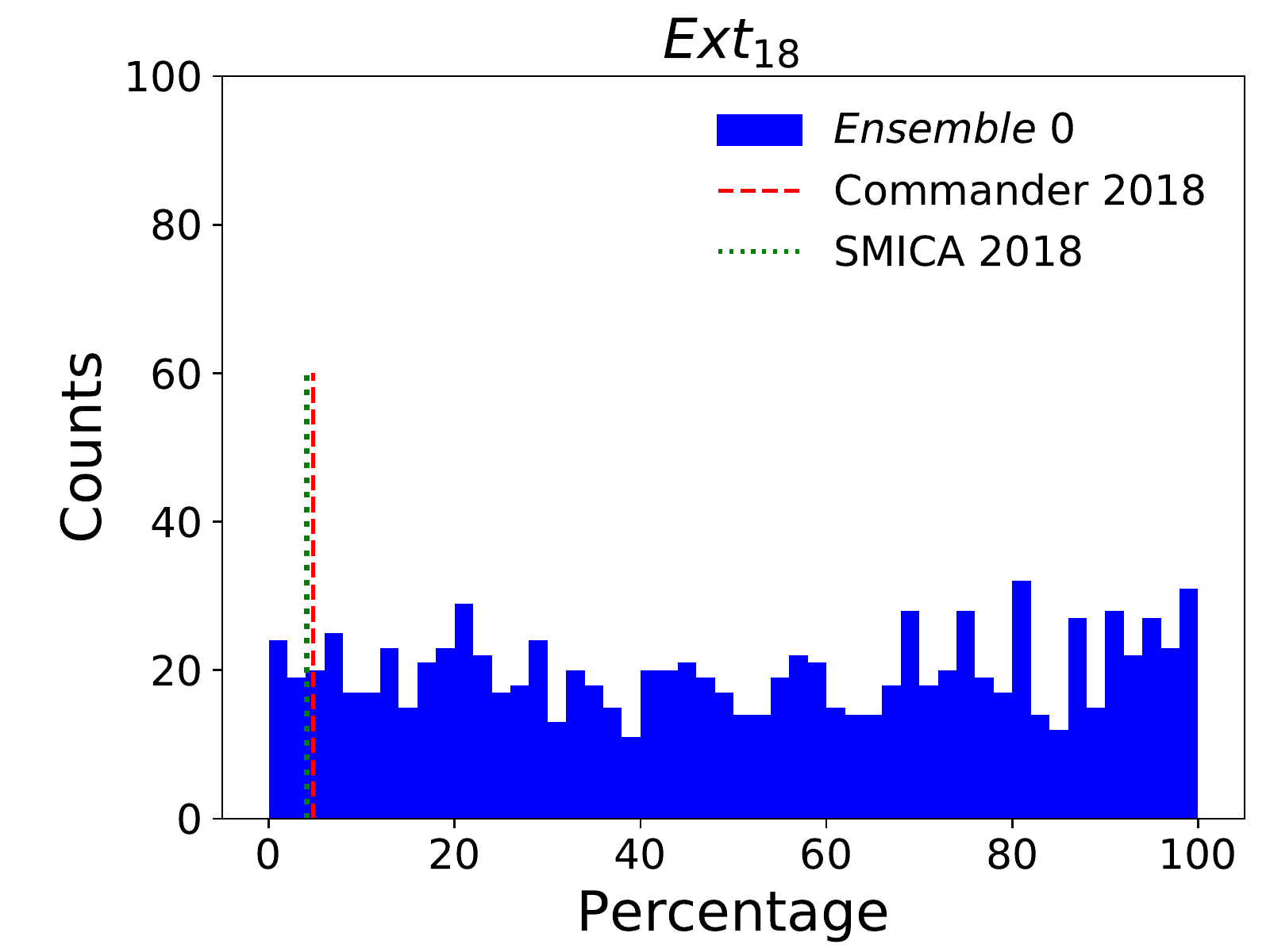}} \\
\subfloat{\includegraphics[width=.35\textwidth]{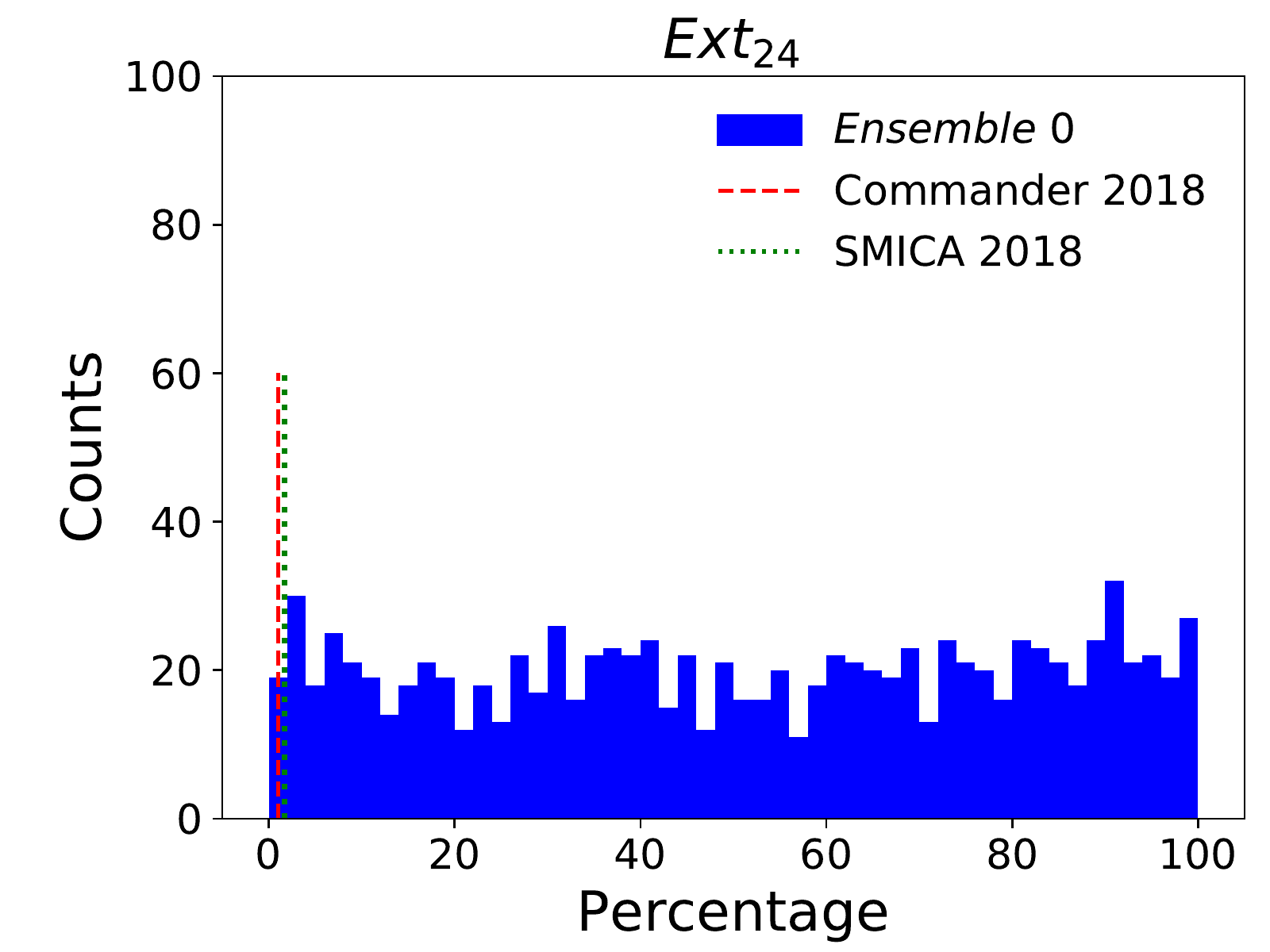}}
\subfloat{\includegraphics[width=.35\textwidth]{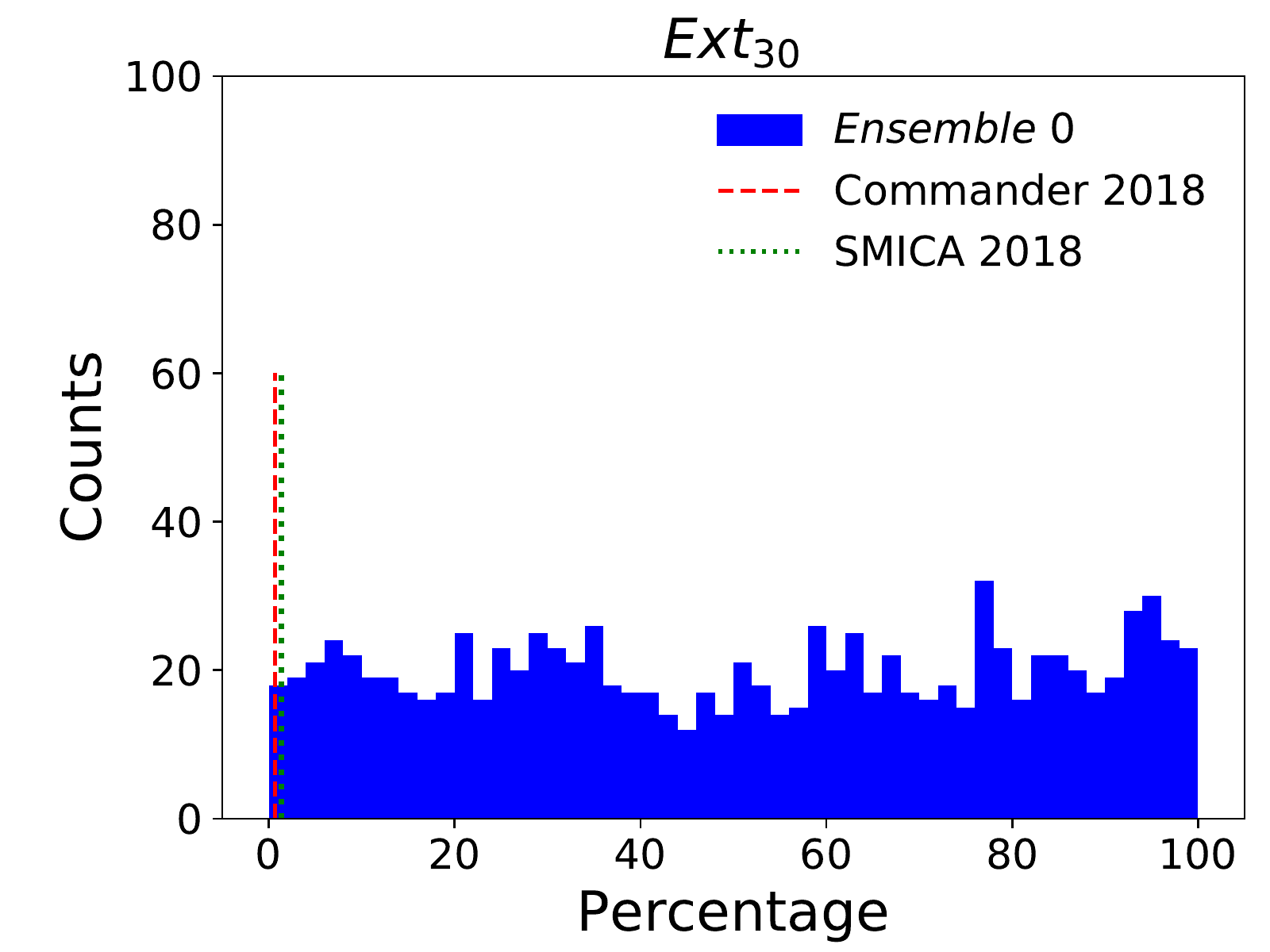}}
\caption{Histograms of the LTP of finding a rotated map of the \ensemble\ 0 with $V^{rot}<V$, where $V$ is the variance of the corresponding unrotated map. Each panel shows the results obtained using a different mask. Red dashed and green dotted vertical bars are the LTP for \texttt{Commander} and \texttt{SMICA} respectively.}\label{fig:p_value_rot_1000_case_separate_LCDM}
\end{figure}  

For each map $\textbf{m}_i$ belonging to \ensemble\ 0 we build the histogram of $V_i$ obtained through 10$^3$ random rotations of that map. Hence, we compute the LTP of that map $\textbf{m}_i$, denoted with LTP$_i$, with respect to the corresponding set of rotations. This can be repeated for $i=1,...,10^3$, i.e. for all the maps of the \ensemble\ 0 and for all the considered masks. Thus, for each mask, we obtain a MC of 10$^3$ values of LTP representing the distribution of probabilities expected in a $\Lambda$CDM model. Since the variance does not depend on the orientation, the distribution of LTP is expected to be uniform, that is, each LTP is equiprobable. 
The empirical distribution of the LTP-estimator for each considered mask shown in Fig.~\ref{fig:p_value_rot_1000_case_separate_LCDM} confirms our expectations.
In the same Figure we also show the LTP obtained from {\it Planck} data as vertical bars, red for \texttt{Commander} and green for \texttt{SMICA}. 
The corresponding values are reported in left panel of Table \ref{tab:p_value_comm_SMICA_ruotati_LCDM}. 
When we consider higher Galactic latitude, we find that the probability of observing 
a LTP with respect to its rotations lower than the corresponding LTP of  \texttt{Commander} (\texttt{SMICA}) 2018 is anomalous at $\sim 2.8\,\sigma$ ($\sim2.5\,\sigma$). 
Indeed, in the Ext$_{30}$ case, only 5 (13) out of 10$^3$ maps of the \ensemble\ 0 have a lower LTP than the \texttt{Commander} (\texttt{SMICA}) 2018 map, i.e. only in the $0.5\%$ (1.3\%) of the cases the anomaly associated to the power localisation around the Galactic plane is higher than data (see right panel of Tables \ref{tab:p_value_comm_SMICA_ruotati_LCDM}).

\begin{table}[t]
	\centering
\begin{minipage}{.5\linewidth}
\begin{tabular}{ l | c | c  }
	\hline
	\hline
	& \multicolumn{2}{c}{\textbf{LTP} [\%]} \\
	\cline{2-3}
	Mask & $V^{(\text{rot})}_\textrm{c}<V_\textrm{c}$& $V^{(\text{rot})}_\textrm{s}<V_\textrm{s}$\\
	\hline
	Std 2018 & 22.7 & 56.8\\
	
	Ext$_{12}$ & 5.7 & 3.5\\
	
	Ext$_{18}$ & 4.8 & 4.1\\
	
	Ext$_{24}$ & 1.0 & 1.7\\
	
	Ext$_{30}$ & 0.7 & 1.4\\
	\hline 
	\hline
\end{tabular}
\end{minipage}%
\begin{minipage}{.5\linewidth}
\begin{tabular}{ l | c | c  }
	\hline
	\hline
	& \multicolumn{2}{c}{\textbf{LTP} [\%]} \\
	\cline{2-3}
	Mask & $\textrm{LTP}_i<\textrm{LTP}_\textrm{c}$ & $\textrm{LTP}_i<\textrm{LTP}_\textrm{s}$\\
	\hline
	Std 2018 & 18.8 & 49.1\\
	
	Ext$_{12}$ & 6.7 & 4.5\\
	
	Ext$_{18}$ & 4.7 &4.4\\
	
	Ext$_{24}$ & 1.0 & 1.5\\
	
	Ext$_{30}$ & 0.5 & 1.3\\
	\hline 
	\hline
\end{tabular}
\end{minipage} 
\caption{Left table: The probability of obtaining a value of the variance of the rotated \texttt{Commander} map (second row), $ V^{(\text{rot})}_\textrm{c}$, and rotated \texttt{SMICA} map (third row), $ V^{(\text{rot})}_\textrm{s}$, smaller than the unrotated one, $V_\textrm{c}$ and $V_\textrm{s}$ respectively. Right table: LTP of obtaining a simulation of the \ensemble\ 0 with LTP lower than the one obtained with the \texttt{Commander} map, $\textrm{LTP}_\textrm{c}$, or \texttt{SMICA} map, $\textrm{LTP}_\textrm{s}$.}\label{tab:p_value_comm_SMICA_ruotati_LCDM}
\end{table}


\subsubsection{$r$-estimator}\label{sec:r-estimator0}
\begin{figure}[t]
	\centering
	{\includegraphics[scale=0.46]{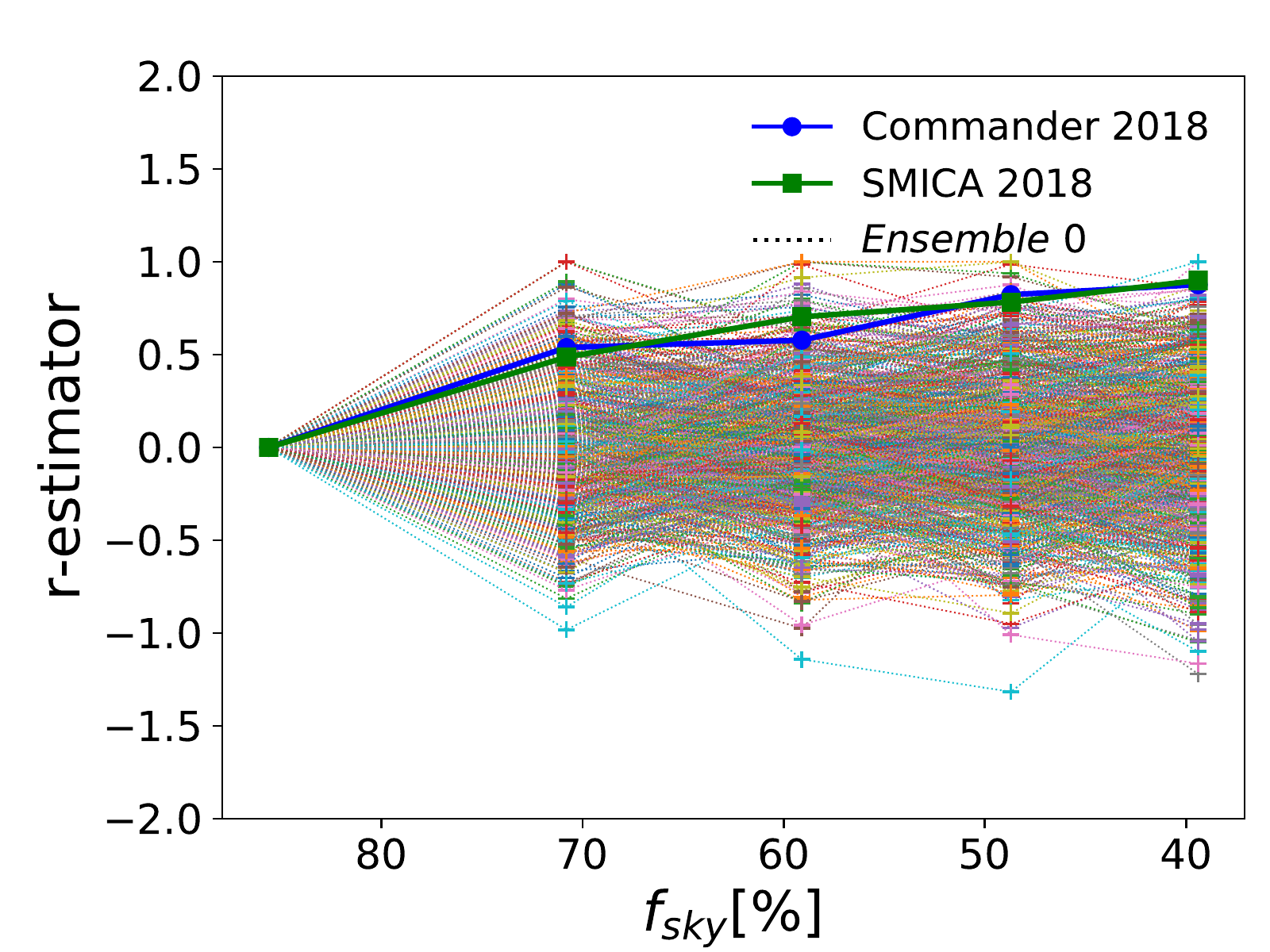}}
	{\includegraphics[scale=0.46]{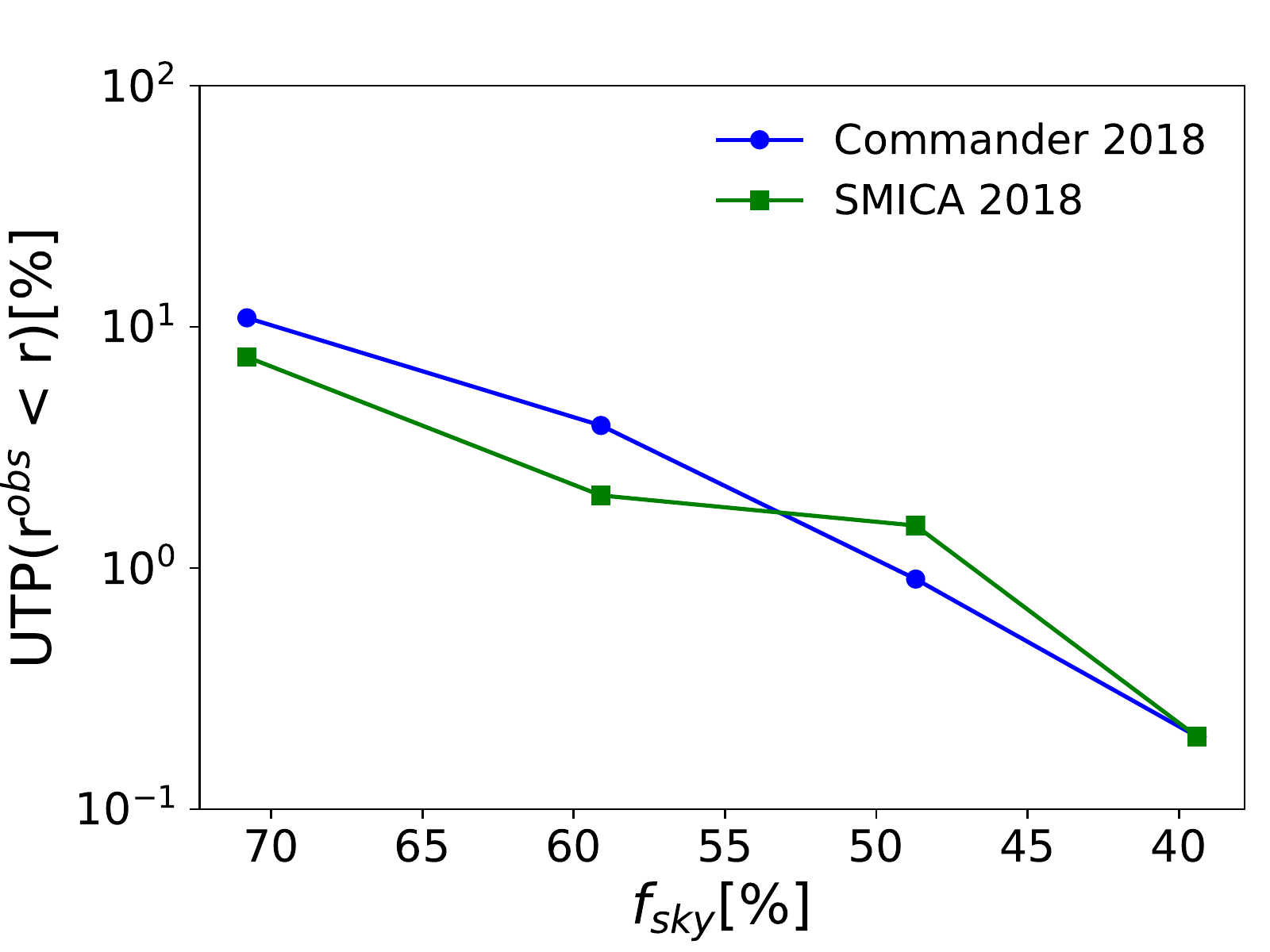}}
	\caption{Left panel: $r$-estimator computed with Eq. (\ref{eqn:r_estimator}) versus the sky fraction. The coloured dotted lines stand for the $r$ value obtained from the \ensemble\ 0. Blue and green solid lines stand for \texttt{Commander} and \texttt{SMICA} respectively. Right panel: UTP of obtaining a simulation with $r$ larger than the one obtained with \texttt{Commander} (blue line) or \texttt{SMICA} (green line) as a function of the sky fraction.} \label{fig:r_v_std_meno_v_j_LCDM}
\end{figure}

We use here the $r$-estimator defined as
\begin{equation}\label{eqn:r_estimator}
r\equiv\frac{ V_{std}-V_{mask}}{\underset{\mathrm{j \in rotations}}{\mathrm{max}} \left\{V^{(j)}_{std}-V^{(j)}_{mask} \right\}} \, ,
\end{equation}
where $V_{std}$ is the variance computed in the Std 2018 mask, while $V_{mask}$ is the variance computed in one of the other four extended masks. 
The numerator of Eq.~(\ref{eqn:r_estimator}) fixes the sign of the $r$-estimator as determined by the decrease ($r > 0$), or increase ($r< 0$), of the variance as we widen the Galactic mask. This behaviour is normalised by the denominator, which picks up the maximum decrease among all the rotations\footnote{In the denominator of $r$ we include also the unrotated case, denoted here as the 0$^{th}$ rotation.}. The $r$-estimator is therefore upper bounded by 1, but it can become lower than -1. 
In other words, the $r$-estimator represents the fractional change of $V$, computed in an extended mask with respect to the Std 2018 mask value, relative to the maximum decrease across rotations. 
For example, $r=0.5$ means that, we are dealing with a map which, in a given mask, has a variance difference with respect to the standard mask equal to exactly half of the maximum difference which can be found among all rotations. 
In the left panel of Fig.~\ref{fig:r_v_std_meno_v_j_LCDM} we show the $r$-estimator for all the considered cases. Dotted lines connect the MC values of $r$ represented with a plus symbol. 
Solid blue line connects the \texttt{Commander} 2018 values (dot symbols) and the solid green line connects the \texttt{SMICA} 2018 values (square symbols).
For this estimator we consider the upper tail probability, UTP, defined as the fraction of simulations with larger values of $r$ than the observed one. They are shown in the right panel of Fig.~\ref{fig:r_v_std_meno_v_j_LCDM} and quoted in Table \ref{tab:r_value_LCDM}.
Notice that both \texttt{Commander} and \texttt{SMICA} present an increase of $r$ for higher and higher Galactic latitudes and in the Ext$_{30}$ case, they are close to 1, being $r^c=0.88$ for \texttt{Commander} and $r^s=0.90$ for \texttt{SMICA}.
This means that the observed maps in the Ext$_{30}$ case are almost aligned to the direction which maximizes the lowering of $V$ obtainable through rotations. 
The probability corresponding to this event is $0.2 \%$ for both \texttt{Commander} and \texttt{SMICA}.
This leads to an anomalous value of $r$ at a level of $3.1 \,\sigma$.

\begin{table}[t]
\begin{center}
\begin{tabular}{ l | c | c  }
	\hline
	\hline
	& \multicolumn{2}{c}{\textbf{UTP} [\%]} \\
	\cline{2-3}
	Mask & $r^{\textrm{c}}<r$& $r^\textrm{s}<r$\\
	\hline
	Ext$_{12}$ & 10.9 & 7.5\\
	
	Ext$_{18}$ & 3.9 & 2.0 \\
	
	Ext$_{24}$ & 0.9 & 1.5\\
	
	Ext$_{30}$ & 0.2& 0.2\\
	\hline 
	\hline
\end{tabular}
\caption{UTP of obtaining a simulation of the \ensemble\ 0 with $r$ larger than the one obtained from the data. Second column shows the UTP for \texttt{Commander}, third column the UTP for \texttt{SMICA}.}\label{tab:r_value_LCDM}
\end{center}
\end{table}

\section{Analysis of $\Lambda$CDM simulations with low variance}
\label{analysis}

\begin{figure}[t]
	\hspace{-0.85cm}
	\centering
	\subfloat{\includegraphics[width=.35\textwidth]{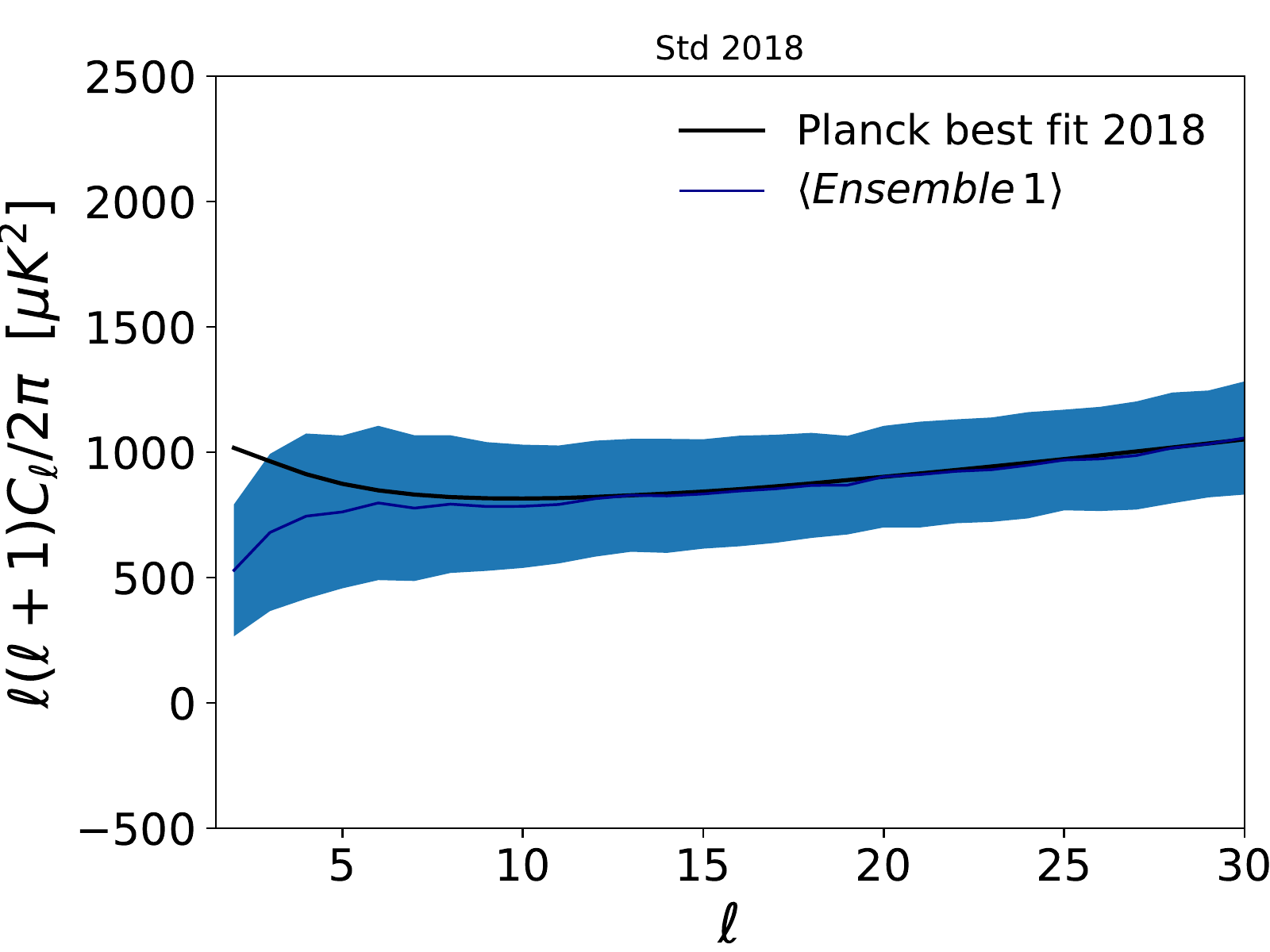}}
	\subfloat{\includegraphics[width=.35\textwidth]{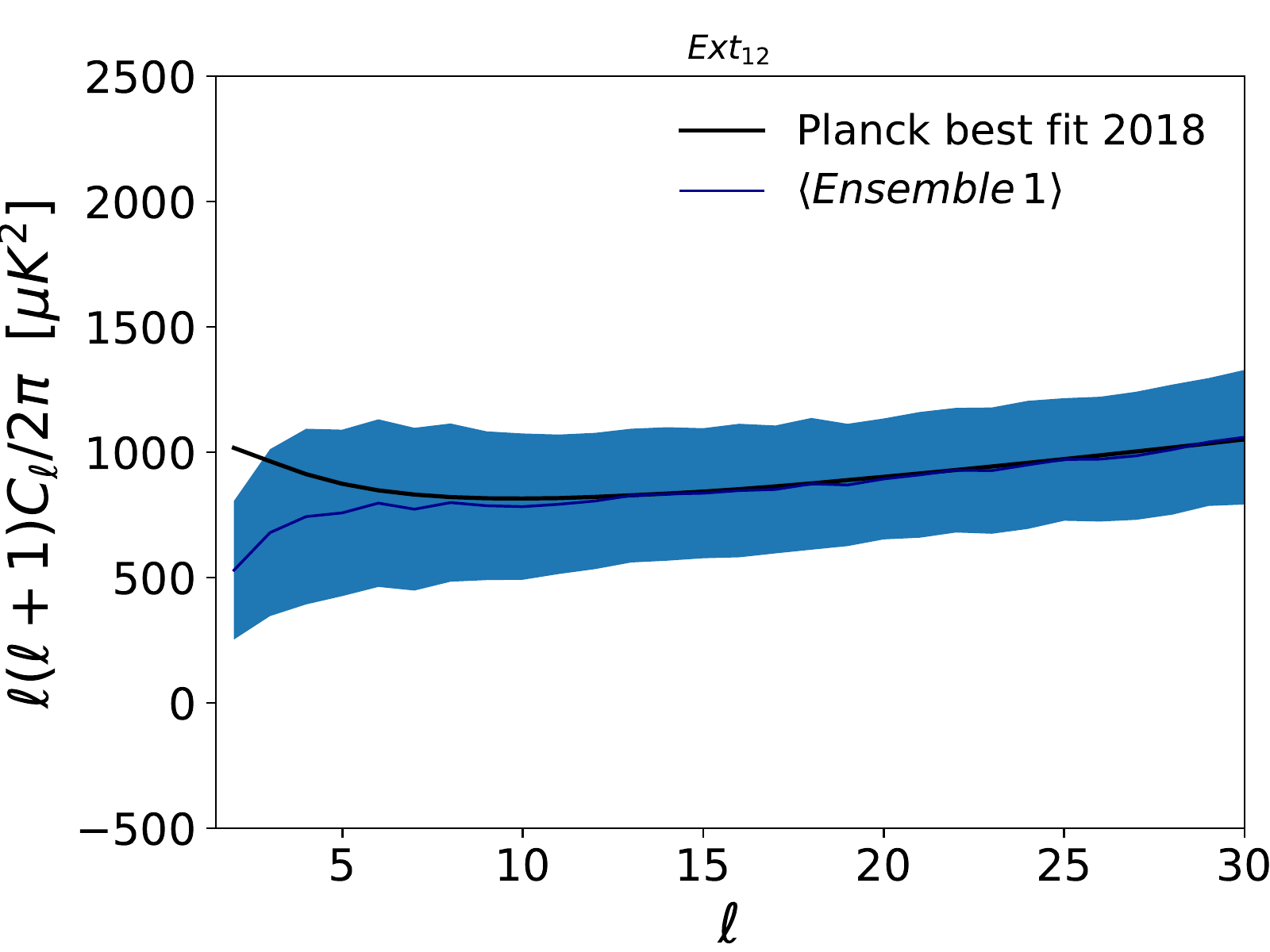}}
	\subfloat{\includegraphics[width=.35\textwidth]{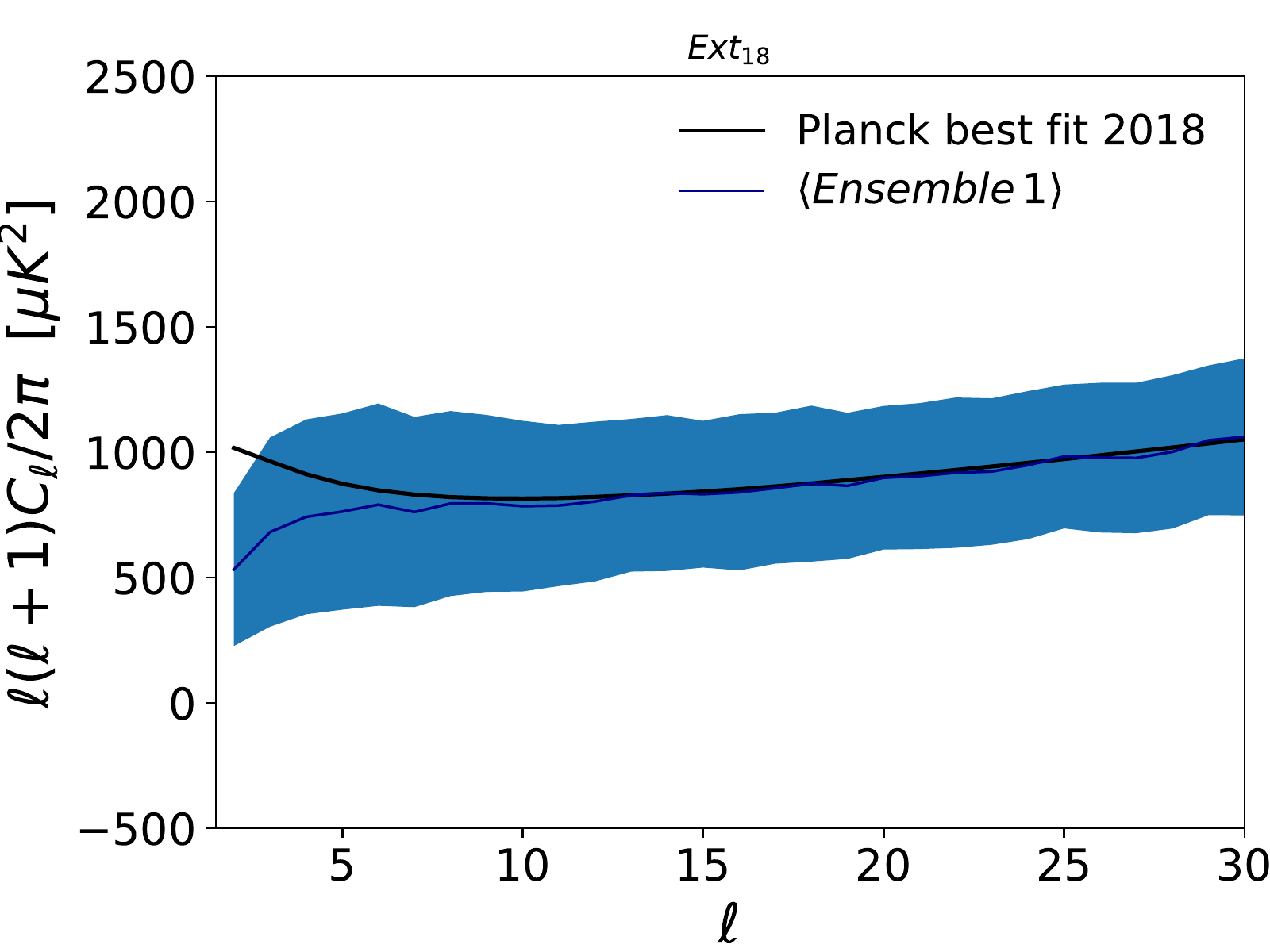}} \\
	\subfloat{\includegraphics[width=.35\textwidth]{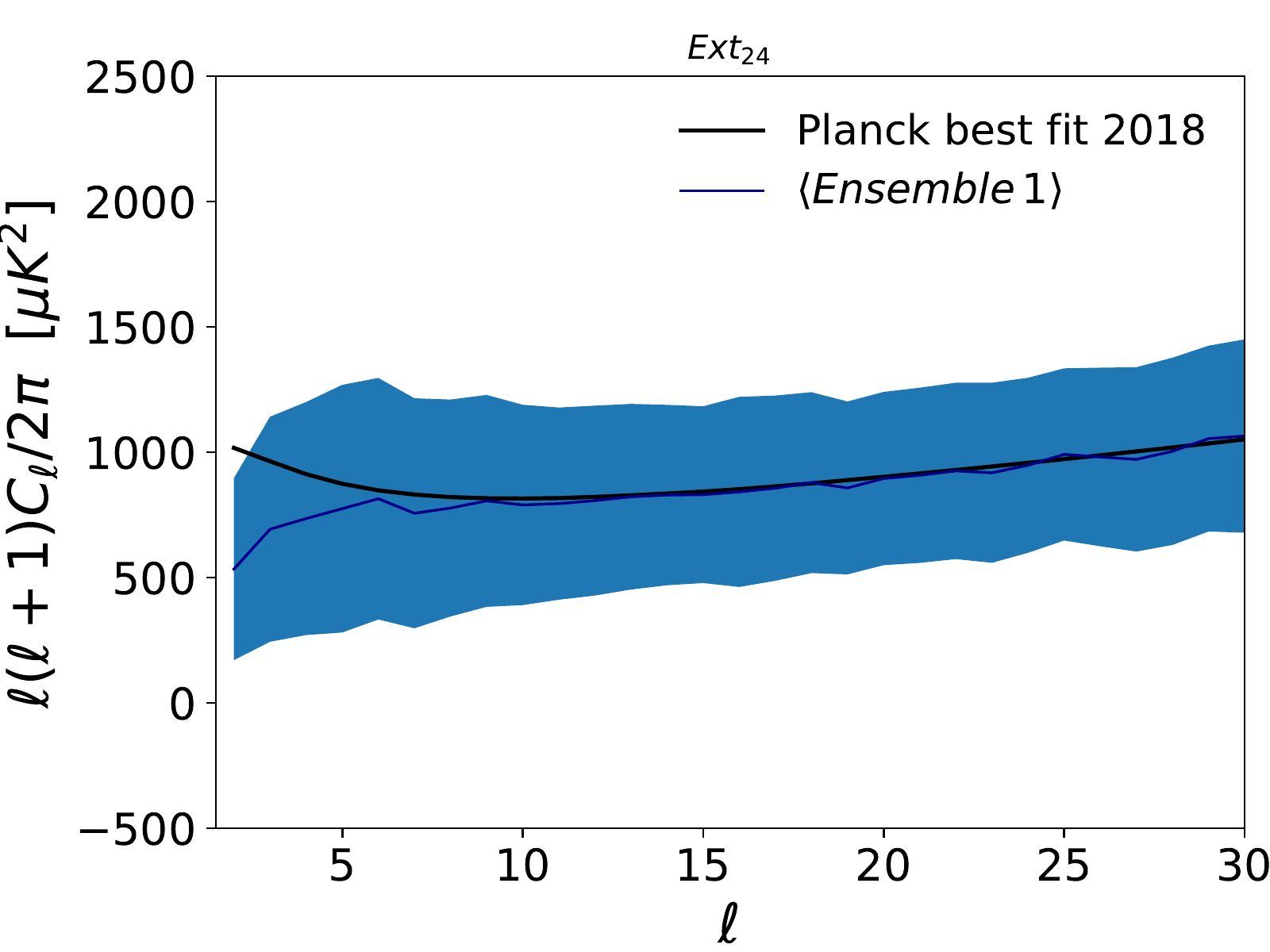}}
	\subfloat{\includegraphics[width=.35\textwidth]{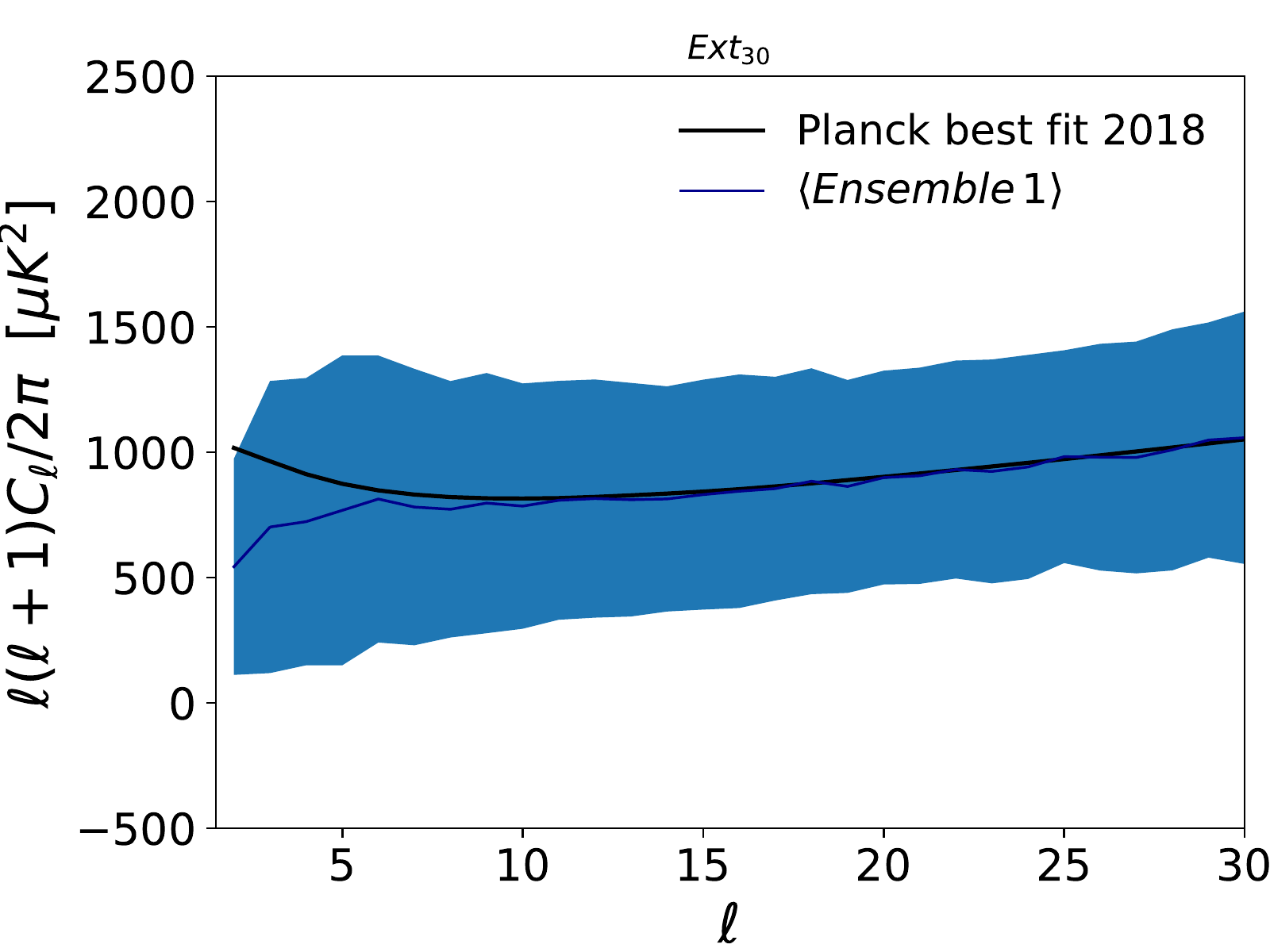}}
	\caption{Each panel shows the \textit{Planck} 2018 best-fit model (black solid line) and the average APS of \ensemble\ 1 (blue line), with its 1$\sigma$ dispersion (blue region) for all the considered masks.}
	\label{fig:MC_ensamble1}
\end{figure}

In this section we repeat the analysis performed in Section \ref{preliminary} but now considering simulated maps which have almost the same variance $V$ as the one observed by the CMB solutions (\texttt{Commander} and \texttt{SMICA}) of the {\it Planck} 2018 release. These are collected in the \ensemble\ 1, as described in Section \ref{dataset}.
The aim of this analysis is to check whether the previous results still hold when the variance is constrained to be low also across the simulations.
In other words we would like to exclude the possibility that the observed trend of a lowering variance when extending the Galactic mask, is connected to the low value of the variance measured in the Standard mask.
In Fig.~\ref{fig:MC_ensamble1} we display the \textit{Planck} 2018 best-fit model (black solid line) and the average of \ensemble\ 1 (blue line), with its standard deviation (blue region) for all the considered masks. 
Notice the increase of the statistical uncertainty as the observed sky fraction decreases. This figure shows that \ensemble\ 1 behaves differently from the fiducial power spectrum only at low-$\ell$.
In other words, selecting a subset of $\Lambda$CDM realisations with low variance is in fact equivalent to choosing maps with suppressed $C_{\ell}$ at low 
multipoles\footnote{Note that we recover empirically the well-known correlation between low-$V$ and low-$C_2$ anomalies \cite{Muir:2018hjv}.}.

\begin{figure}[t]
	\hspace{-0.85cm}
	\centering
	\subfloat{\includegraphics[width=.35\textwidth]{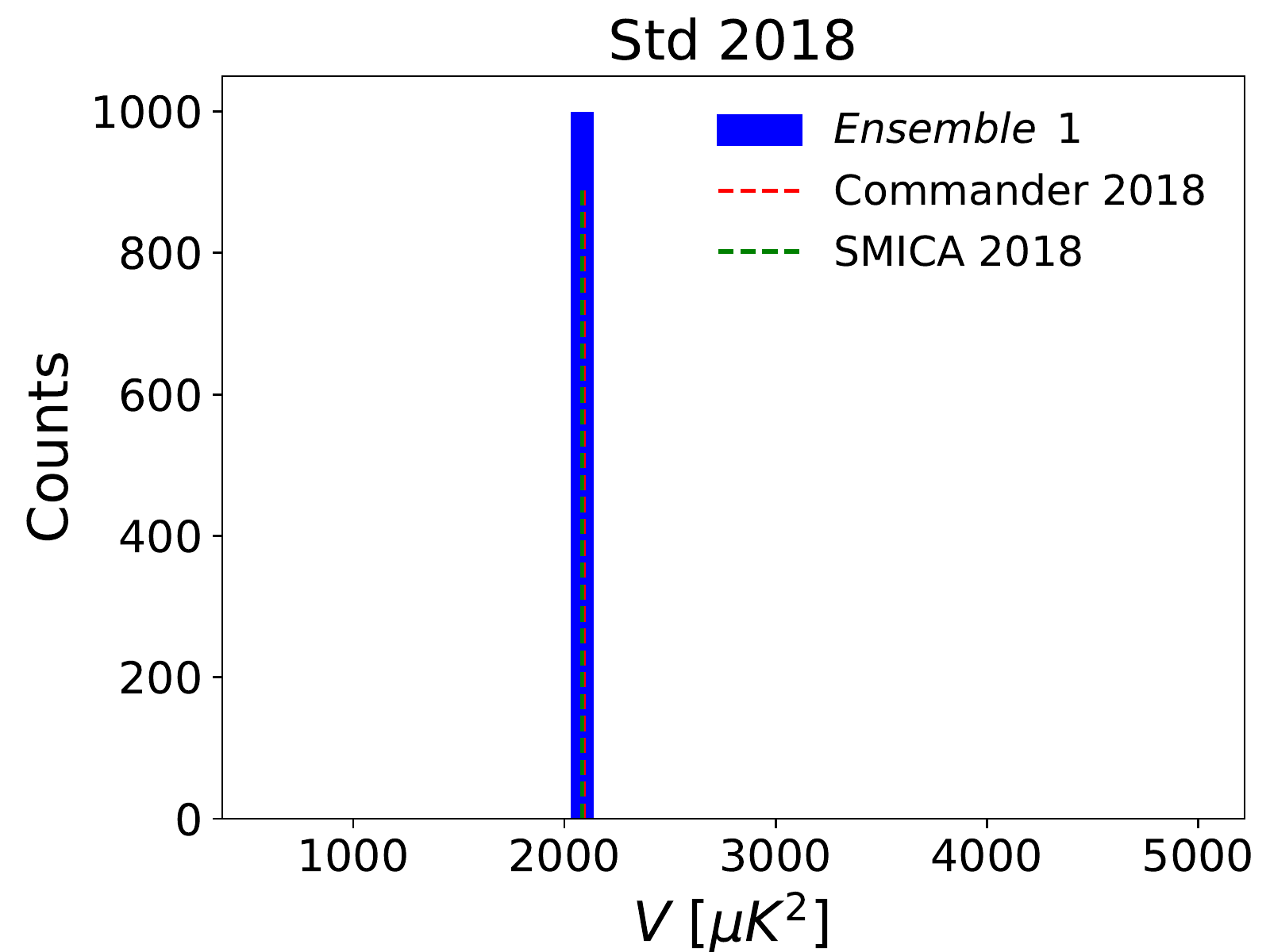}}
	\subfloat{\includegraphics[width=.35\textwidth]{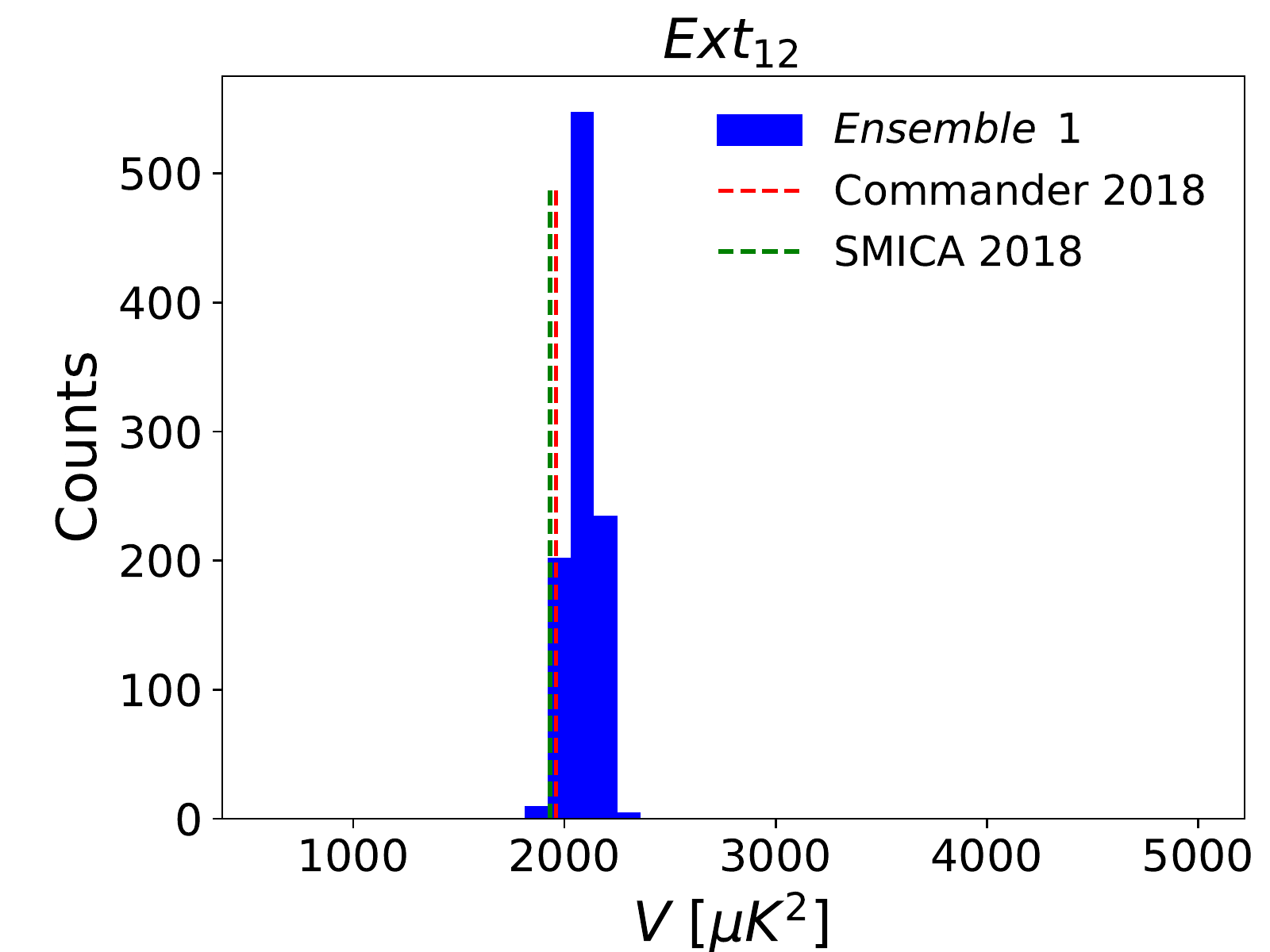}}
	\subfloat{\includegraphics[width=.35\textwidth]{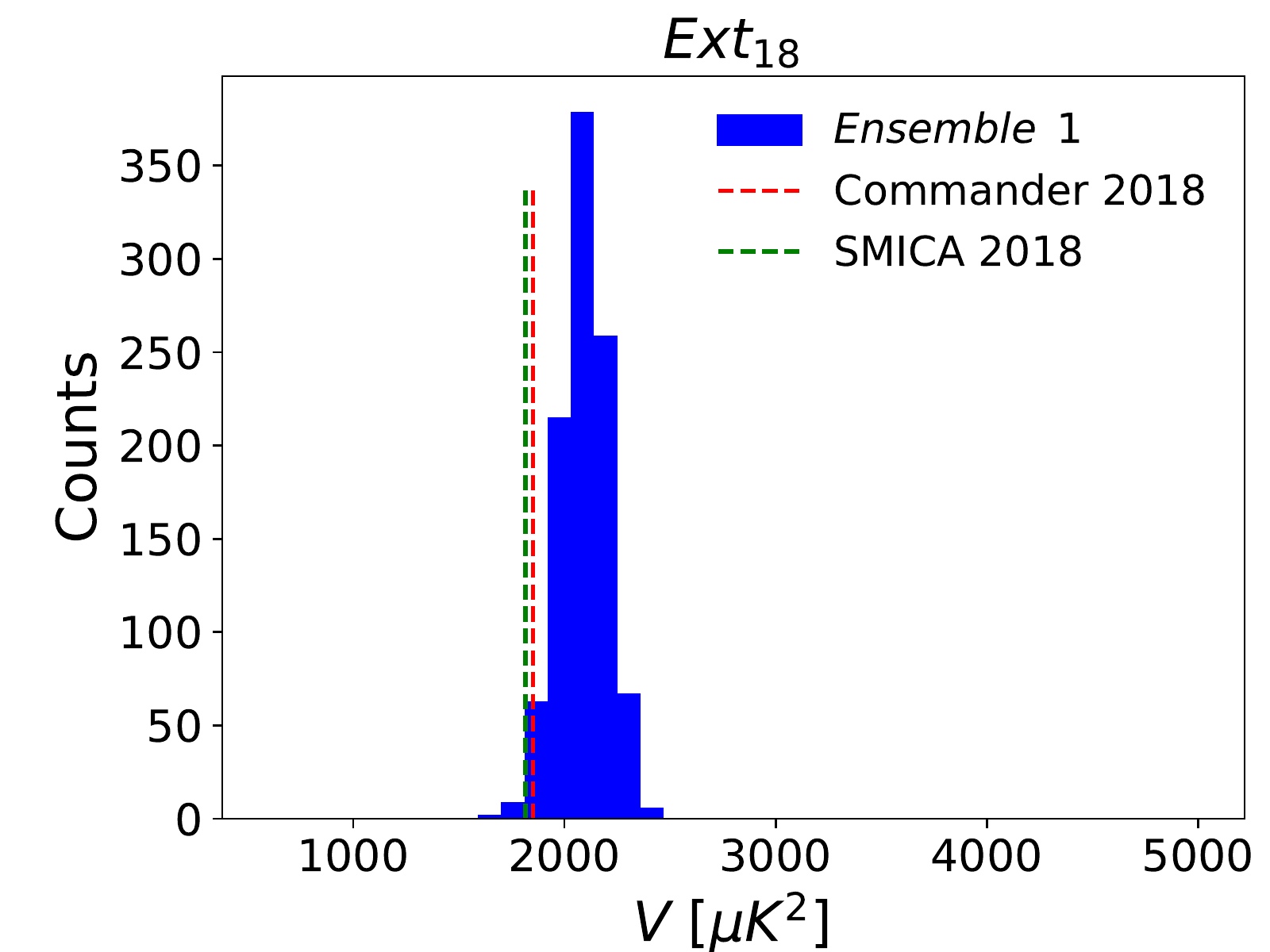}} \\
	\subfloat{\includegraphics[width=.35\textwidth]{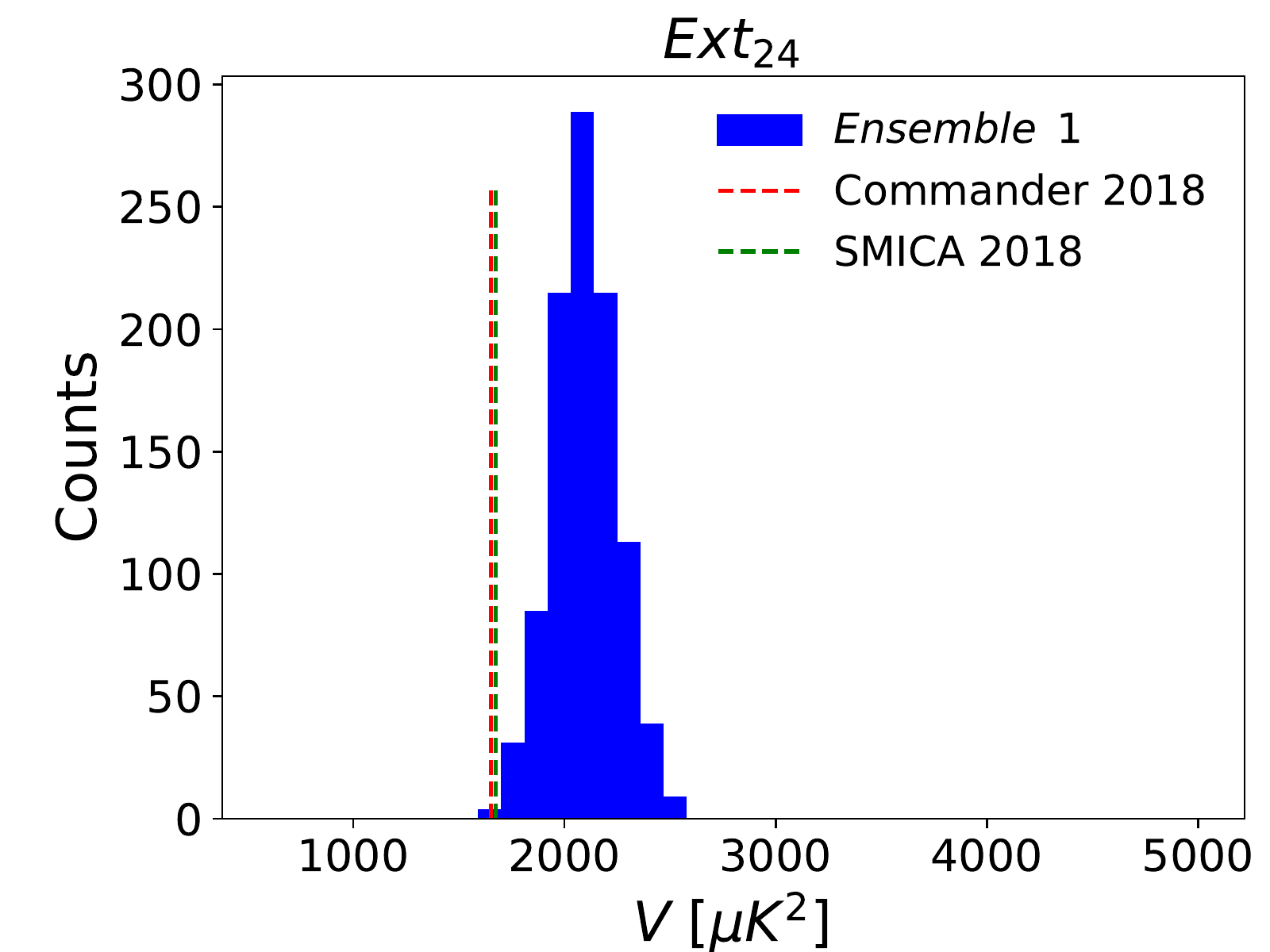}}
	\subfloat{\includegraphics[width=.35\textwidth]{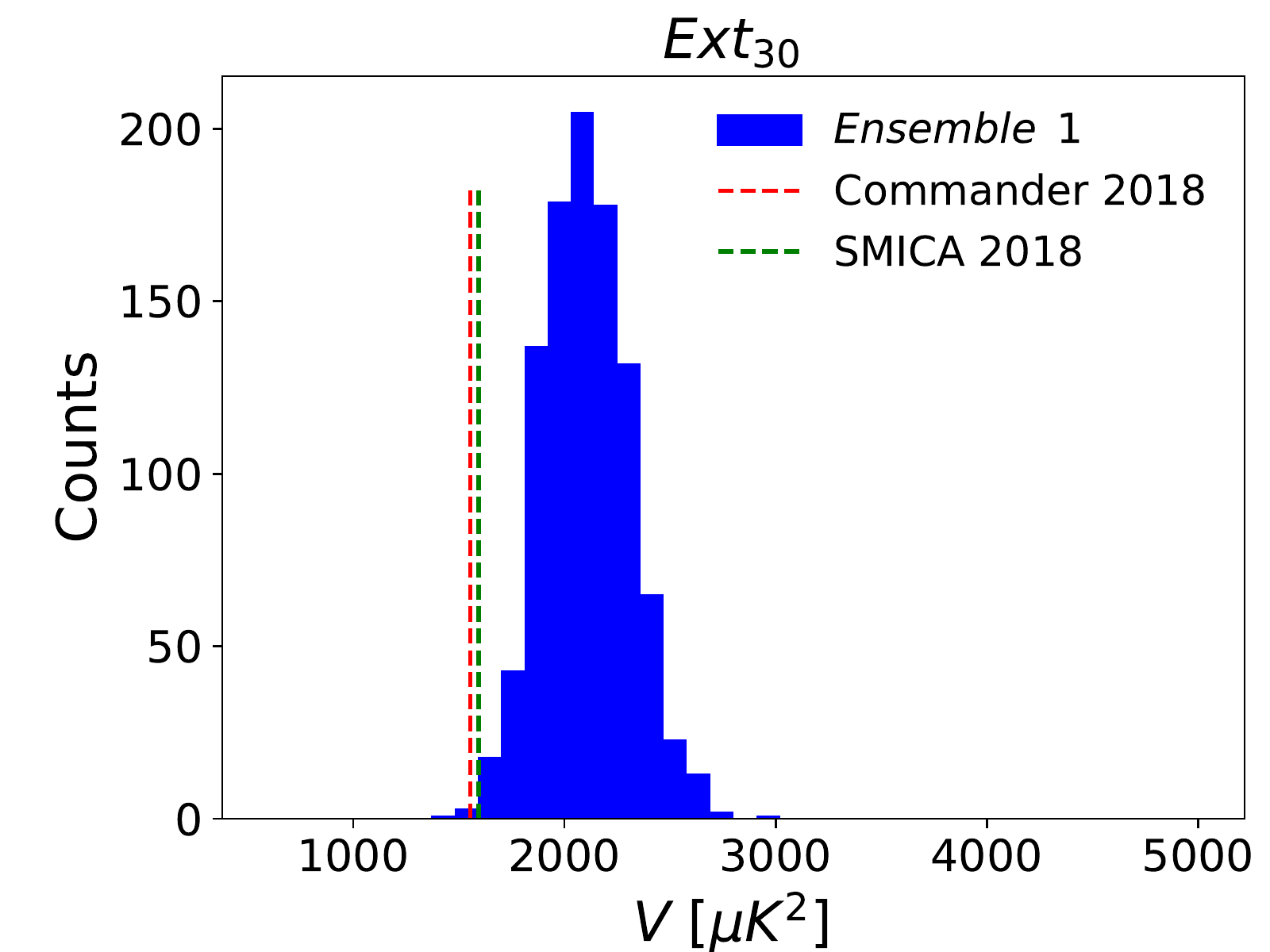}}
	\caption{Histograms of the variances $V$ of the maps belonging to the \ensemble\ 1 computed with the masks Std 2018, Ext$_{12}$, Ext$_{18}$, Ext$_{24}$ and Ext$_{30}$. The red dashed line identifies the variance of the \texttt{Commander} map, $V_{\textrm{c}}$. The green dashed line identifies the variance of the \texttt{SMICA} map, $V_{\textrm{s}}$.}\label{fig:var_comm_vs_var_1000_maps_same_variance}
\end{figure}

\begin{figure}[t]
	\centering
	\includegraphics[scale=0.46]{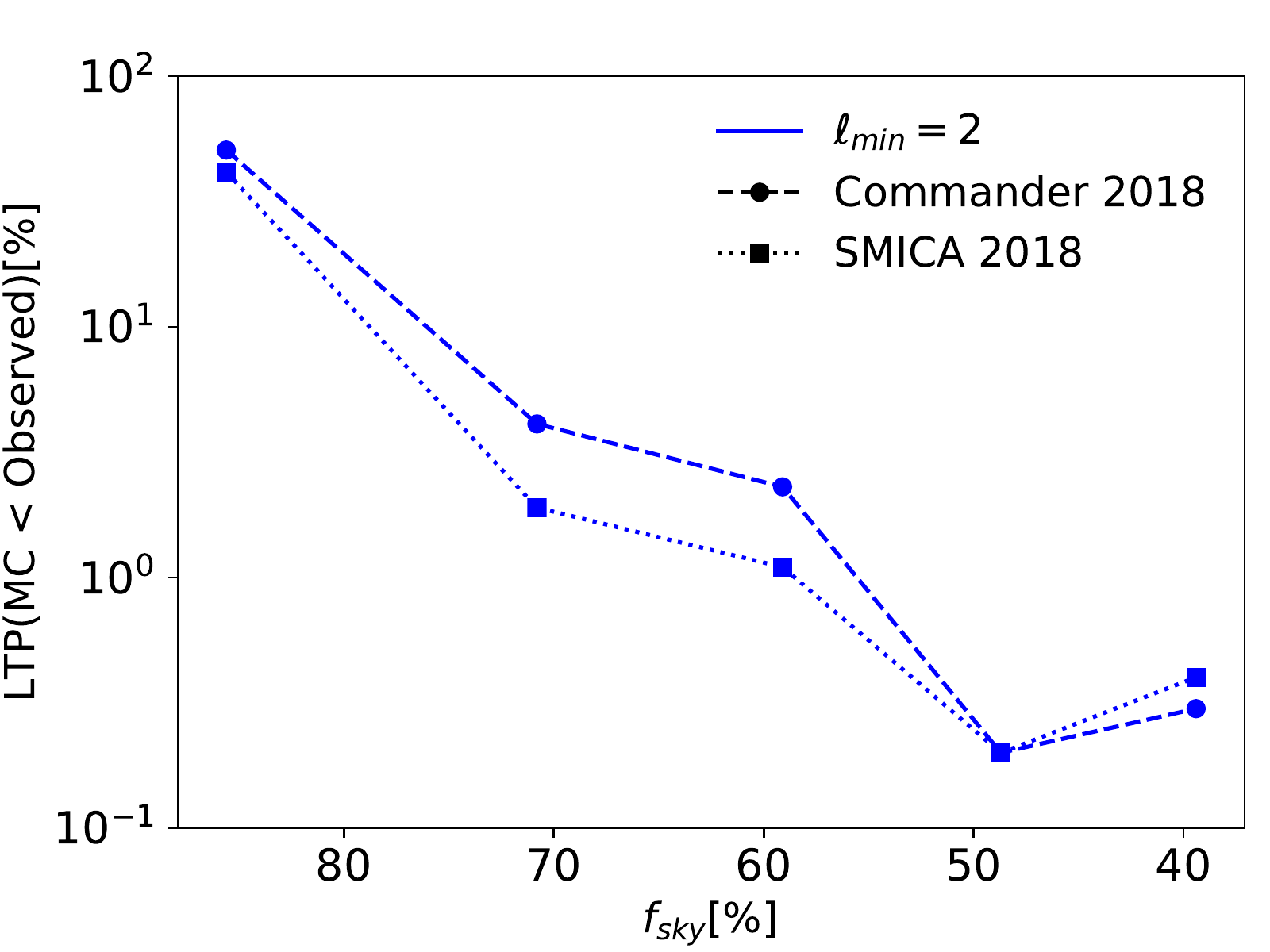} 
	\includegraphics[scale=0.46]{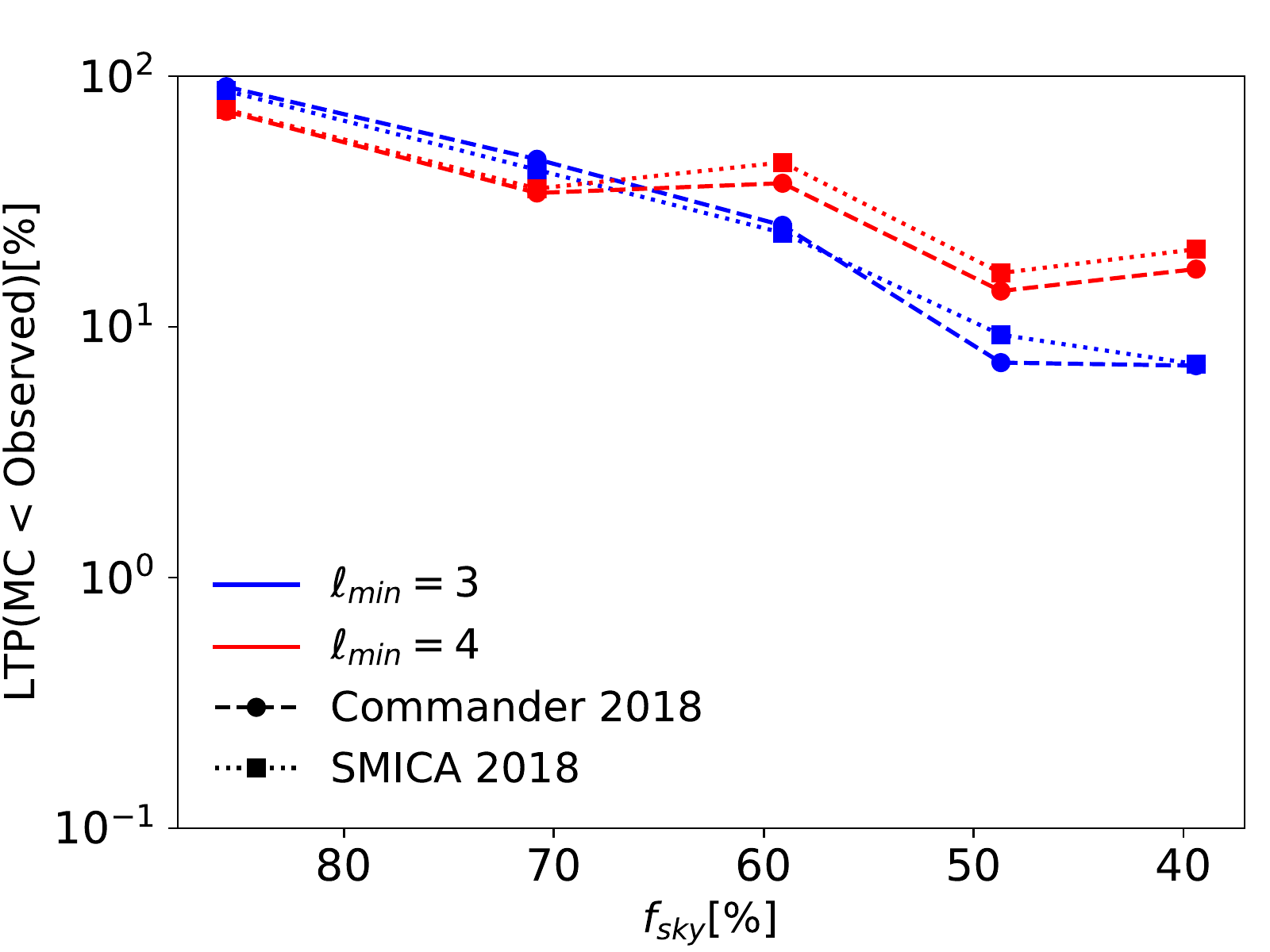}
	\caption{Right panel: LTP of the variance estimator for the {\it Planck} 2018 data in percentage as a function of the sky fraction. Left panel: the same as in right panel but with $\ell_{min}=3$ (blue line) or $\ell_{min}=4$ (red line).}\label{fig:var_comm_vs_var_1000_maps_same_variance_l_min}
\end{figure}

We evaluate the variance $V$ for each element of \ensemble\ 1 and for each of the considered masks. 
Results are shown in Fig.~\ref{fig:var_comm_vs_var_1000_maps_same_variance} where each panel provides the histogram of $V$ for each mask.
Dashed red line represents $V$ as measured from \texttt{Commander}, and the dashed green line stands for $V$ of \texttt{SMICA}.
In the left panel of Fig.~\ref{fig:var_comm_vs_var_1000_maps_same_variance_l_min} we display the LTP of the {\it Planck} 2018 data in percentage as a function of the sky fraction. 
They are also reported in Table \ref{tab:p_value_variance_maps_same_variance} for convenience.
We find that the monotonic behaviour shown in Fig.~\ref{fig:p_value_MC_seme} for the $10^5$ $\Lambda$CDM simulations is 
almost\footnote{Note that for the \texttt{Commander} case the difference between the two last cases, i.e. Ext$_{24}$ and  Ext$_{30}$ case, is of the order 
of the numerical sensitivity of the \ensemble\ 1, since it composed of 10$^3$ simulations.}
recovered for the \ensemble\ 1: $V$ still decreases at high Galactic latitudes with a percentage of compatibility at the level of $0.3-0.4 \%$ in the Ext$_{30}$ case.
This means that a ``low variance'' model (low as the one observed by {\it Planck}) is not enough to explain this behaviour at high Galactic latitude.
\begin{table}
\begin{center}
\begin{tabular}{ l | c | c  }
	\hline
	\hline
	 & \multicolumn{2}{c}{\textbf{LTP} [\%]} \\
	\cline{2-3}
	Mask & $V<V_\textrm{c}$ & $V<V_\textrm{s}$\\
	\hline
	Std 2018 & 50.7 & 41.5\\
	
	Ext$_{12}$ & 4.1 & 1.9\\

	Ext$_{18}$ & 2.3 &1.1 \\

	Ext$_{24}$ & 0.2 & 0.2\\

	Ext$_{30}$ & 0.3 & 0.4\\
	\hline 
	\hline
\end{tabular}
\caption{The probability of obtaining a value for the variance $V$ smaller than that of \texttt{Commander} (second column), $V_\textrm{c}$, or \texttt{SMICA} (third column), $V_\textrm{s}$, for a map of the \ensemble\ 1. Note that the difference between the Ext$_{24}$ and  Ext$_{30}$ case is of the order of the numerical sensitivity of the \ensemble\ 1, since it is made of 10$^3$ simulations.}\label{tab:p_value_variance_maps_same_variance}
\end{center}
\end{table}
Notice also that this effect is largely dominated by the quadrupole and the octupole. 
This is shown in the right panel of Fig.~\ref{fig:var_comm_vs_var_1000_maps_same_variance_l_min}, where the LTP vs the observed sky fraction is shown
when we exclude only the quadrupole (blue dashed lines) or both the quadrupole and the octupole (red dashed lines) in the computation of $V$.


\subsection{Variance analyses including rotations}

As for the \ensemble\ 0 we now include random rotations in the analysis of the \ensemble\ 1. 
We still use the LTP-estimator and the $r$-estimator defined above. 

\subsubsection{LTP estimator}
\label{sec:LTP_estimator}

\begin{figure}[t]
		\hspace{-0.85cm}
	\centering
\subfloat{\includegraphics[width=.35\textwidth]{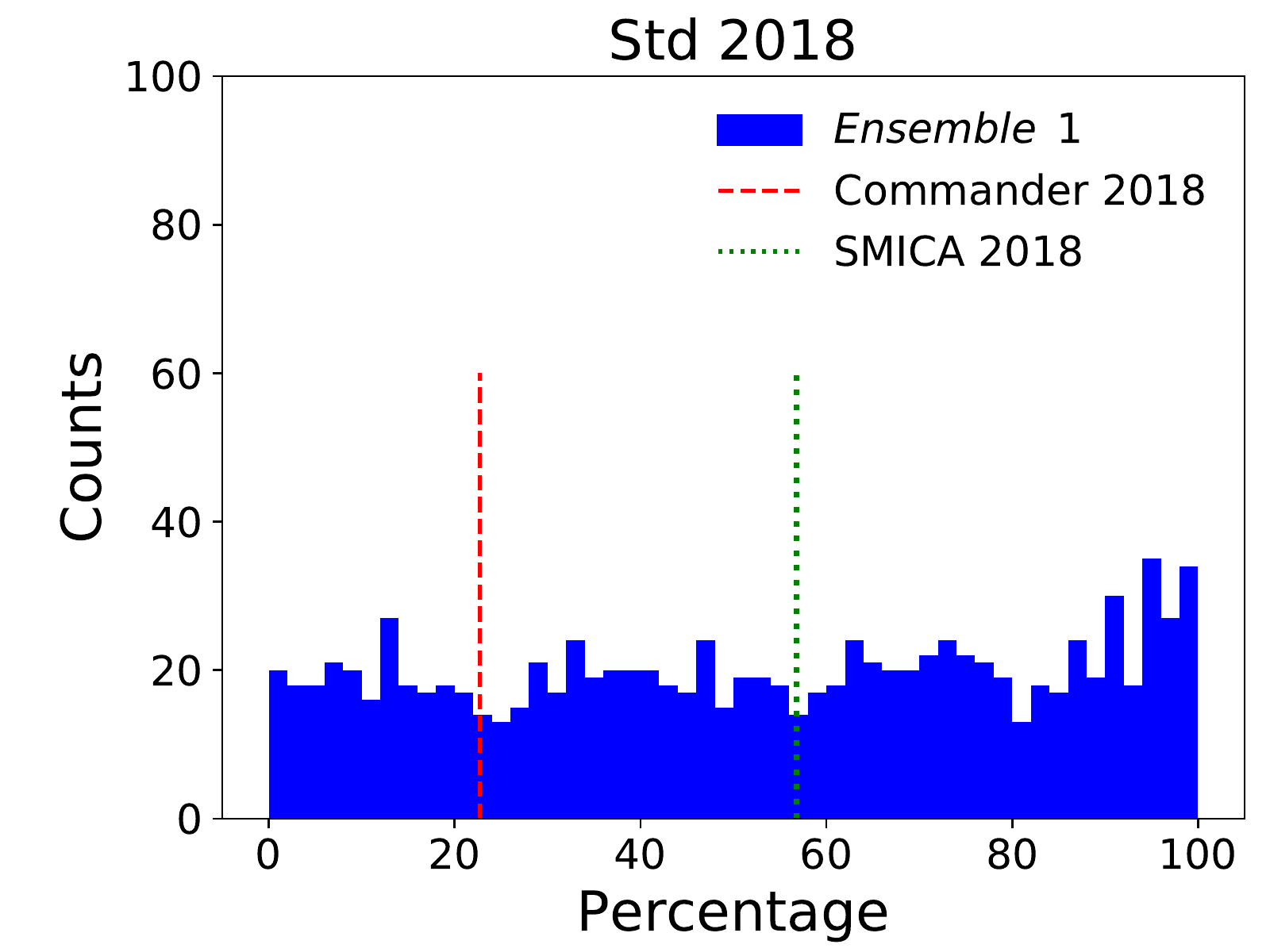}}
\subfloat{\includegraphics[width=.35\textwidth]{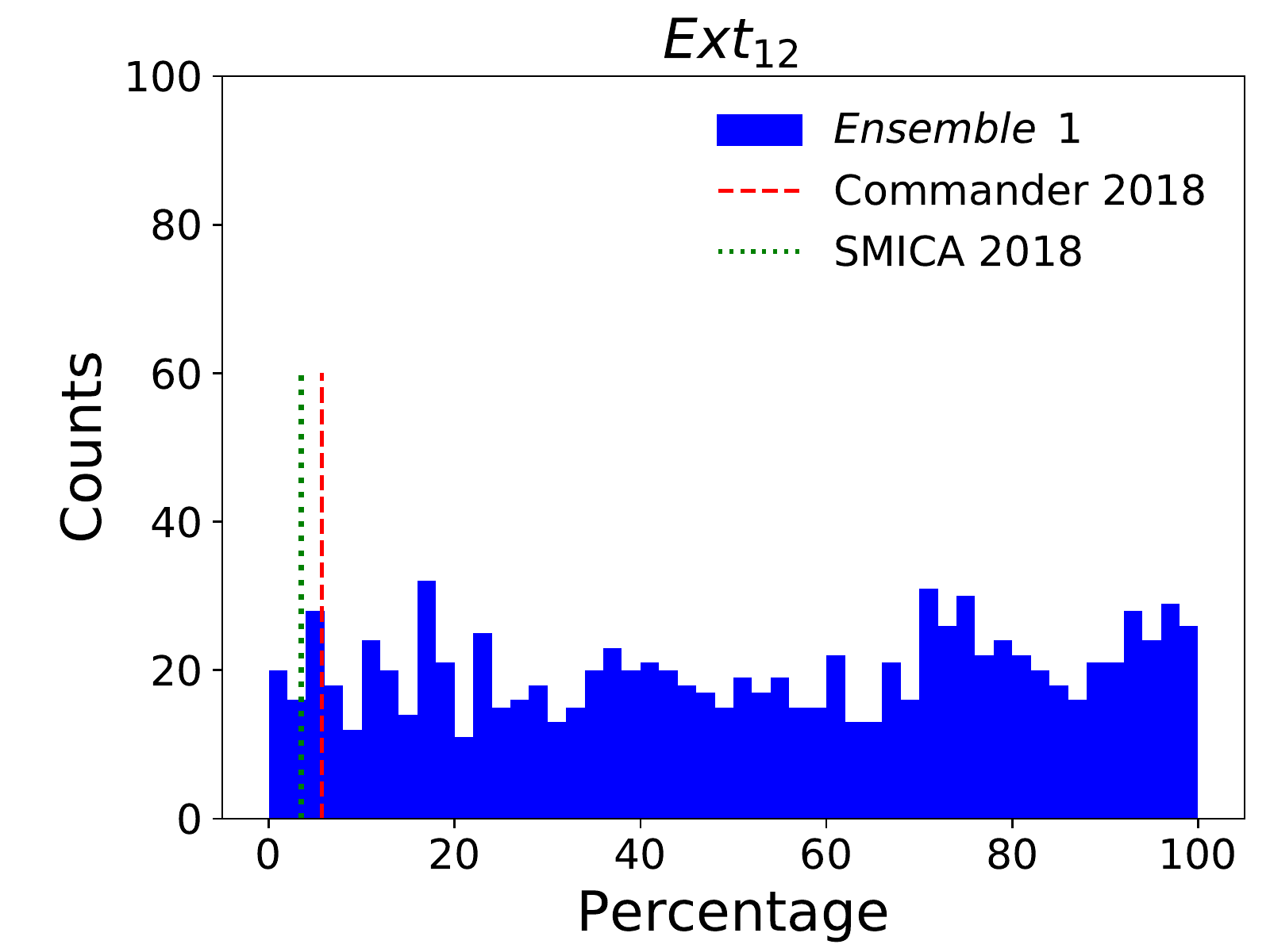}}
\subfloat{\includegraphics[width=.35\textwidth]{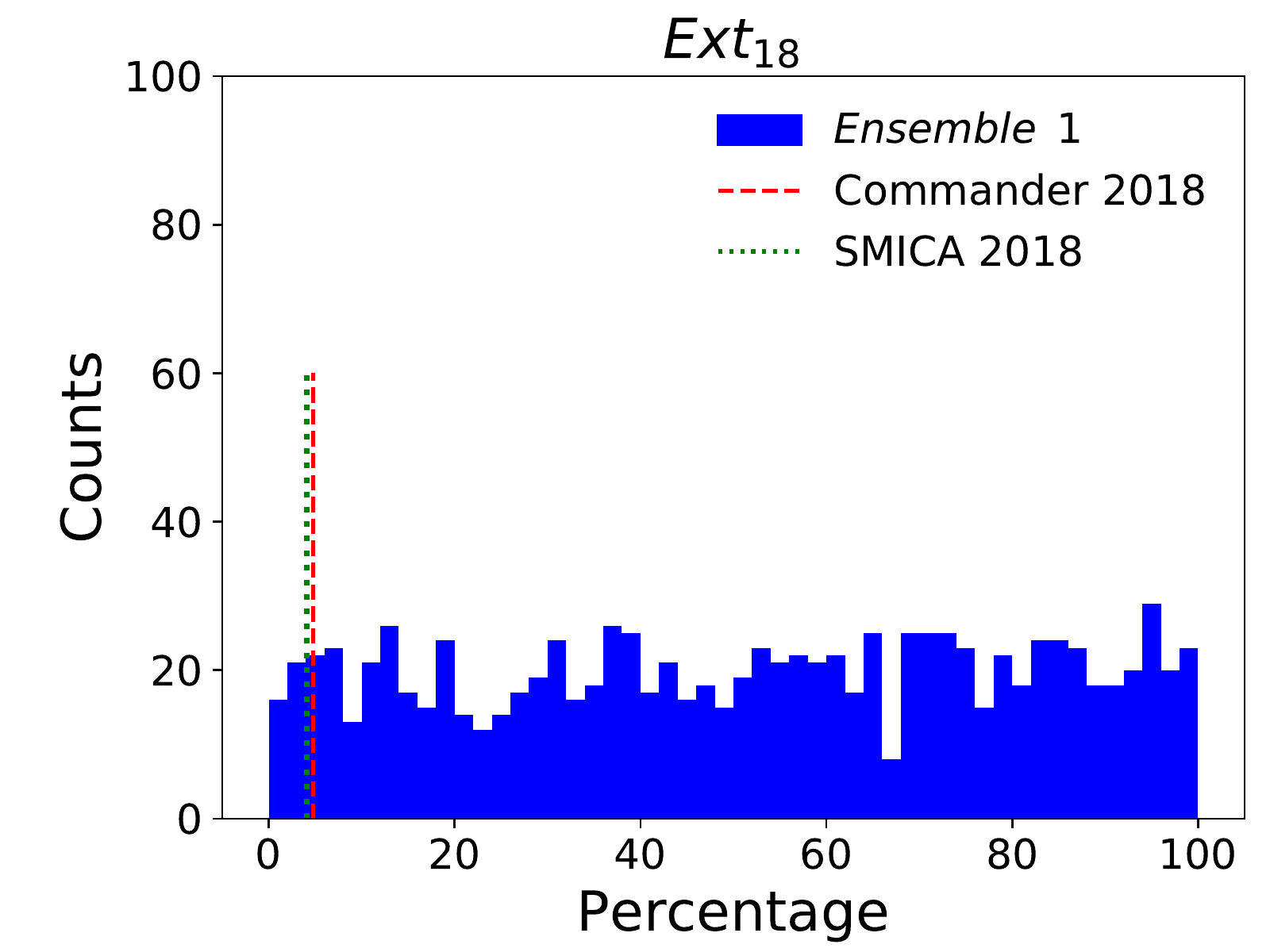}} \\
\subfloat{\includegraphics[width=.35\textwidth]{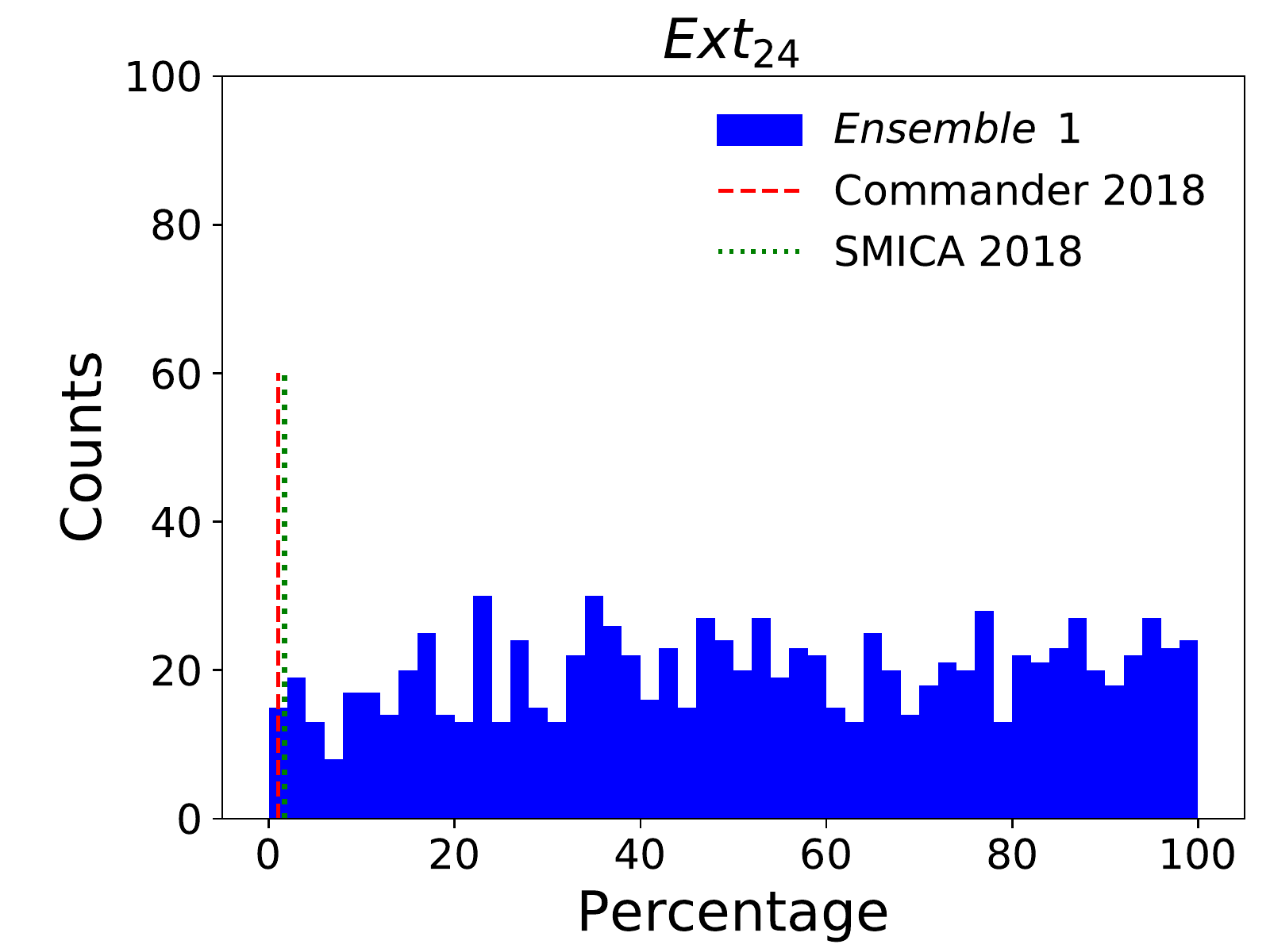}}
\subfloat{\includegraphics[width=.35\textwidth]{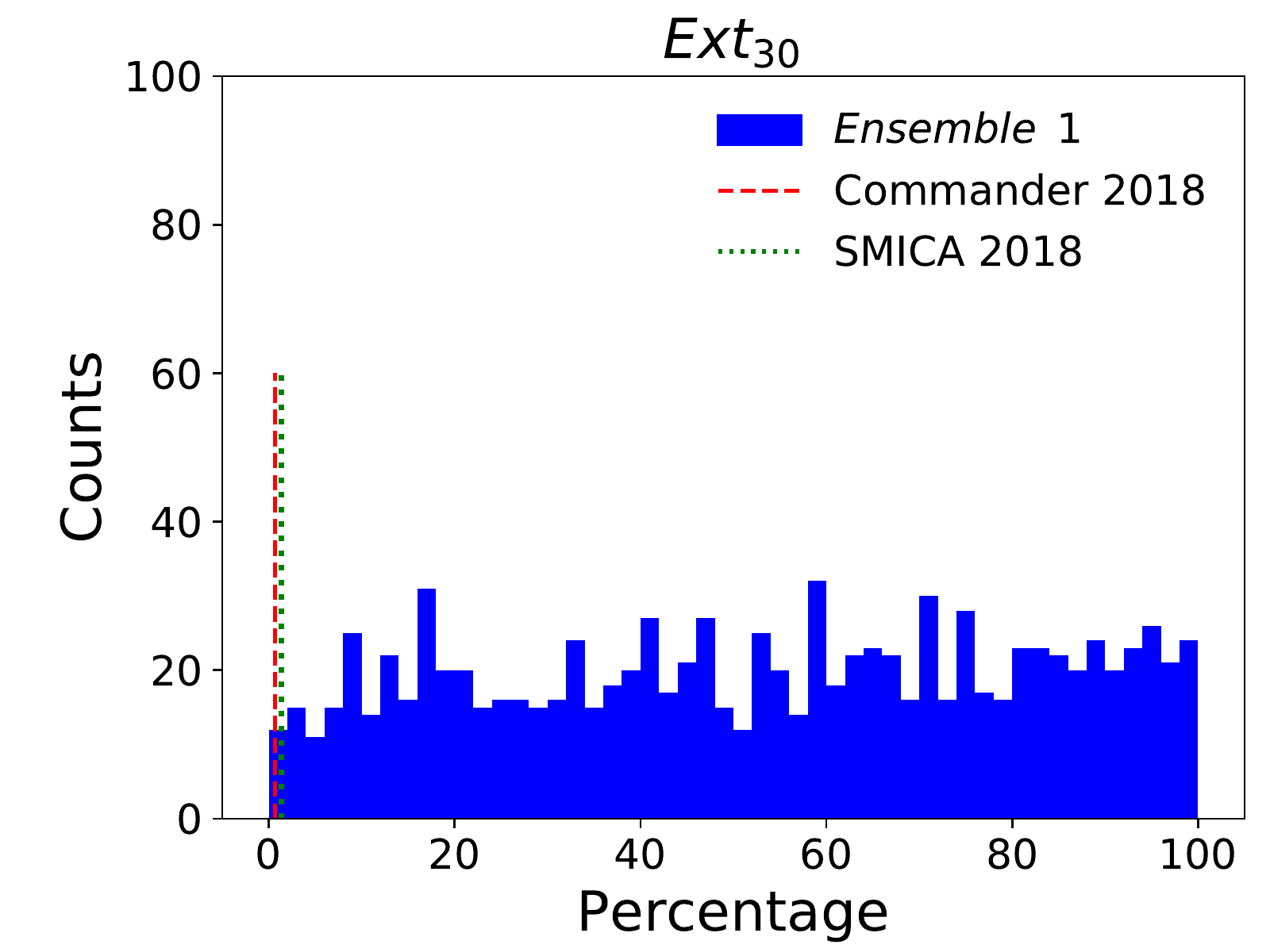}}
\caption{Histograms of the LTP of finding a rotated map of the \ensemble\ 1 with $V^{rot}<V$, where $V$ is the variance of the corresponding unrotated map. Each panel shows the results obtained using a different mask. Red dashed and green dotted vertical bars are the LTP for \texttt{Commander} and \texttt{SMICA} respectively.}\label{fig:p_value_rot_1000_case_separate}
\end{figure}

For each map $\textbf{m}_i$ belonging to \ensemble\ 1 and its rotations we obtain the MC of 10$^3$ values of LTP$_i$. In Fig.~\ref{fig:p_value_rot_1000_case_separate} we show the histograms of such LTP$_i$ for each considered mask. The observed LTP (i.e. those obtained from {\it Planck} data and shown in left panel of Table \ref{tab:p_value_comm_SMICA_ruotati_LCDM}) are also shown in the same figure as vertical bars, red for \texttt{Commander} and green for \texttt{SMICA}.   
Notice that, by construction, even in a $\Lambda$CDM model constrained to have a low-variance as \ensemble\ 1, the variance does not depend on the orientation.
Therefore the distribution of LTP is still uniform as it is found in the histograms of Fig.~\ref{fig:p_value_rot_1000_case_separate}. 
In this case we find for \ensemble\ 1 a very similar behaviour to \ensemble\ 0. 
For \texttt{Commander} (\texttt{SMICA}) the LTP estimator gives a $\sim 2.8\,\sigma$ ($\sim2.6\,\sigma$) anomaly at high Galactic latitude, see Table \ref{tab:p_value_comm_SMICA_ruotati}.

\begin{table}[!htb]
	\centering
\begin{tabular}{ l | c | c  }
	\hline
	\hline
	& \multicolumn{2}{c}{\textbf{LTP} [\%]} \\
	\cline{2-3}
	Mask & $\textrm{LTP}_i<\textrm{LTP}_\textrm{c}$ & $\textrm{LTP}_i<\textrm{LTP}_\textrm{s}$\\
	\hline
	Std 2018 & 21.6 & 52.9\\
	
	Ext$_{12}$ & 5.7 & 3.2\\
	
	Ext$_{18}$ & 4.5 &3.9 \\
	
	Ext$_{24}$ & 0.7 & 1.3\\
	
	Ext$_{30}$ & 0.5 & 0.9\\
	\hline 
	\hline
\end{tabular}
\caption{LTP of obtaining a simulation of the \ensemble\ 1 with LTP lower than the one obtained with the \texttt{Commander} map, $\textrm{LTP}_\textrm{c}$, and \texttt{SMICA} map, $\textrm{LTP}_\textrm{s}$.}\label{tab:p_value_comm_SMICA_ruotati}
\end{table}

\subsubsection{$r$-estimator}\label{sec:r-estimator}

\begin{figure}[t]
	\centering
	{\includegraphics[scale=0.46]{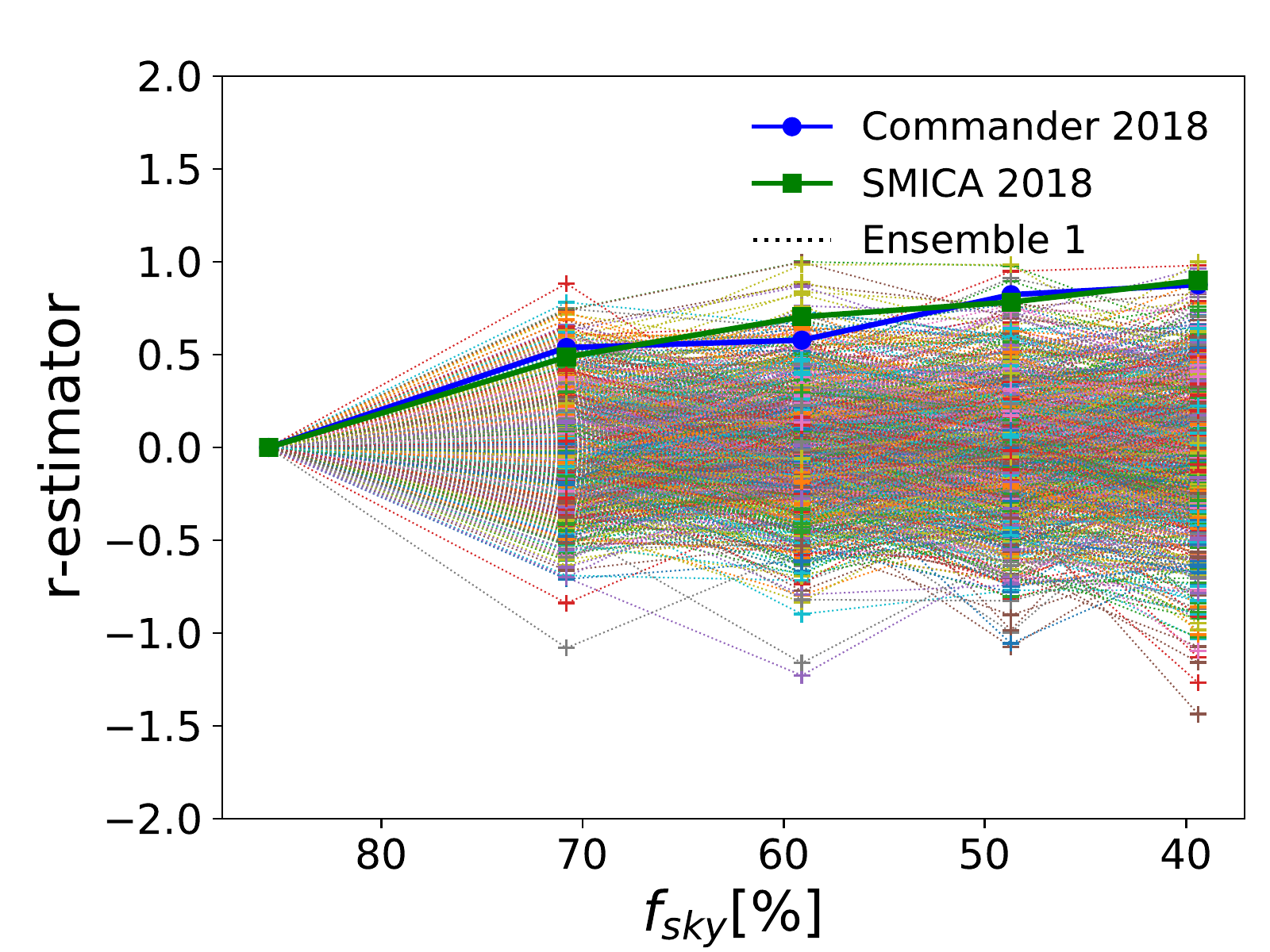}}
	{\includegraphics[scale=0.46]{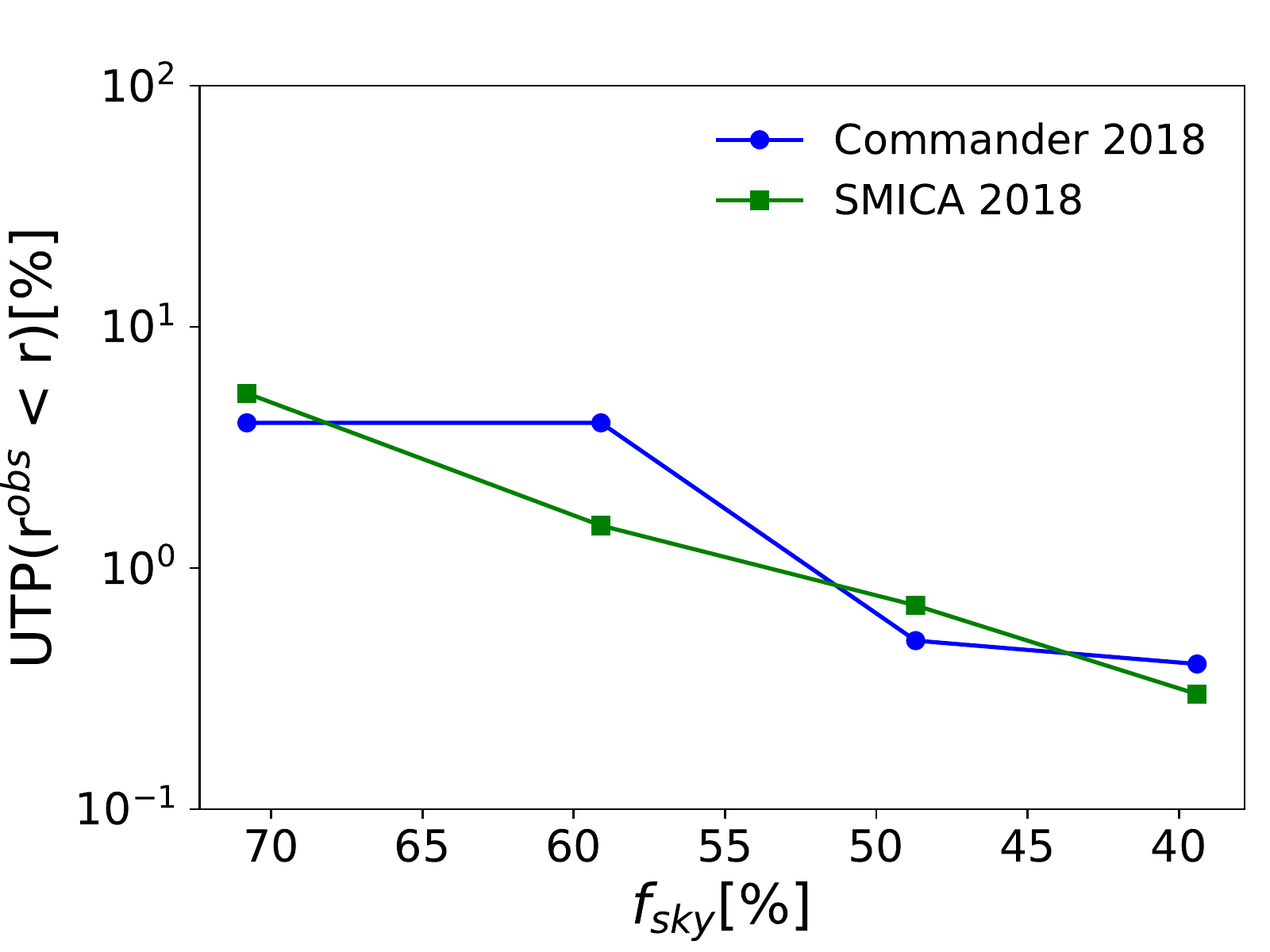}}
	\caption{Left panel: $r$-estimator computed with Eq. (\ref{eqn:r_estimator}) versus the sky fraction. The coloured dotted lines stand for the $r$ value obtained from the \ensemble\ 1. Blue and green solid lines stand for \texttt{Commander} and \texttt{SMICA} respectively. Right panel: UTP of obtaining a simulation with $r$ larger than the one obtained with \texttt{Commander} (blue line) or \texttt{SMICA} (green line) as a function of the sky fraction.} \label{fig:r_v_std_meno_v_j}
\end{figure}

We apply here the $r$-estimator to the \ensemble\ 1 simulations.
In Fig.~\ref{fig:r_v_std_meno_v_j} we show the results for all the considered cases. Dotted lines connect the MC values of $r$ represented with a plus symbol. 
Solid blue line connects the \texttt{Commander} values (dot symbols) and the solid green line connects the \texttt{SMICA} values (square symbols). The UTP are shown in the right panel of Fig.~\ref{fig:r_v_std_meno_v_j} and quoted in Table \ref{tab:r_value}.
At high Galactic latitude we find an anomalous value for $r$ at the level of $\sim 2.9\,\sigma$ with a UTP of $0.4 \%$ for \texttt{Commander} and $0.3 \%$ for \texttt{SMICA}.
In conclusions the results for the \ensemble\ 1 are similar to those of \ensemble\  0 even when rotations are considered.

\begin{table}[!h]
\begin{center}
\begin{tabular}{ l | c | c  }
	\hline
	\hline
	& \multicolumn{2}{c}{\textbf{UTP} [\%]} \\
	\cline{2-3}
	Mask & $r^{\textrm{c}}<r$& $r^\textrm{s}<r$\\
	\hline
	Ext$_{12}$ & 4.0 & 5.3\\
	
	Ext$_{18}$ & 4.0 &1.5 \\
	
	Ext$_{24}$ & 0.5 & 0.7\\
	
	Ext$_{30}$ & 0.4 & 0.3\\
	\hline 
	\hline
\end{tabular}
\caption{UTP of obtaining a simulation with $r$ larger than the one obtained from the data. Second column shows the UTP for \texttt{Commander}, third column the UTP for \texttt{SMICA}.}\label{tab:r_value}
\end{center}
\end{table}


\section{Conclusions}
\label{conclusions}

In this paper we analysed the lack-of-power anomaly, a well known characteristic of the CMB temperature anisotropy pattern showing up at large angular scales. 
In particular, we focused on the intriguing fact that this feature is statistically more significant (at a $\sim 3 \sigma$) when only high Galactic latitude data are taken into account. The latter observations suggests that most of the large scale anisotropy power happens to be mainly localised around the Galactic plane. This might sound bizzarre because the early universe should not know anything about the ``direction'' of the disk of our Galaxy. To tackle the issue, we evaluated how often a $\Lambda$CDM realisation happens to have most of its power localised at low Galactic latitude. 

To support the analysis, we generated a $\Lambda$CDM Monte Carlo set of $10^5$ CMB maps from the {\it Planck} 2018 best-fit model. By analysing this set, we first showed that the {\it Planck} 2018 data exhibits the same trend of decreasing  CMB field variance while increasing the Galactic mask, which was found previously in the literature. 
We then proceeded to randomly rotate the simulated maps (denoted as \ensemble\ 0), as well as the data, $10^3$ times. The rotated maps are employed to compute the empirical distribution function of two estimators, based on the CMB field variance (Section \ref{preliminary}).  
With the LTP-estimator (Section \ref{sec:LTP_estimator0}) we test to what extent the low CMB anisotropy power in the data depends on the orientation of the Galactic plane. 
With the $r$-estimator (Section \ref{sec:r-estimator0})  we assess instead the behaviour against rotation of the decreasing trend of the CMB variance at increasing Galactic latitude. 
The introduction of random rotations is a key-element to evaluate whether the lack of power anomaly is indeed correlated with Galactic latitude.  

To further investigate this behaviour we also selected from the $10^5$ $\Lambda$CDM simulations set a smaller set, of $10^3$ maps, which exhibits the same low-variance as the one observed in the \texttt{Commander} and \texttt{SMICA} 2018 maps. We called this set \ensemble\ 1 and repeated the analyses performed on the \ensemble\ 0. 

We find that even when performing random rotations, our CMB sky is anomalous in power at about  $2.8 - 2.5 \, \sigma$ depending on the considered component separation method when employing the LTP estimator. Specifically, only $5$ maps out of $10^5$ have a LTP at high Galactic latitude (in the Ext$_{30}$ mask) smaller than the Planck \texttt{Commander} data. For the $r$-estimator we evaluate that only the $0.2\%$ of the maps  show a larger value of $r$ between Std 2018 and Ext$_{30}$ masks, again with respect to \texttt{Commander}. Results are substantially stable if we employ \texttt{SMICA} in place of \texttt{Commander}. Finally, using the low-variance constrained simulation of  \ensemble\ 1 yields simular results, showing that having a low-variance field in the first place is not enough to justify the observed trend with Galactic latitude.
		
In conclusion, the introduction of rotations do not spoil the lack of power anomaly at high Galactic latitude which turns out to be quite stable against the ``look-elsewhere effect'' spawned by random rotations of the reference frame.

\acknowledgments
This work is based on observations obtained with {\it Planck} (http://www.esa.int/Planck), an ESA science mission with instruments and contributions directly funded by ESA Member States, NASA, and Canada.
We acknowledge the use of computing facilities at NERSC and Cineca. 
Some of the results in this paper have been derived using the \texttt{HEALPix} \cite{Gorski:2004by} package. Authors acknowledge financial support by ASI Grant No. 2016-24-H.0 (COSMOS).

\appendix

\section{Generating the rotations}
\label{sec:rotations}

\begin{figure}[h]
	\centering
	\includegraphics[scale=0.3]{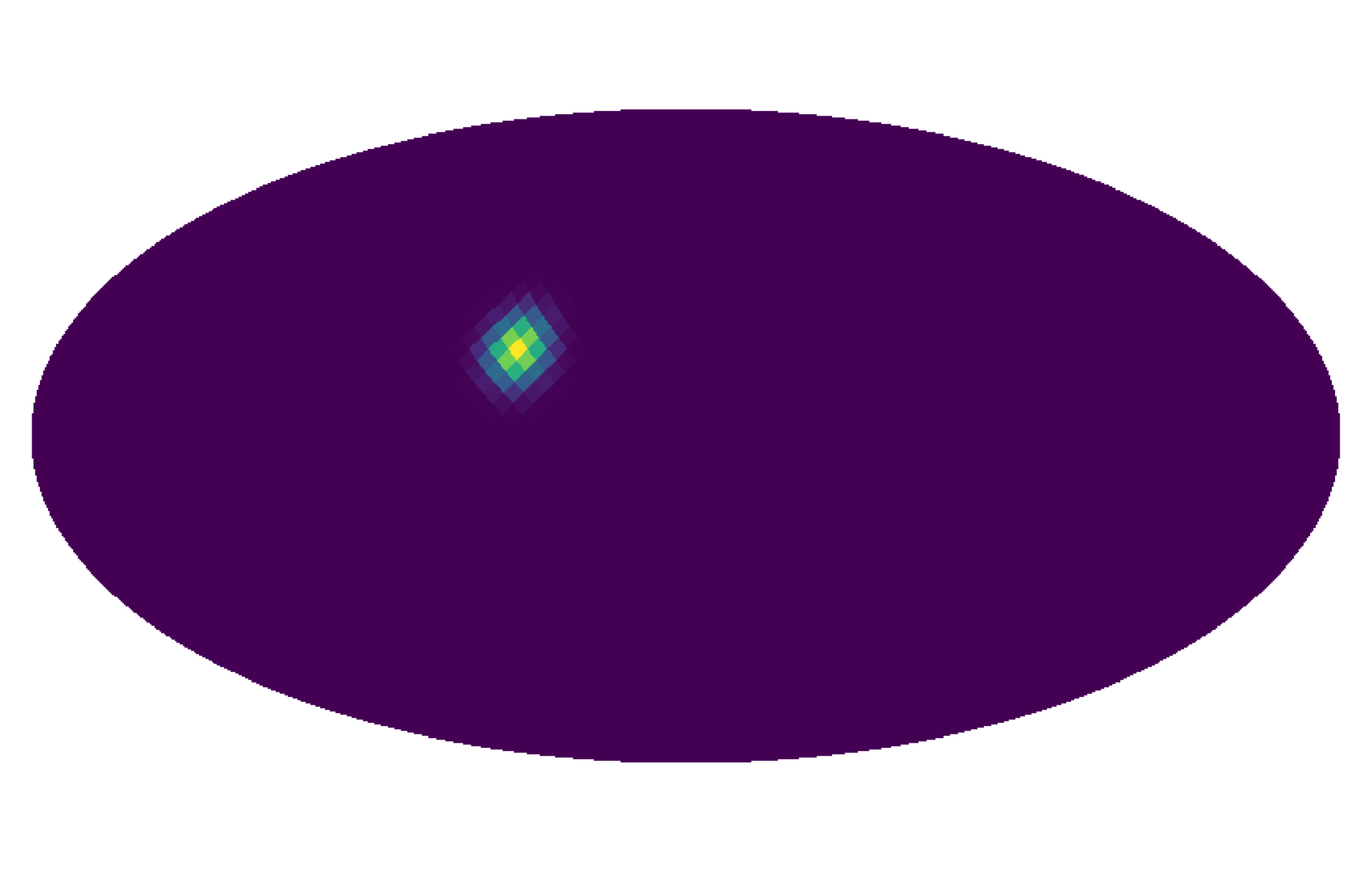}
	\caption{A test map  at \texttt{HEALPix} resolution $N_{side}=16$ with all pixels zero except for 9 pixels set to 1 and after convolution with a Gaussian beam of $9^\circ$.}\label{fig:pixel_gaussiano}
\end{figure}

\begin{figure}[t]
	\centering
	\includegraphics[scale=.3]{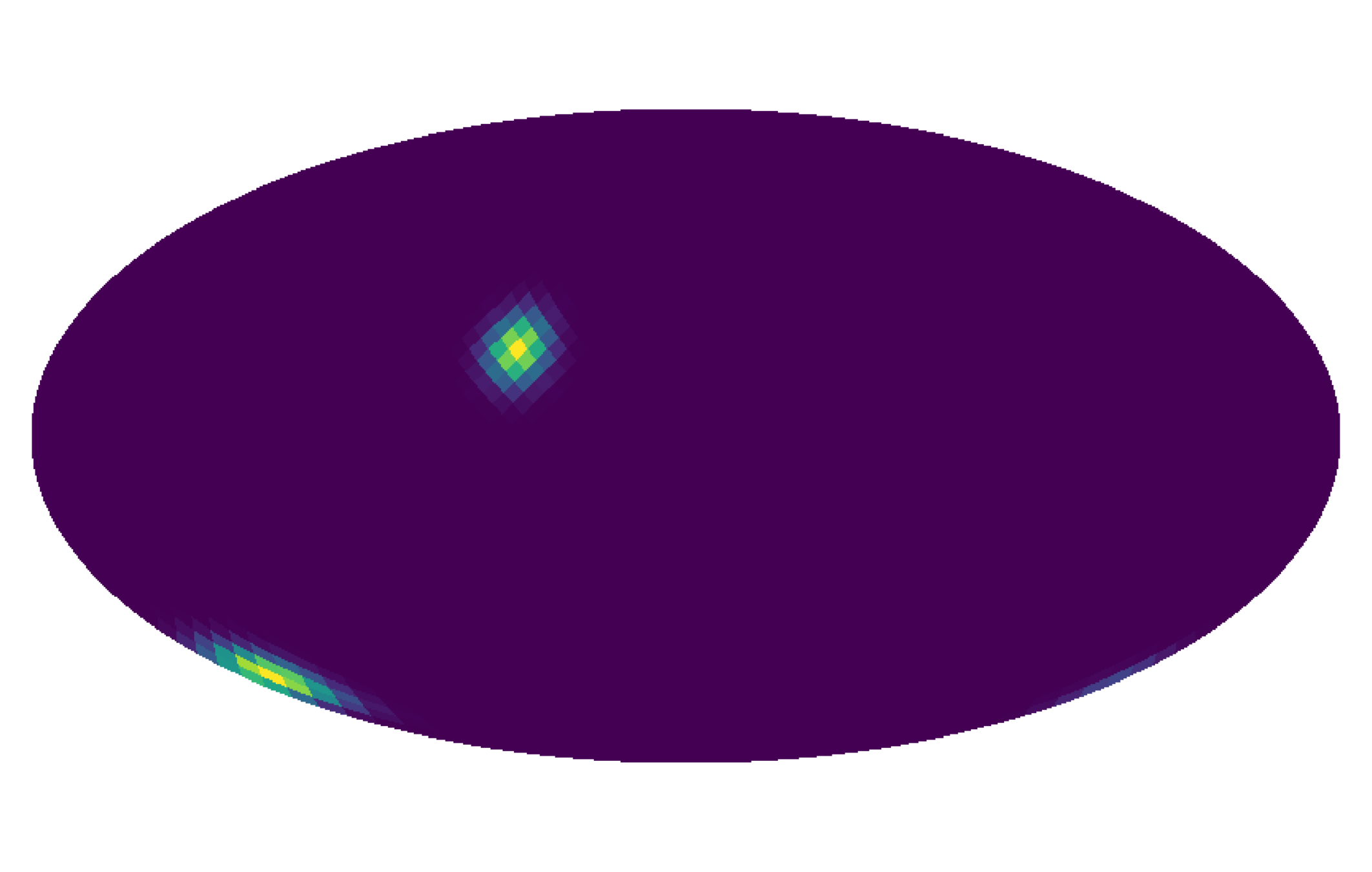}
	\includegraphics[scale=.3]{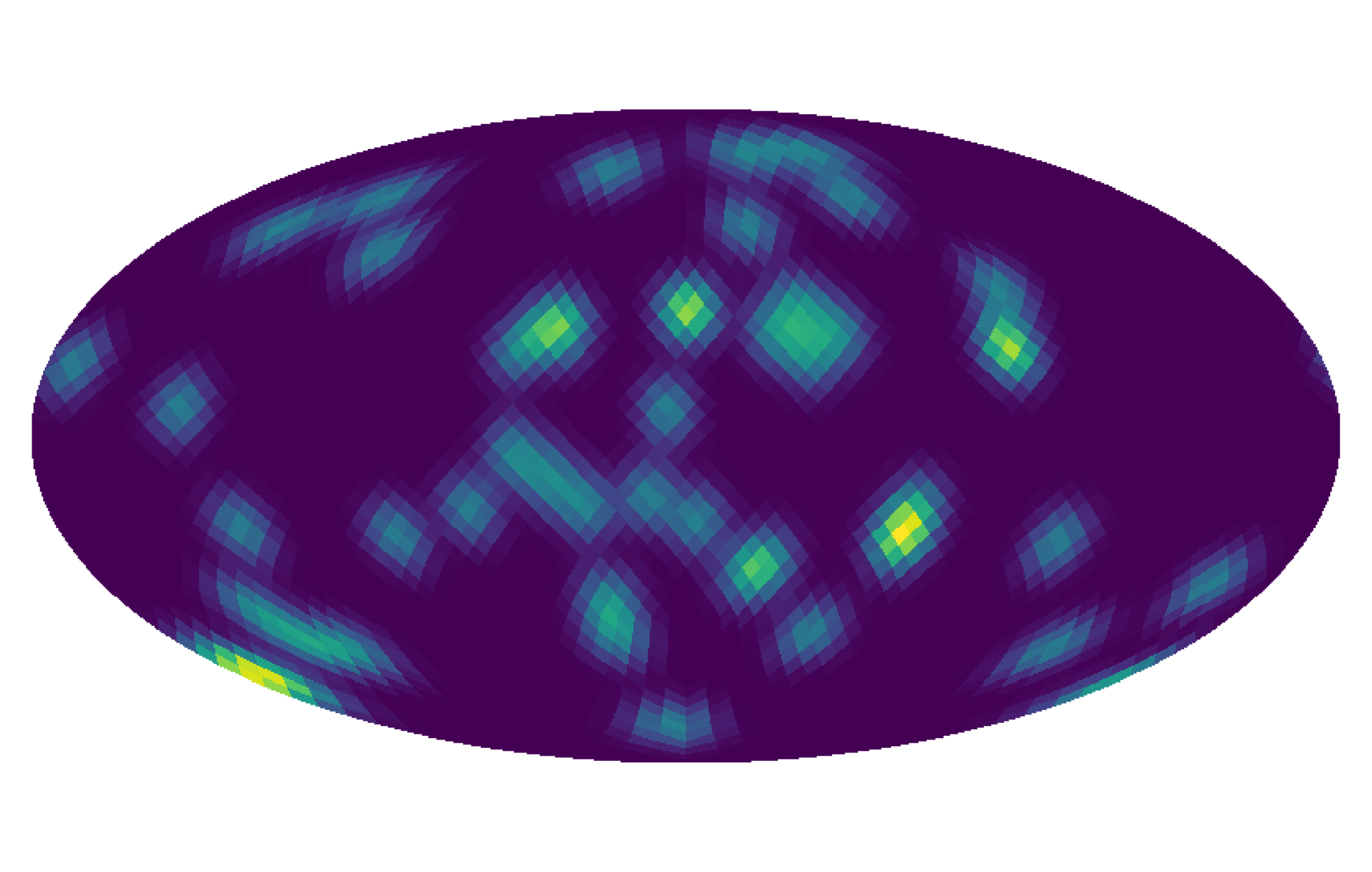}
	\includegraphics[scale=.3]{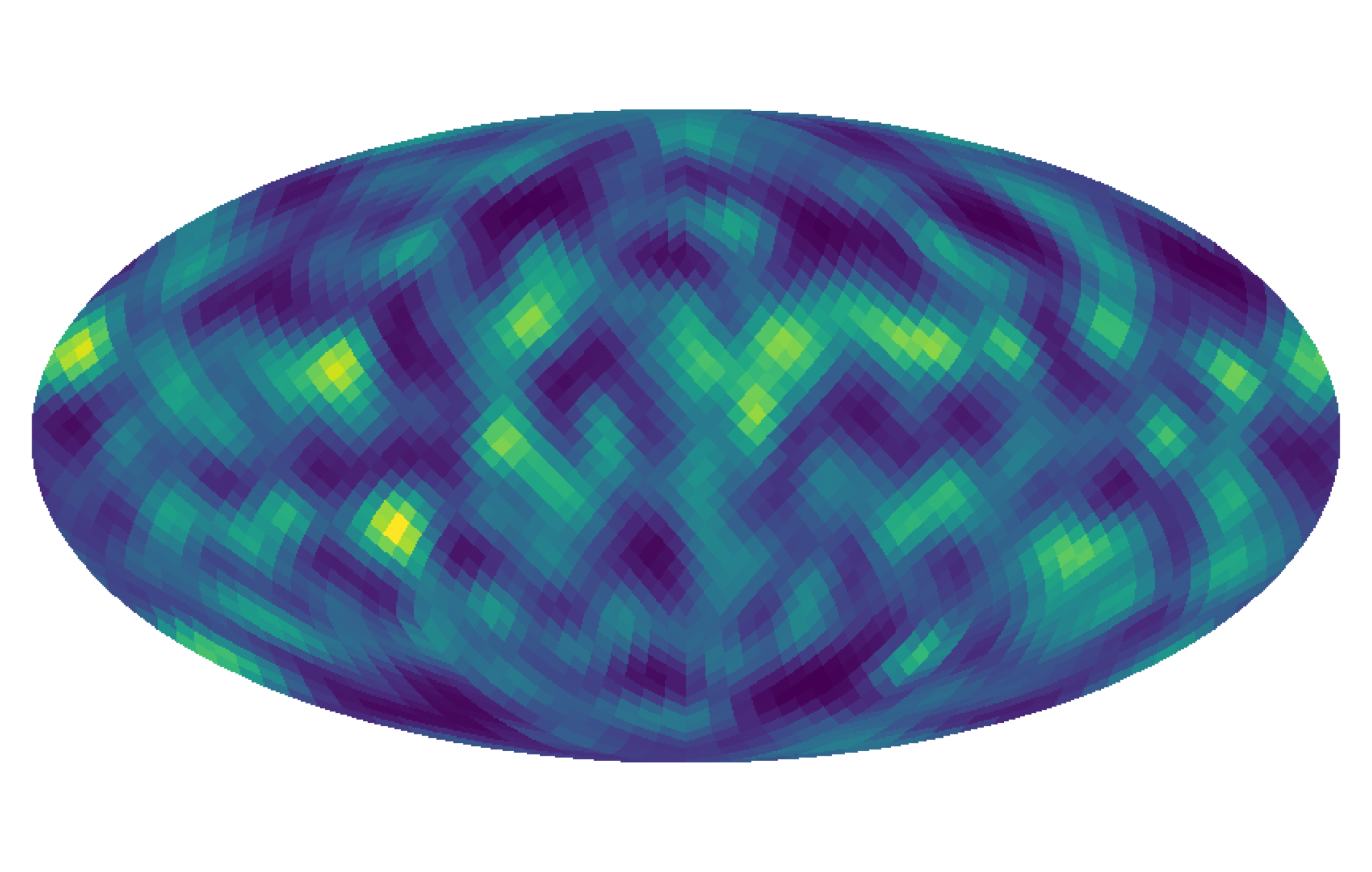}
	\caption{Total map $\textbf{m}^{tot}$ computed through Eq. \ref{eqn:sum_map} at \texttt{HEALPix} resolution $N_{side}=16$ for $N_{rot}=2$ (top left panel), $N_{rot}=50$ (top right panel) and for $N_{rot}=500$ (bottom panel).}\label{fig:total_map}
\end{figure}

\begin{figure}[t]
	\centering
	\includegraphics[scale=0.46]{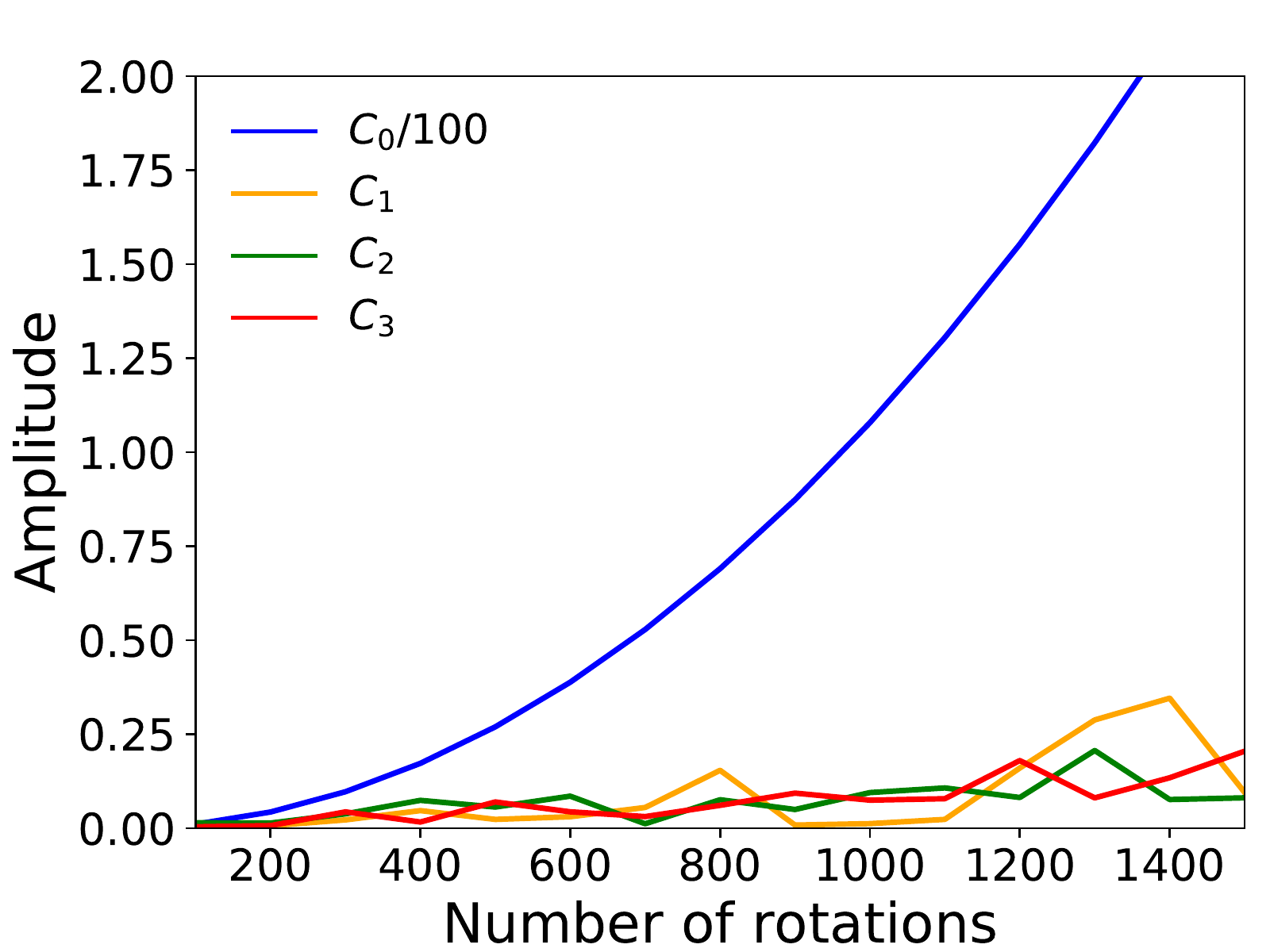}
	{\includegraphics[scale=0.46]{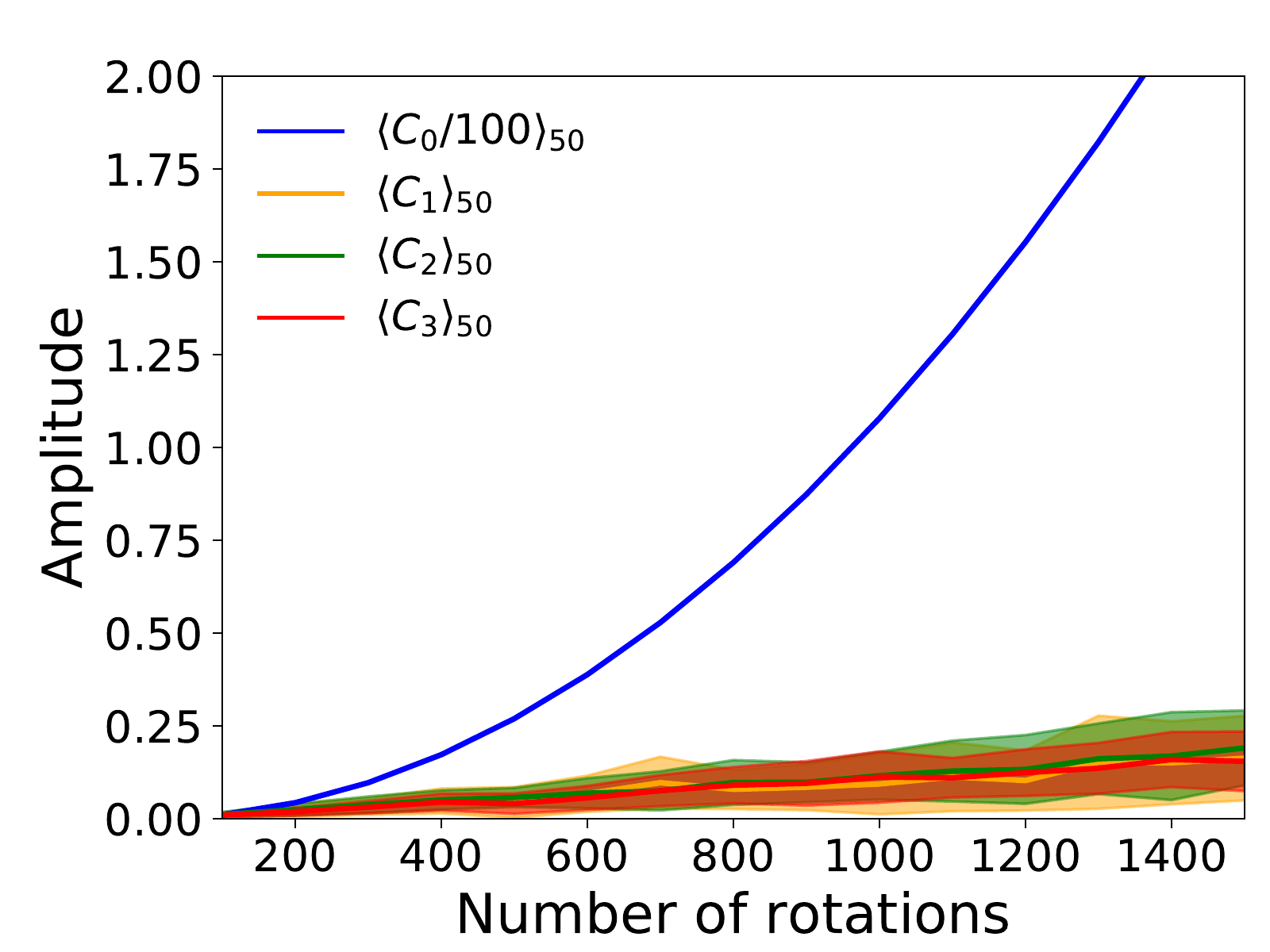}}
	\caption{Left panel: amplitude of the first low-$\ell$ components of the APS of the test map after each rotation. Right panel: the average over 50 repetitions of the machinery described in Sec. \ref{sec:rotations}. The filled regions correspond to the 1 $\sigma$ dispersion of $C_\ell$ components.}\label{fig:mono_dip_quad_oct_mean50}
\end{figure}

Random rotations of temperature CMB maps are generated following an harmonic-based approach through a \texttt{Python} algorithm. We consider maps at \texttt{HEALPix} resolution $N_{side}=16$ which are harmonic-expanded to obtain the initial $a^{in}_{\ell m}$ coefficients. These coefficients are then rotated through the Wigner rotation matrices, $\textbf{R}(\vartheta,\varphi,\psi)$, whose rotations angles $(\vartheta,\varphi,\psi)$ (also known as Euler angles), are randomly extracted from uniform distributions. Technically this is performed thanks to the \texttt{healpy} subroutine \texttt{rotate\_alm}.
After the rotation, the final map, or simply the rotated map, $\textbf{m}^R$ can be written as
\begin{equation}
\textbf{m}^R=\sum_{\ell m}\left(\sum_{m'}\textbf{R}_{m m'}(\vartheta,\varphi,\psi)a_{\ell m'}^{in}\right)Y_{\ell,m}(\theta,\phi)\,.
\label{formularot}
\end{equation}
To validate the procedure which implements random rotations, we consider a map which is zero except for a spot of $9^\circ$, see Fig. \ref{fig:pixel_gaussiano}. This is done simply setting to 1, nine neighboring pixels and then 
smoothing\footnote{The smoothing is applied in order to minimise aliasing effects when going from real to harmonic space and vice versa.} the map with a Gaussian beam with a FWHM$ =9^\circ$. For convenience we call $\textbf{m}_0$ this initial map.
Starting from $\textbf{m}_0$ we perform $N_{rot}$ rotations\footnote{In other words, we apply $N_{rot}$ times Eq.~(\ref{formularot}).} considering $\textbf{m}_{i-1}$ as the input for $i^{th}$ rotation, with $i=1,...N_{rot}$.
We then compute the following total map, 
\begin{equation}\label{eqn:sum_map}
	\textbf{m}^{tot}=\sum_{i=0}^{N_{rot}}\textbf{m}_i \, ,
\end{equation}
which is shown in Fig.~\ref{fig:total_map}, for $N_{rot}=2, 50$ and $500$. 
The idea is to use $\textbf{m}^{tot}$ to test whether the set of considered rotations is able to ``cover uniformly'' all the possible directions. This is our requirement for validation which is quantified computing the APS of $\textbf{m}^{tot}$ and comparing the monopole with higher order multipoles: when the former dominates over the latter we can safely state that the set of rotations is sufficiently populated to have its isotropic part leading over accidental anisotropies. 
Note that, in turn, this procedure provides the minimum number of rotations which are needed to fullfill the requirement mentioned above.
%
The left panel of Fig.~\ref{fig:mono_dip_quad_oct_mean50} shows the behaviour of the lowest multipoles, namely the monopole $C_0$, the dipole $C_1$, the quadrupole $C_2$, and the octupole $C_3$, against the number of rotations. The monopole component increases its magnitude quadratically versus the number of rotations whereas low-$\ell$ components oscillate around a very slowly monotonic growth. We repeat this procedure 50 times and compute the mean distribution of the same first low-$\ell$ components, see right panel of Fig.\ref{fig:mono_dip_quad_oct_mean50}. The mean behaviour of the different components, and the hierarchy among the multipoles, is substantially unchanged with respect to what obtained with the single realisation. In particular the hierarchy among low-$\ell$ multipole components seems to become stable for $N_{rot} > 900$. 
Most importantly, we find that the magnitude of the ratio $C_0/C_1$ at $N_{rot}=1000$ is of the order $10^3$: therefore we choose this threshold to define the minimal number of rotations needed to cover sufficiently homogeneously the whole sky.

\section{Comparison between 2018 and 2015 {\it Planck} release}
\label{comparison_2018_2015}

\begin{figure}[t]
	\centering
	\includegraphics[scale=0.46]{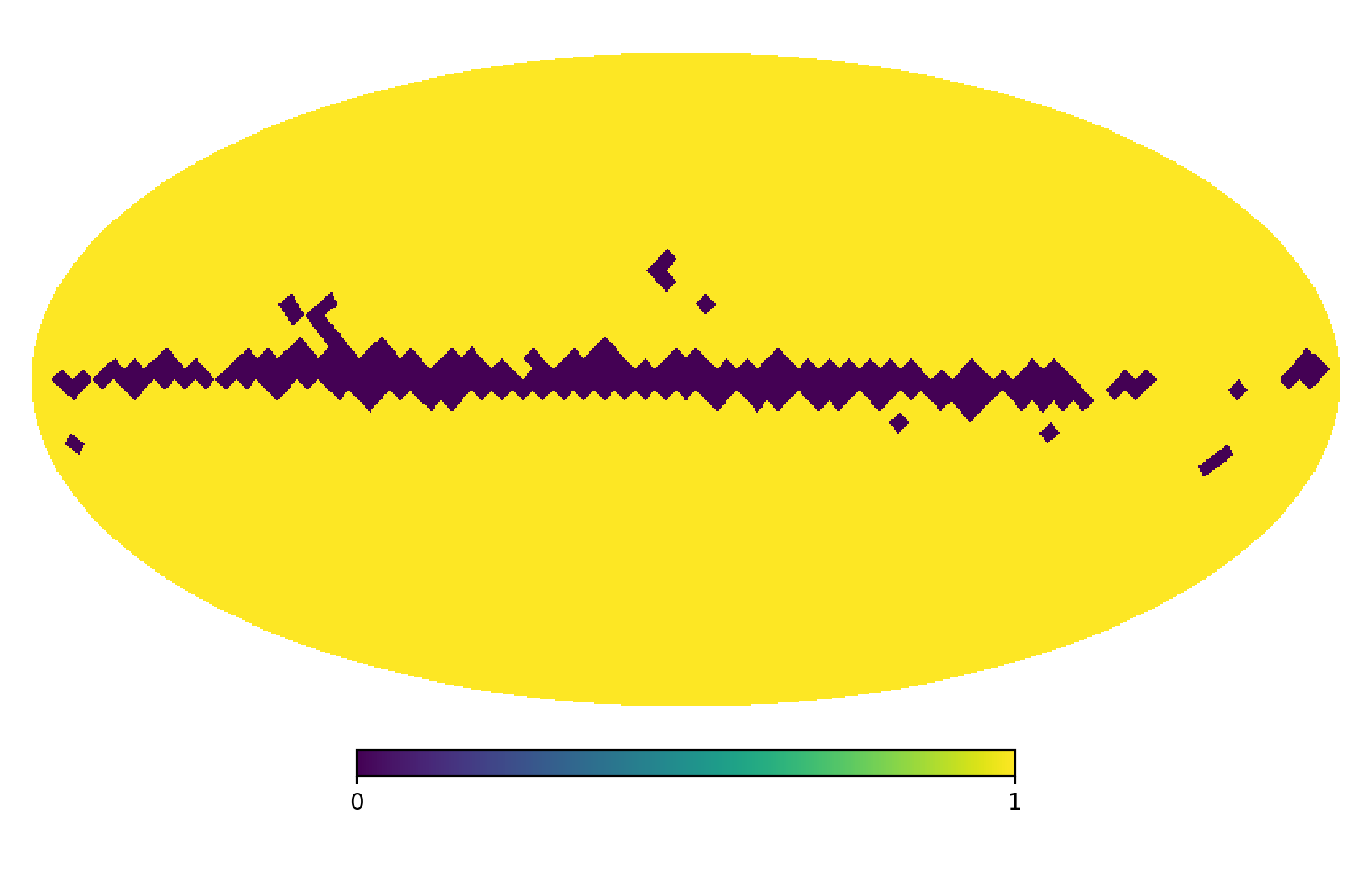}
	\caption{Std 2015 temperature mask.}\label{fig:mask2015}
\end{figure}

\begin{figure}[t]
		\hspace{-0.85cm}
	\centering
	\subfloat{\includegraphics[width=.35\textwidth]{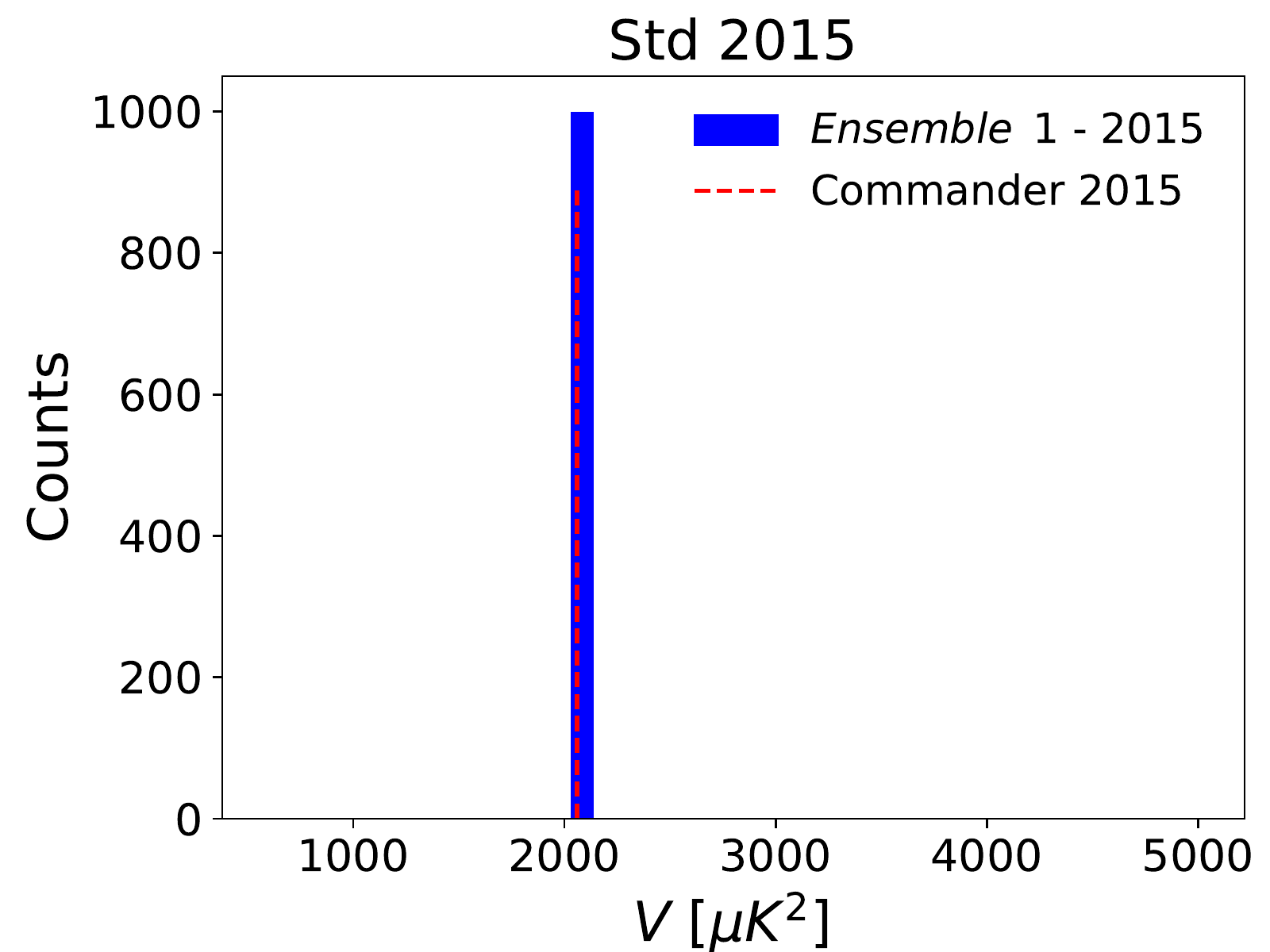}}
	\subfloat{\includegraphics[width=.35\textwidth]{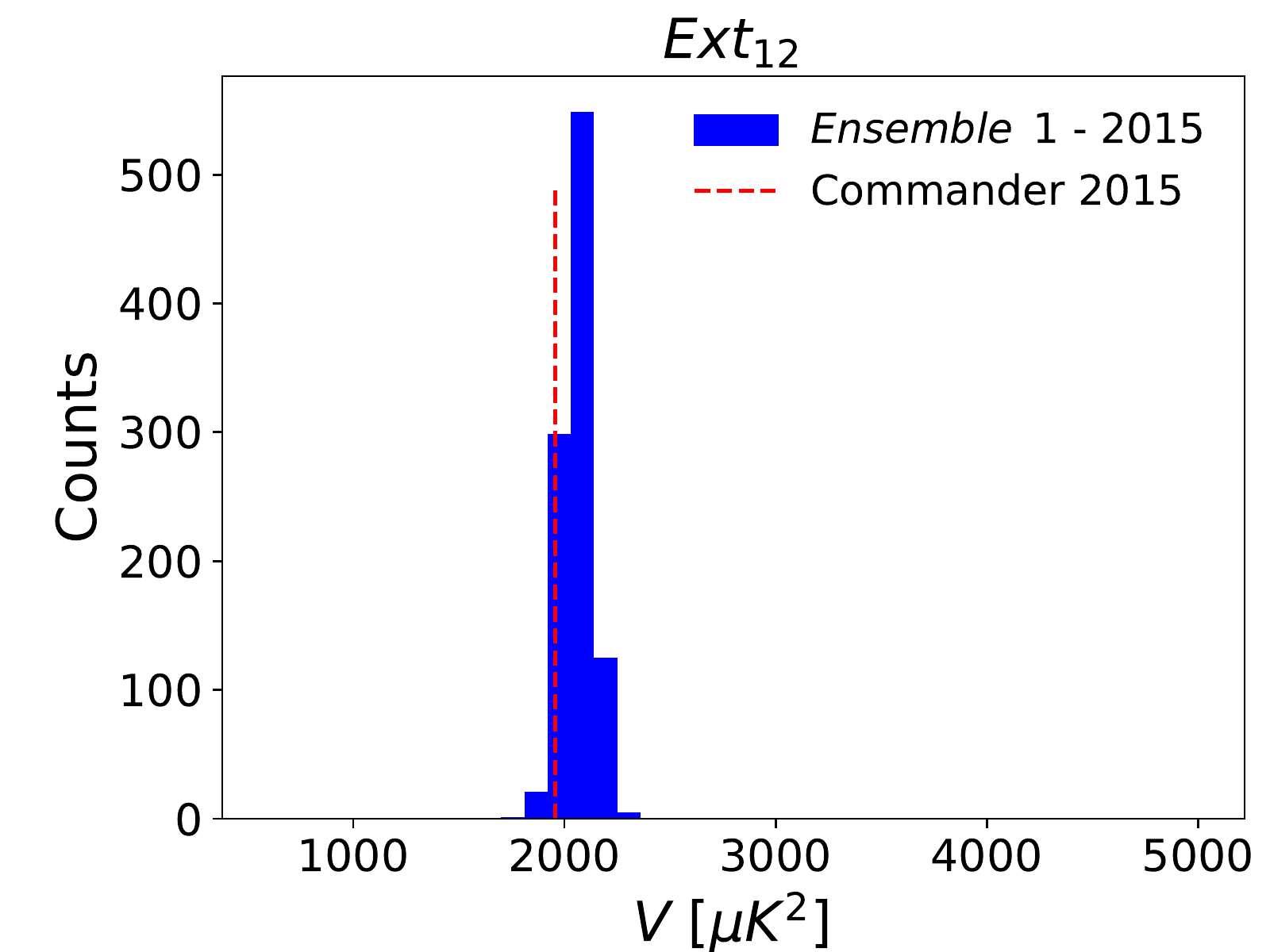}}
	\subfloat{\includegraphics[width=.35\textwidth]{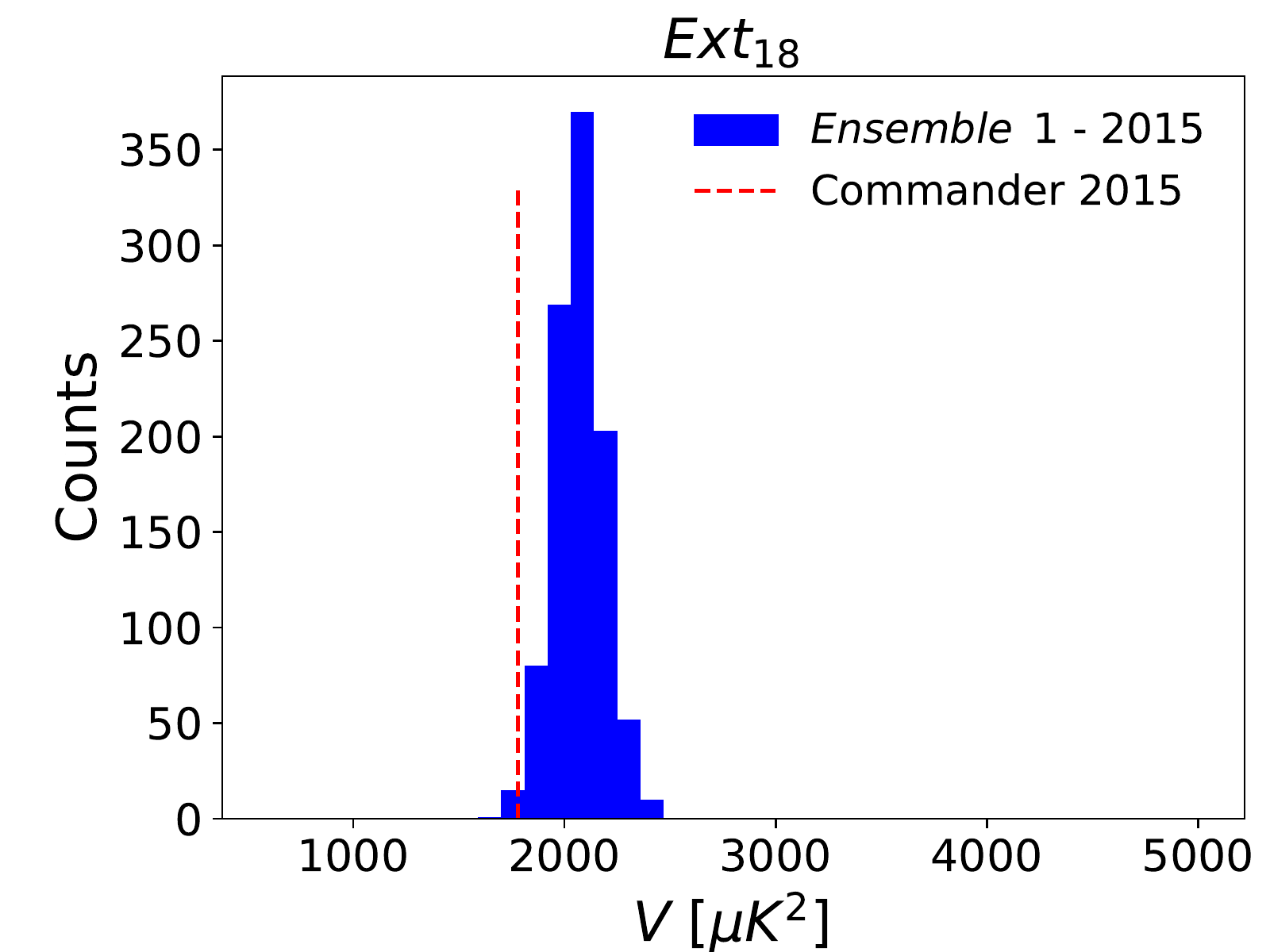}} \\
	\subfloat{\includegraphics[width=.35\textwidth]{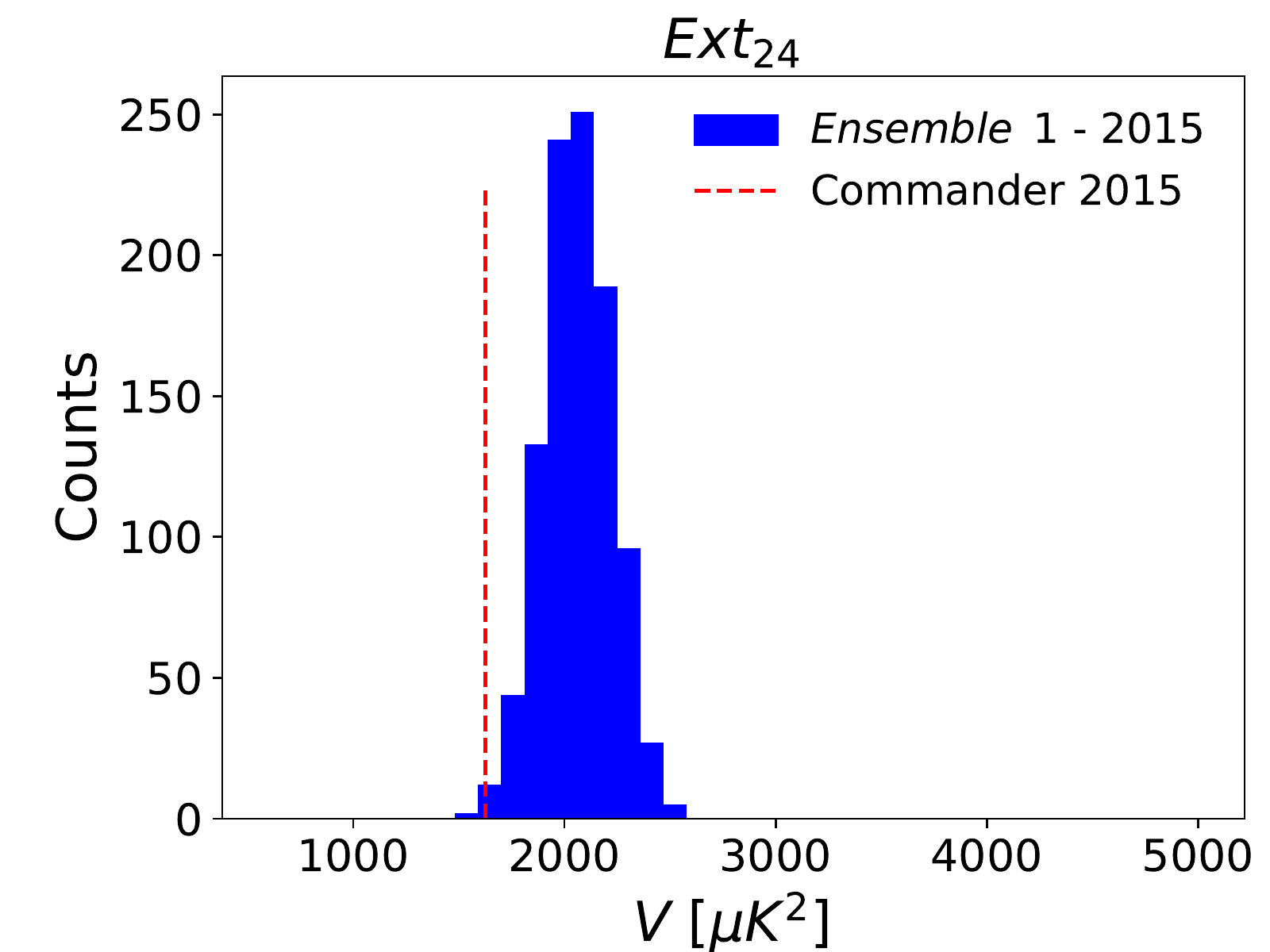}}
	\subfloat{\includegraphics[width=.35\textwidth]{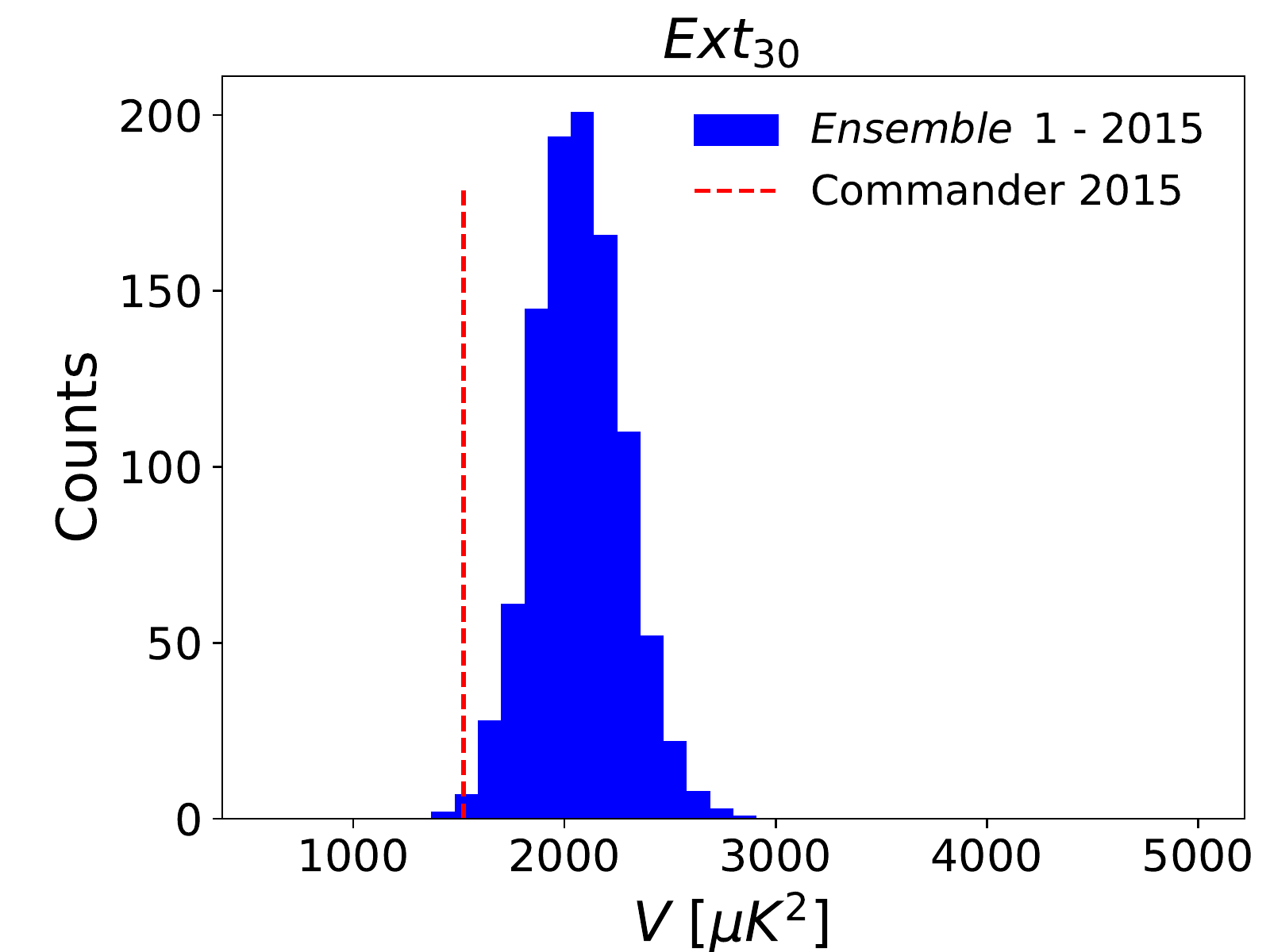}}
	\caption{Histograms of the variance $V$ of the maps belonging to \ensemble\ 1 - 2015 computed for the masks Std 2015, Ext$_{12}$, Ext$_{18}$, Ext$_{24}$ and Ext$_{30}$. The red dashed line identifies the variance of the \texttt{Commander} 2015 map, $V_\textrm{c}$.
}\label{fig:var_comm_vs_var_1000_maps_same_variance_2015}
\end{figure}

In this section we consider the 2015 \textit{Planck} data.
This analysis is performed mainly because the {\it Planck} 2015 standard mask \cite{Aghanim:2015xee}, henceforth called Std 2015, is smaller than the 2018 one. Its observed sky fraction is 93.6$\%$, see Fig.~\ref{fig:mask2015}, versus $85.6\%$ of the Std 2018, see Fig.~\ref{fig:masks} and Table~\ref{tab:masks}.
Hence we employ here the \texttt{Commander} 2015 map used in \cite{Aghanim:2015xee} still at \texttt{HEALPix} resolution $N_{side}=16$ and FWHM of 440 arcmin and consistently to what performed for the 2018 case, we added to this map a regularisation noise of 2~$\mu$K rms. The masks used during this analysis are the same listed in Table \ref{tab:masks}, with the exception of the Std 2018, which has been replaced with the Std 2015.
Similarly to what performed in Section \ref{dataset} for the generation of the \ensemble\ 0, we build here a MC of 10$^4$ maps using the \textit{Planck} 2015 best-fit model. From these maps, we select a subset of 10$^3$ maps with variance $V$ within 20 $\mu$K$^2$ from the value computed with the \texttt{Commander} 2015 map, i.e. $V_\textrm{c}$ = 2060.09 $\mu$K$^2$. This set of simulations is called \ensemble\ 1-2015.
The behaviour of $V$ as a function of the masks obtained with the \ensemble\ 1-2015 is shown in Fig.~\ref{fig:var_comm_vs_var_1000_maps_same_variance_2015} and the corresponding LTP are reported in the first column of
Table~\ref{tab:p_values_2015}. 
We recover a similar monotonic behaviour as for the 2018 case. That is, in the Ext$_{30}$ case, the behaviour of the 2015 data is anomalous at $\sim2.9\,\sigma$. 

\begin{table}
	\begin{center}
		\begin{tabular}{ l | c | c | c  }
			\hline
			\hline
			& \multicolumn{3}{c}{\textbf{LTP} [\%]} \\
			\cline{2-4}
			Mask & $V<V_\textrm{c}$ & $V^{(\text{rot})}_\textrm{c}<V_\textrm{c}$ & $\textrm{LTP}_i<\textrm{LTP}_\textrm{c}$ \\
			\hline
			Std 2015 & 47.2 & 64.2 & 58.3\\
			
			Ext$_{12}$ & 7.2  & 11.1 & 10.0 \\
			
			Ext$_{18}$ & 0.8 & 2.9 & 2.4\\
			
			Ext$_{24}$ & 0.4 & 1.6 & 1.6 \\
			
			Ext$_{30}$ & 0.3 & 0.5 & 0.5\\
			\hline 
			\hline
		\end{tabular}
		\caption{The probability of obtaining a value for the variance $V$ smaller than that of \texttt{Commander} 2015 for a map of the \ensemble\ 1-2015 (first column). The probability of obtaining a value of the variance of the rotated \texttt{Commander} 2015 map, $ V^{(rotated)}_\textrm{c}$, smaller than the unrotated one (second column). LTP of obtaining a simulation with $\textrm{LTP}_i$ lower than the one obtained with the \texttt{Commander} 2015 map, $\textrm{LTP}_\textrm{c}$ (third column). }\label{tab:p_values_2015}
	\end{center}
\end{table}

\begin{figure}[t]
	\hspace{-0.85cm}
	\centering
	\subfloat{\includegraphics[width=.35\textwidth]{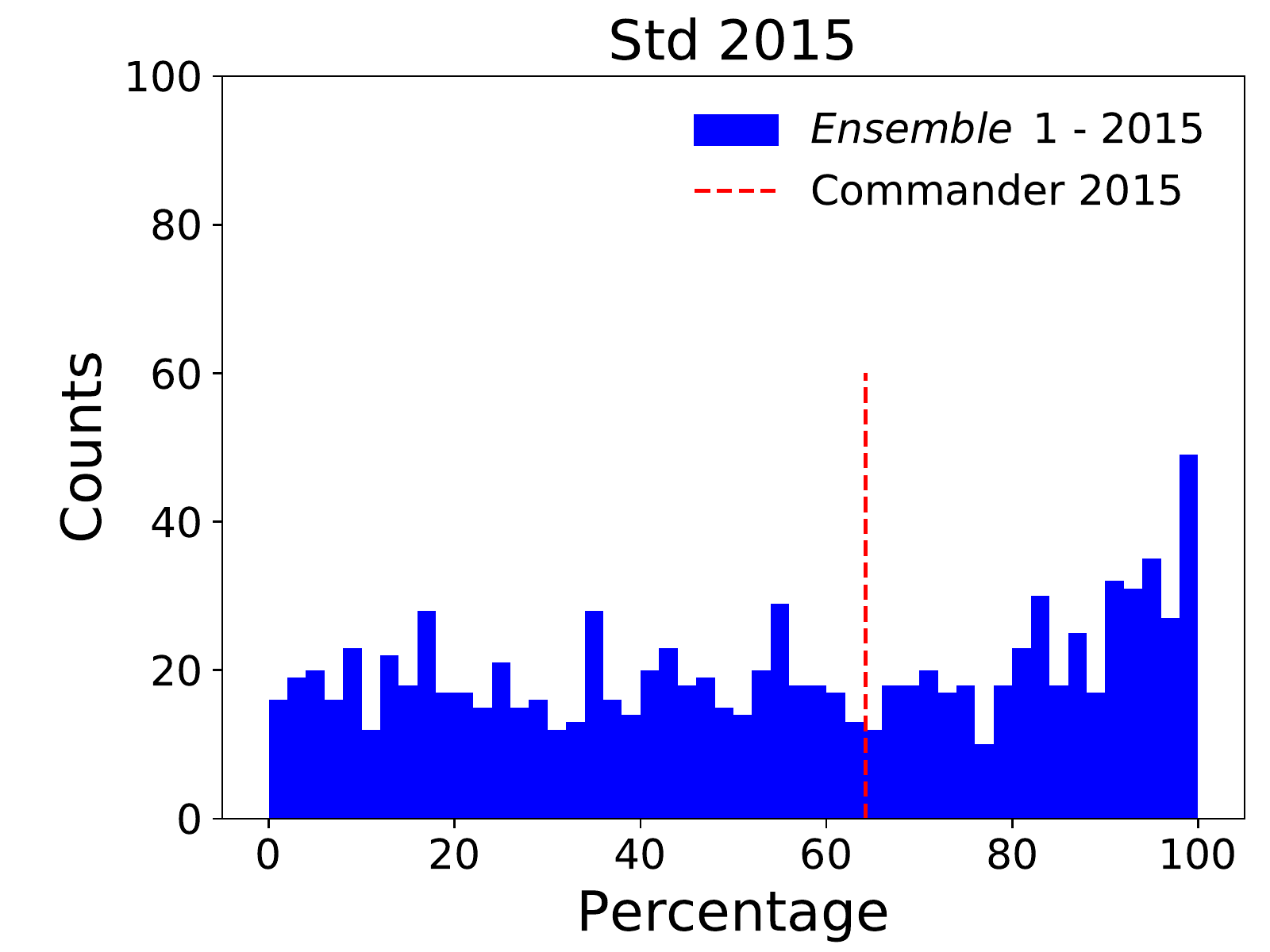}}
	\subfloat{\includegraphics[width=.35\textwidth]{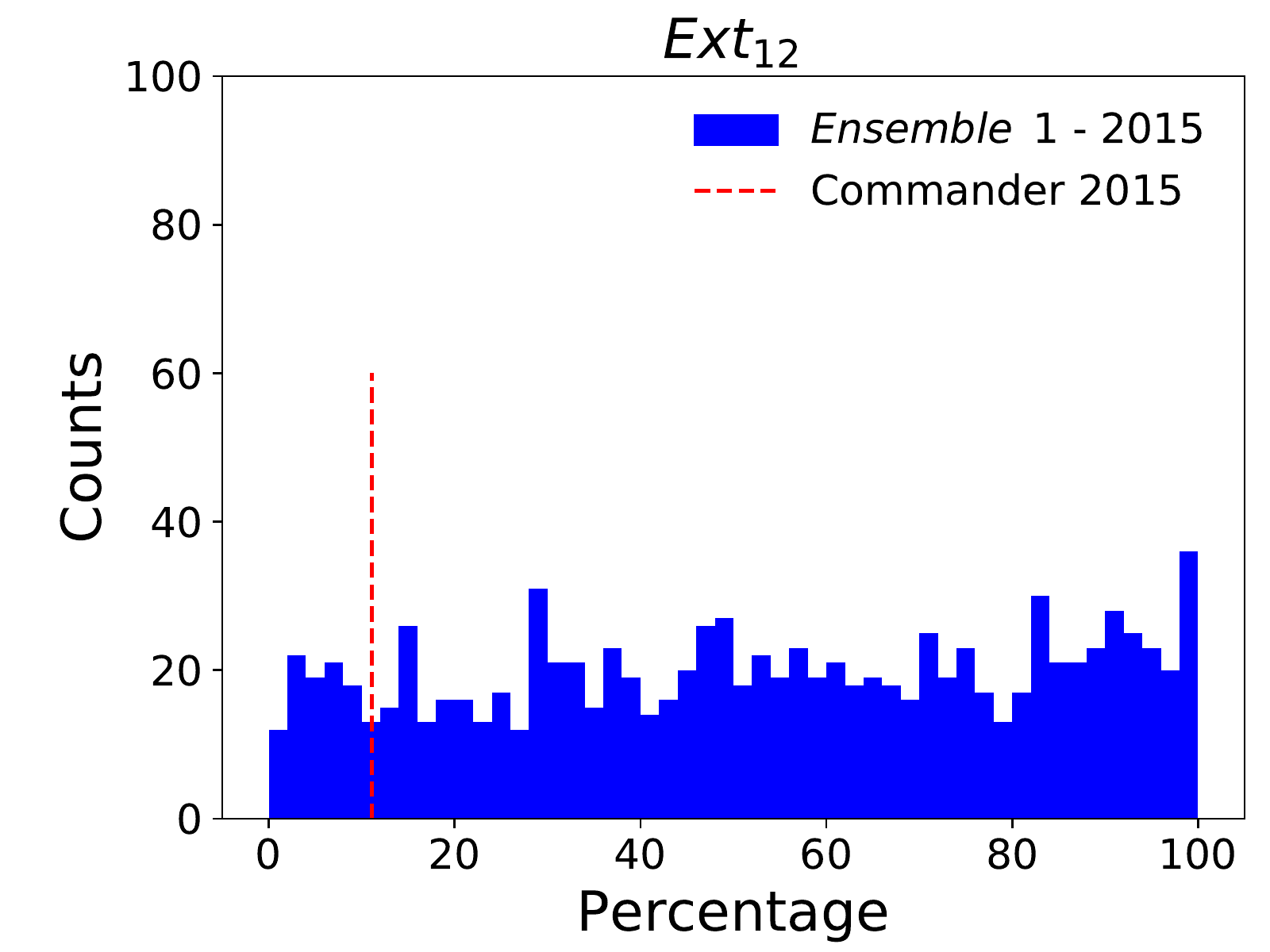}}
	\subfloat{\includegraphics[width=.35\textwidth]{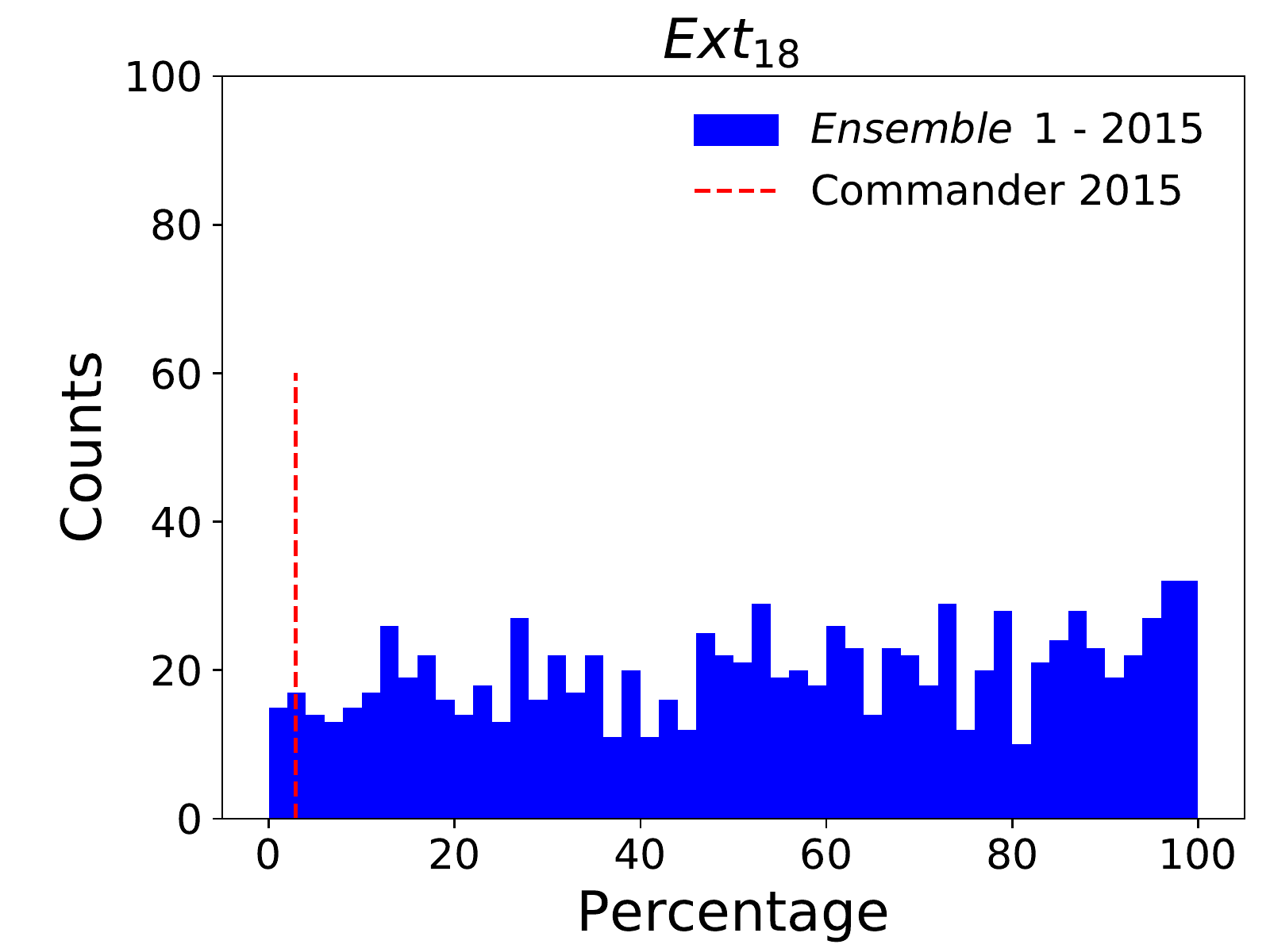}} \\
	\subfloat{\includegraphics[width=.35\textwidth]{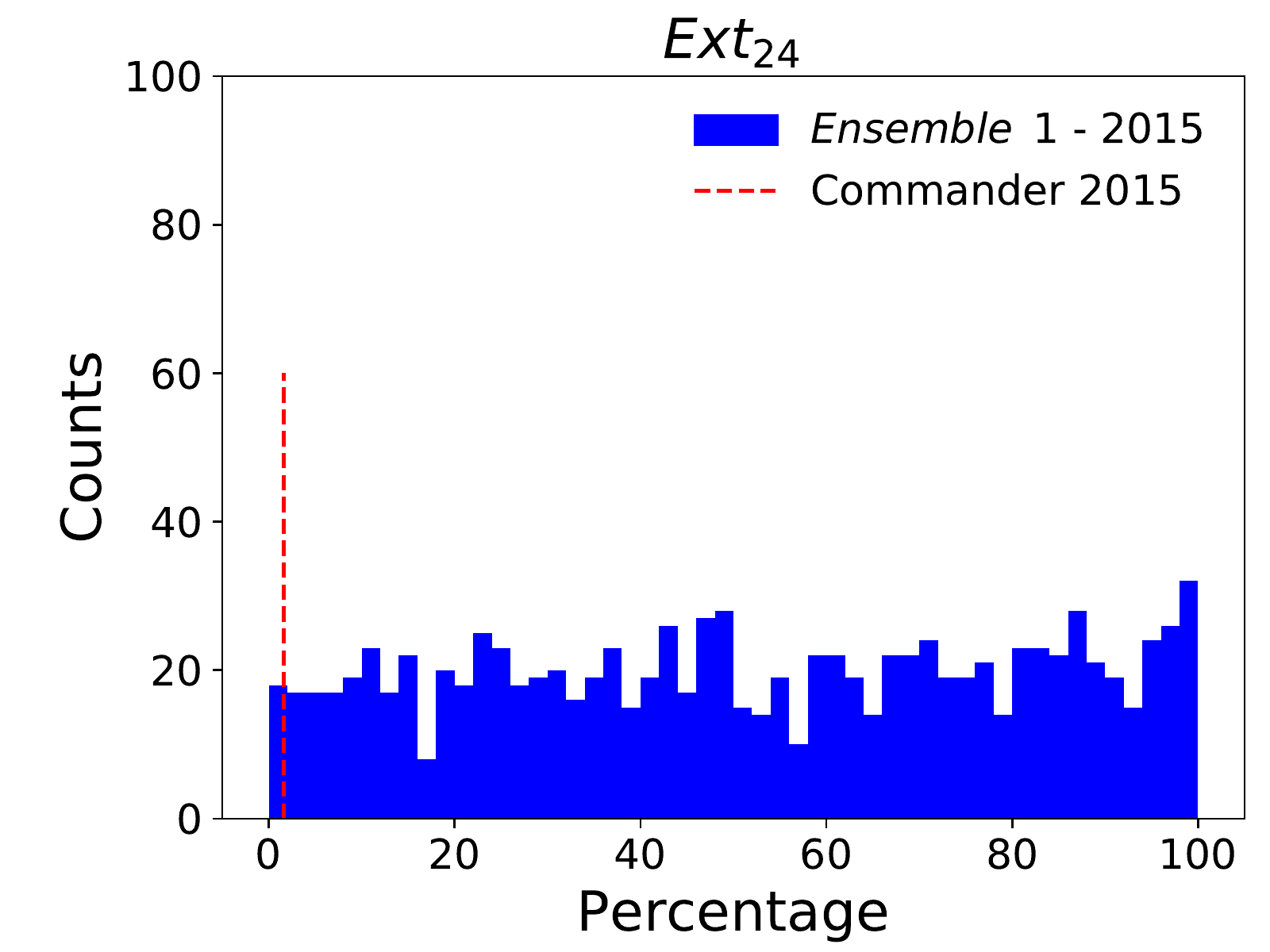}}
	\subfloat{\includegraphics[width=.35\textwidth]{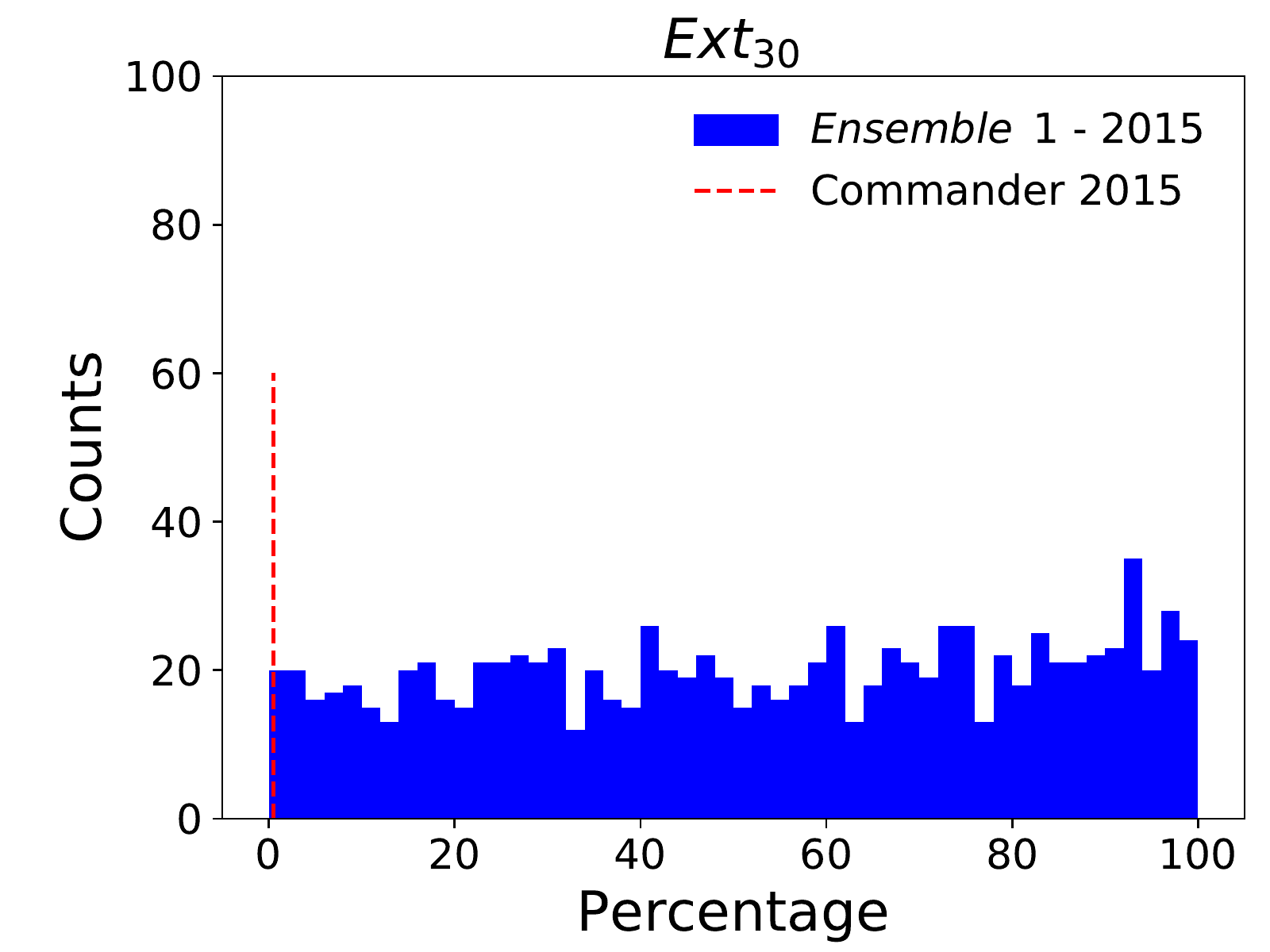}}
	\caption{LTP of finding a rotated map of the \ensemble\ 1 - 2015 with $V^{rot}<V$, where $V$ is the variance of the corresponding unrotated map. Each panel shows the results obtained using a different mask. The dashed vertical bars are the LTP of \texttt{Commander} 2015.}\label{fig:p_value_rot_1000_case_separate_2015}
\end{figure}
\begin{figure}[t]
	\centering
	{\includegraphics[scale=0.46]{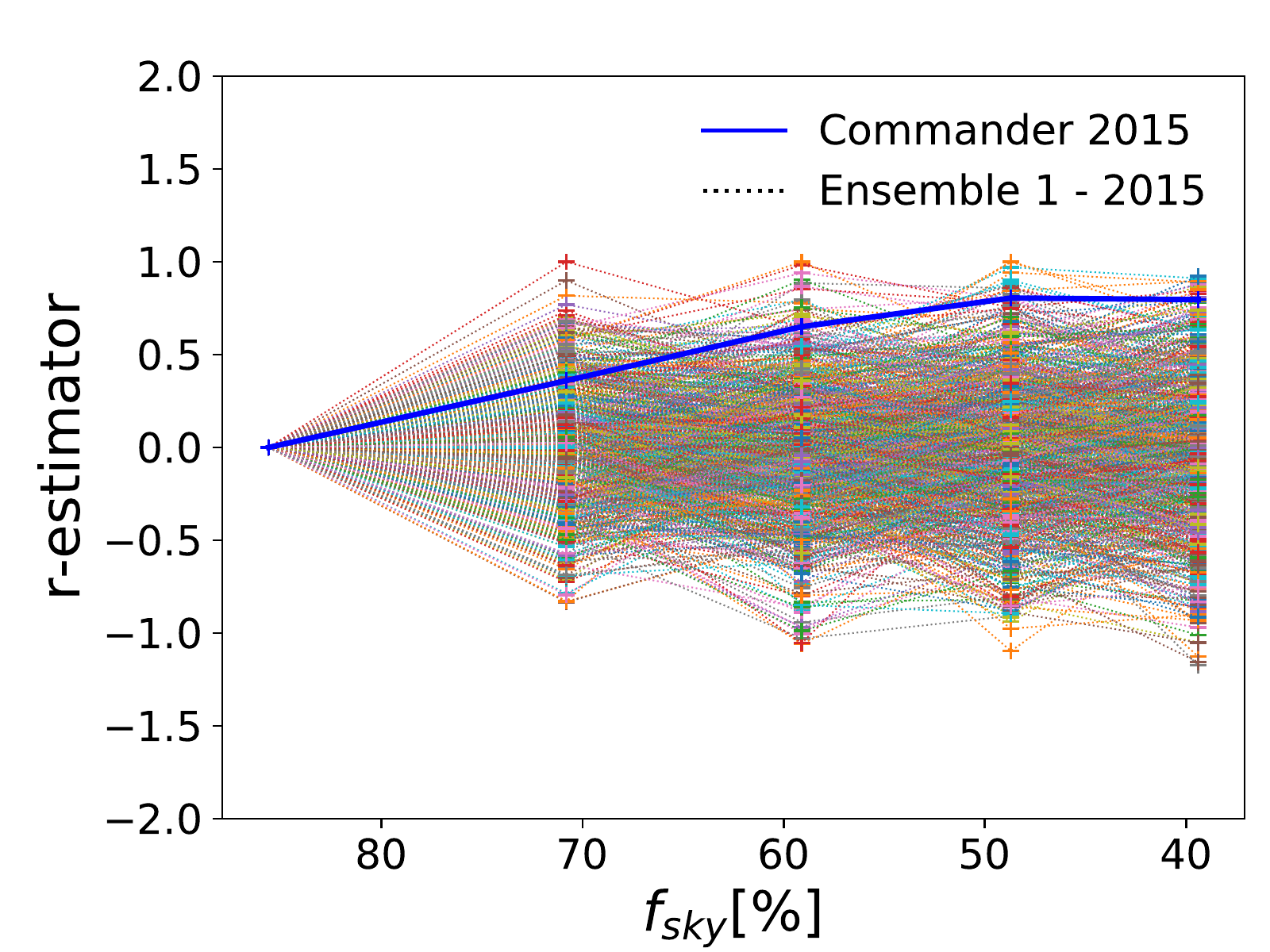}}
	{\includegraphics[scale=0.46]{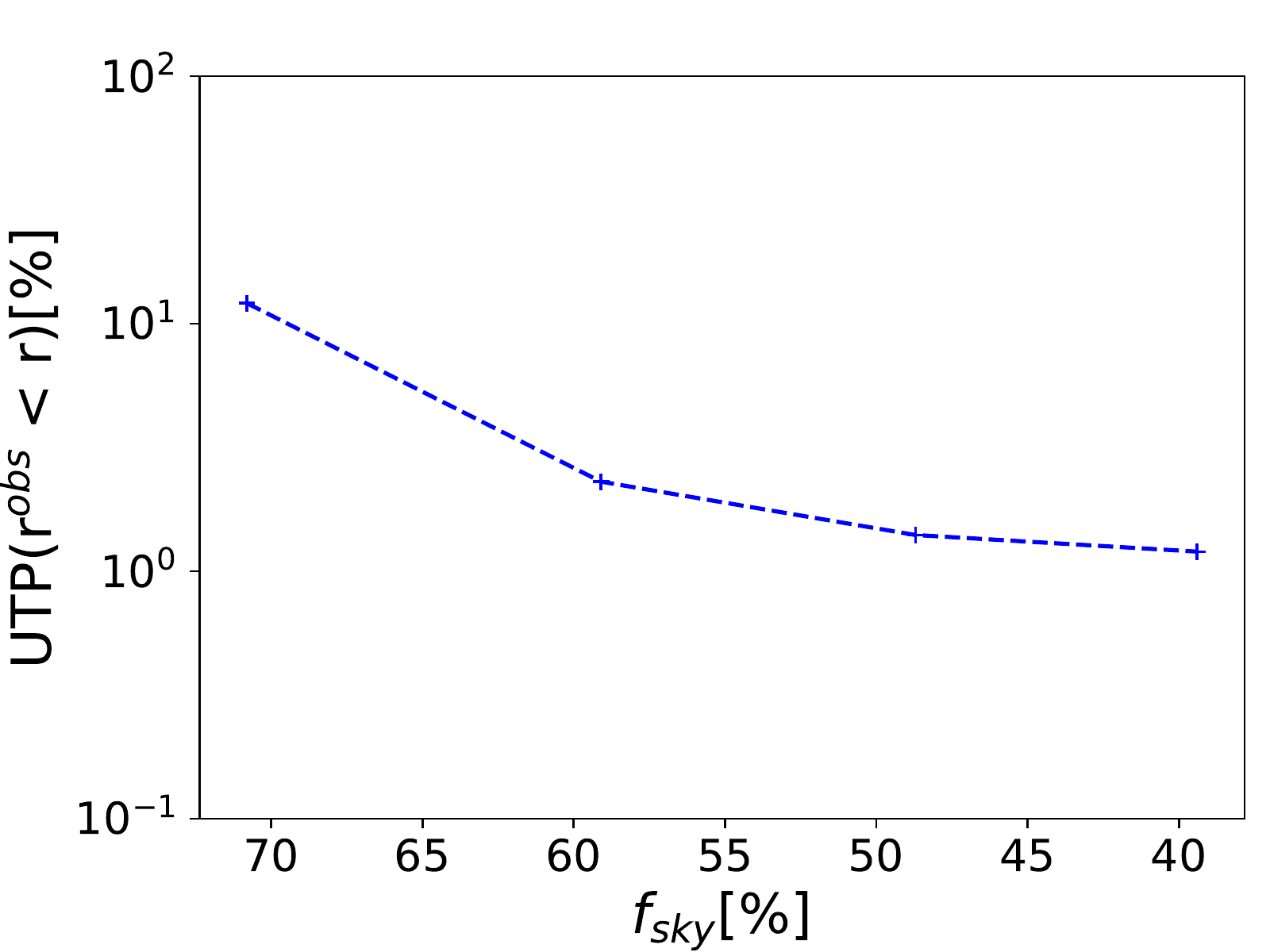}}
	\caption{Left panel: $r$-estimator computed with Eq. (\ref{eqn:r_estimator}) versus sky fraction. Right panel: UTP of obtaining a simulation with $r$ larger than the one obtained with \texttt{Commander} 2015 as a function of the sky fraction.} \label{fig:r_v_std_meno_v_j_2015}
\end{figure}

We take into account now the rotations applied to \ensemble\ 1-2015.
The results for the LTP-estimator and $r$-estimator are shown in Fig.~\ref{fig:p_value_rot_1000_case_separate_2015} and Fig.~\ref{fig:r_v_std_meno_v_j_2015}. For the LTP-estimator we find in the Ext$_{30}$ mask a LTP of $0.5\%$ which is in line with the 2018 analysis. All the LTP for this estimator are reported in Table \ref{tab:p_values_2015}. On the other hand, the $r$-estimator gives $r^c=0.80$ for the Ext$_{30}$ mask with a p-value of 1.2\%, see Table \ref{tab:r_value_2015}. While the general behaviour of $r$ across the mask is recovered here, the probability at high Galactic latitude is slightly higher. 

\begin{table}[!h]
	\begin{center}
		\begin{tabular}{ l | c   }
			\hline
			\hline
			& \multicolumn{1}{c}{\textbf{UTP} [\%]} \\
			\cline{2-2}
			Mask & $r^{\textrm{c}}<r$\\
			\hline
			Ext$_{12}$ & 12.1\\
			
			Ext$_{18}$ & 2.3 \\
			
			Ext$_{24}$ & 1.4\\
			
			Ext$_{30}$ & 1.2\\
			\hline 
			\hline
		\end{tabular}
		\caption{UTP of obtaining a simulation of the \ensemble\ 1 - 2015 with $r$ larger than the one obtained with the \texttt{Commander} 2015 map.}\label{tab:r_value_2015}
	\end{center}
\end{table}


\section{Dependence on threshold}
\label{threshold}

In this section we study the impact on our results of the threshold of $V$ we choose to select the maps of the \ensemble\ 1 from the $10^5$ $\Lambda$CDM simulations.
Specifically, in addition to the threshold of 20 $\mu$K$^2$ used in Section \ref{dataset}, we choose two other thresholds at 10 $\mu$K$^2$ and 30 $\mu$K$^2$. 
These will define two new subsets of 10$^3$ CMB temperature maps which have a variance $V$ close to the value observed by \texttt{Commander} 2018. 
We refer to these two additional subsets as \ensemble\ 2 (E2) and \ensemble\ 3 (E3), respectively. 

\begin{figure}[t]
	\centering
 	\includegraphics[scale=0.46]{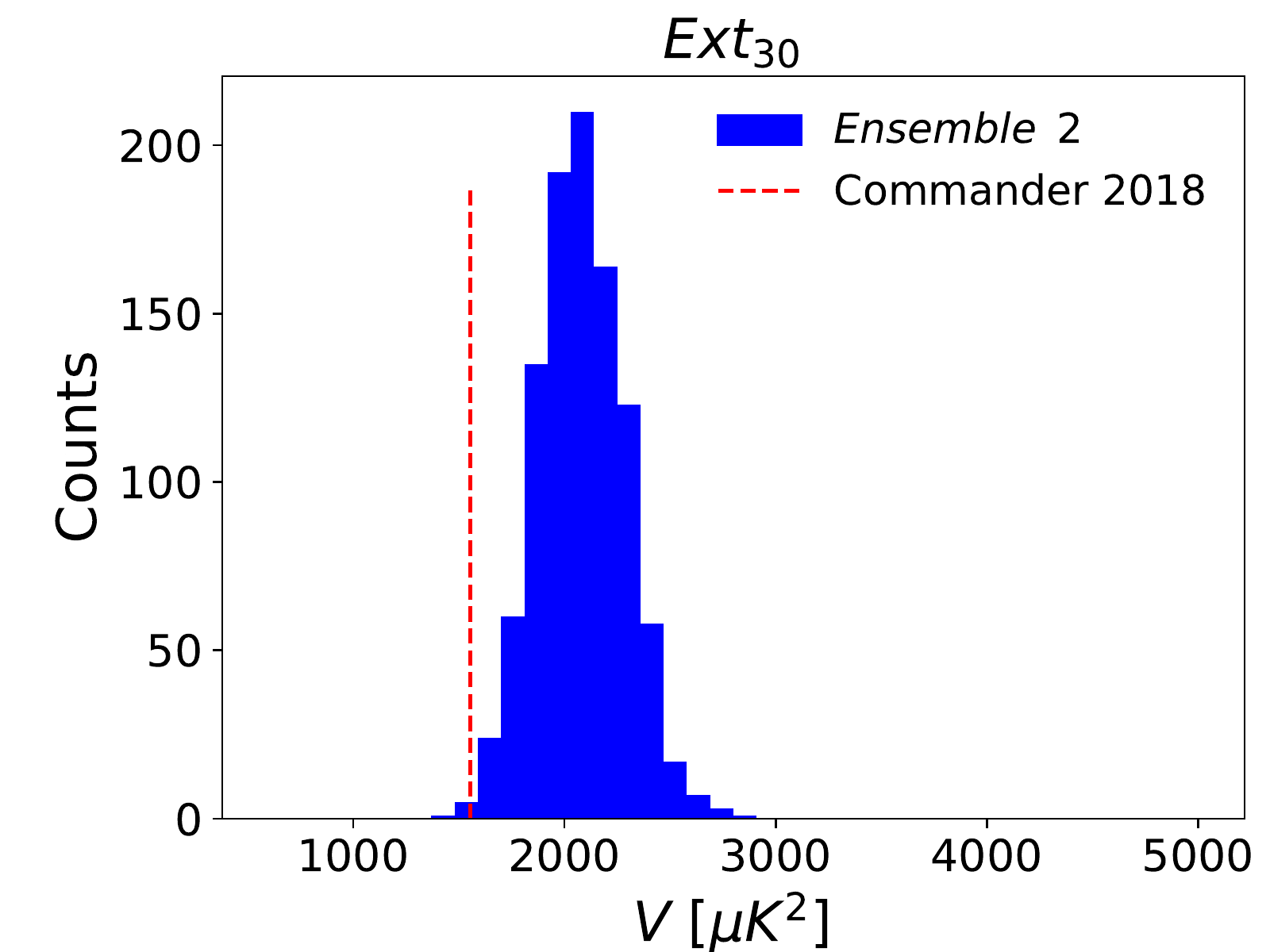}
	\includegraphics[scale=0.46]{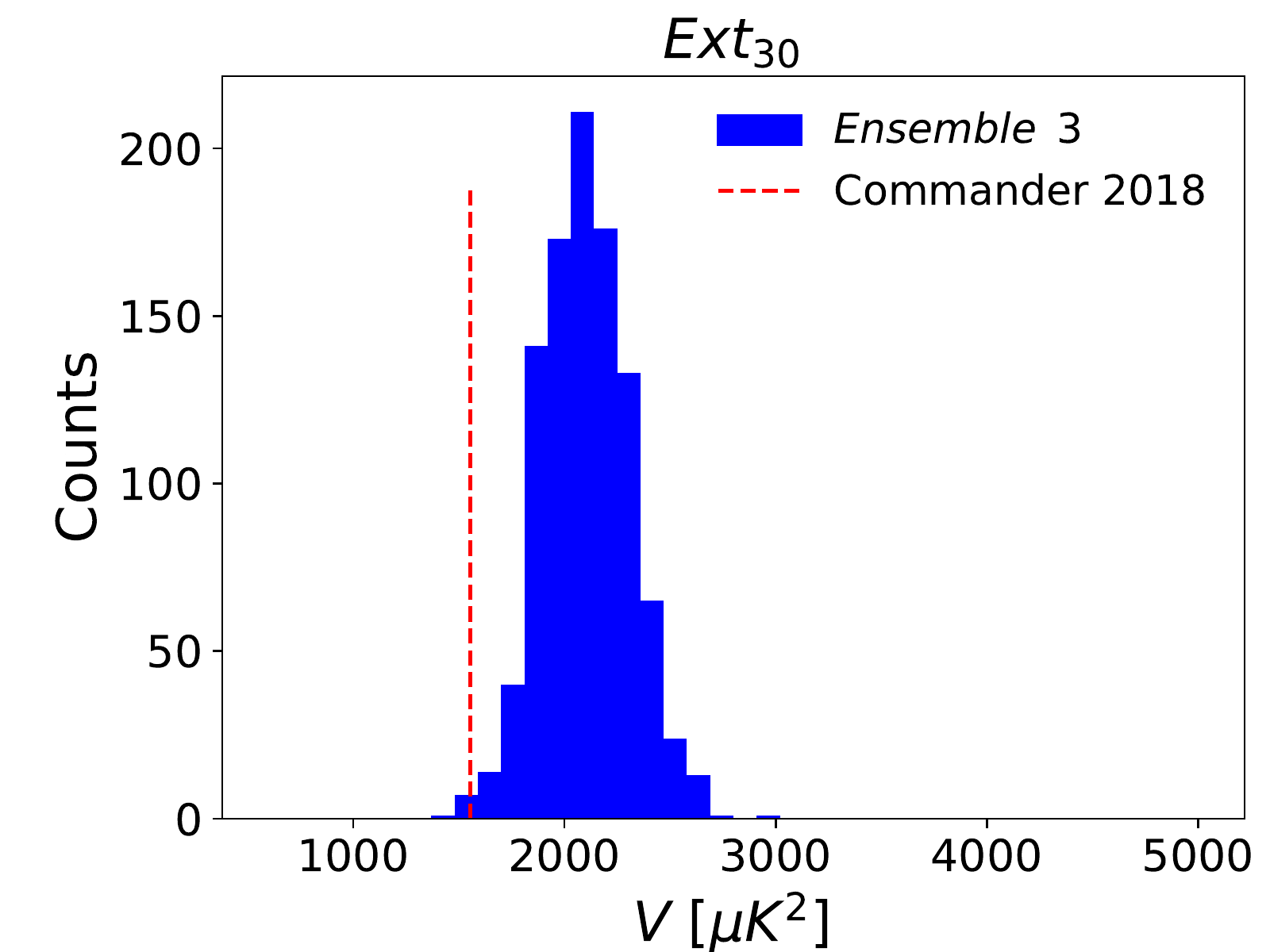}
	\caption{Variance distribution of the \ensemble\ 2 (left panel) and \ensemble\ 3 (right panel) for the Ext$_{30}$ mask. Red dashed line corresponds to the variance of the \texttt{Commander} 2018 map.}\label{fig:variance_ensemble2e3}
\end{figure}

Therefore, we repeat on E2 and E3, the same analysis previously performed on \ensemble\ 1 for both the considered estimators, focusing on the mask Ext$_{30}$.
We start building the distribution of $V$, see Fig.~\ref{fig:variance_ensemble2e3}, where the left panel refers to E2 while the right one to E3. 
The LTP of \texttt{Commander} 2018 are LTP$_{E2}(V_\textrm{c}<V_i)$=0.2$\%$ and LTP$_{E3}(V_\textrm{c}<V_i)$=0.7$\%$, which are consistent with what obtained with \ensemble\ 1.

\begin{figure}[h]
	\centering
	{\includegraphics[scale=.46]{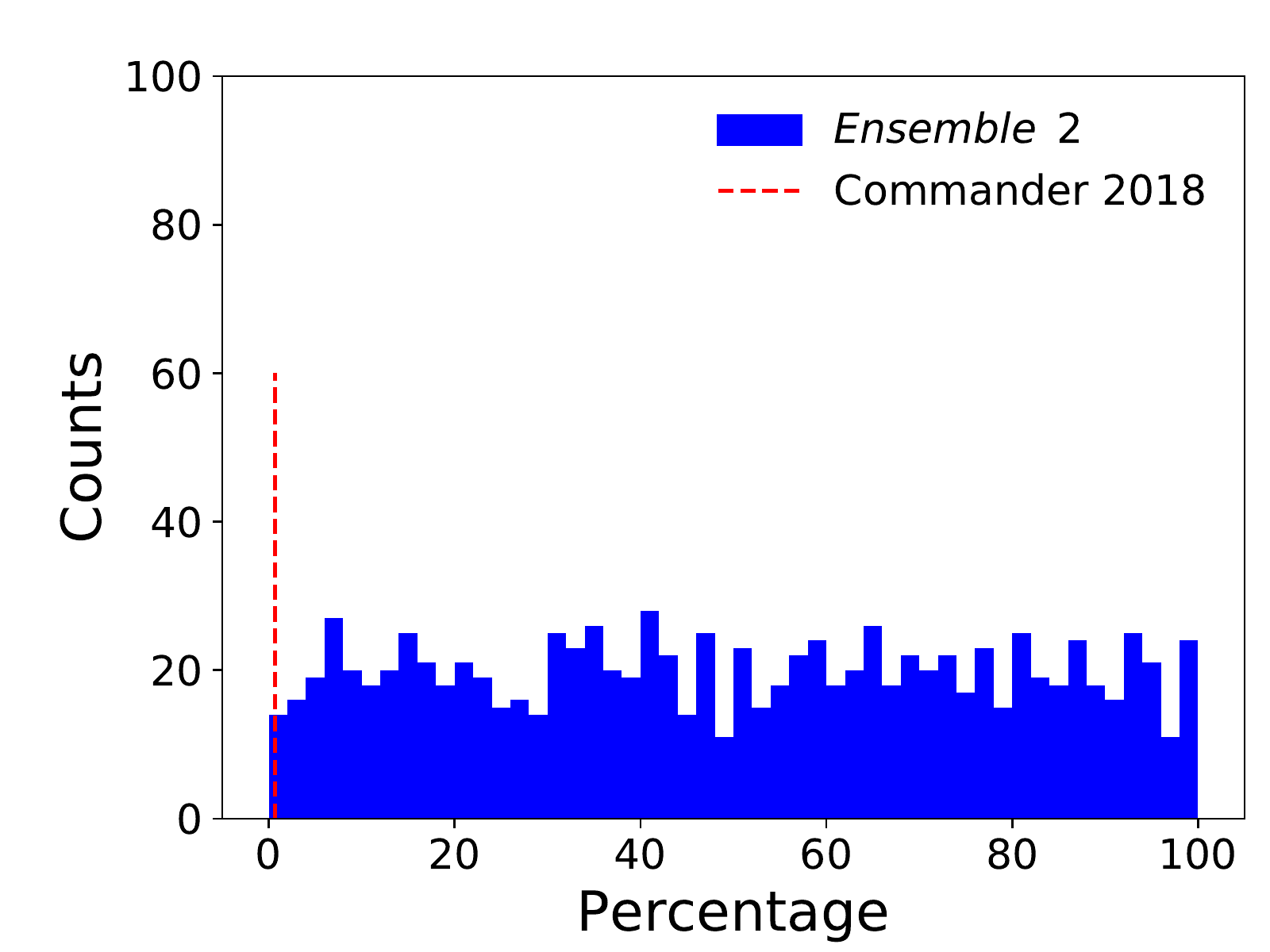}}
	{\includegraphics[scale=.46]{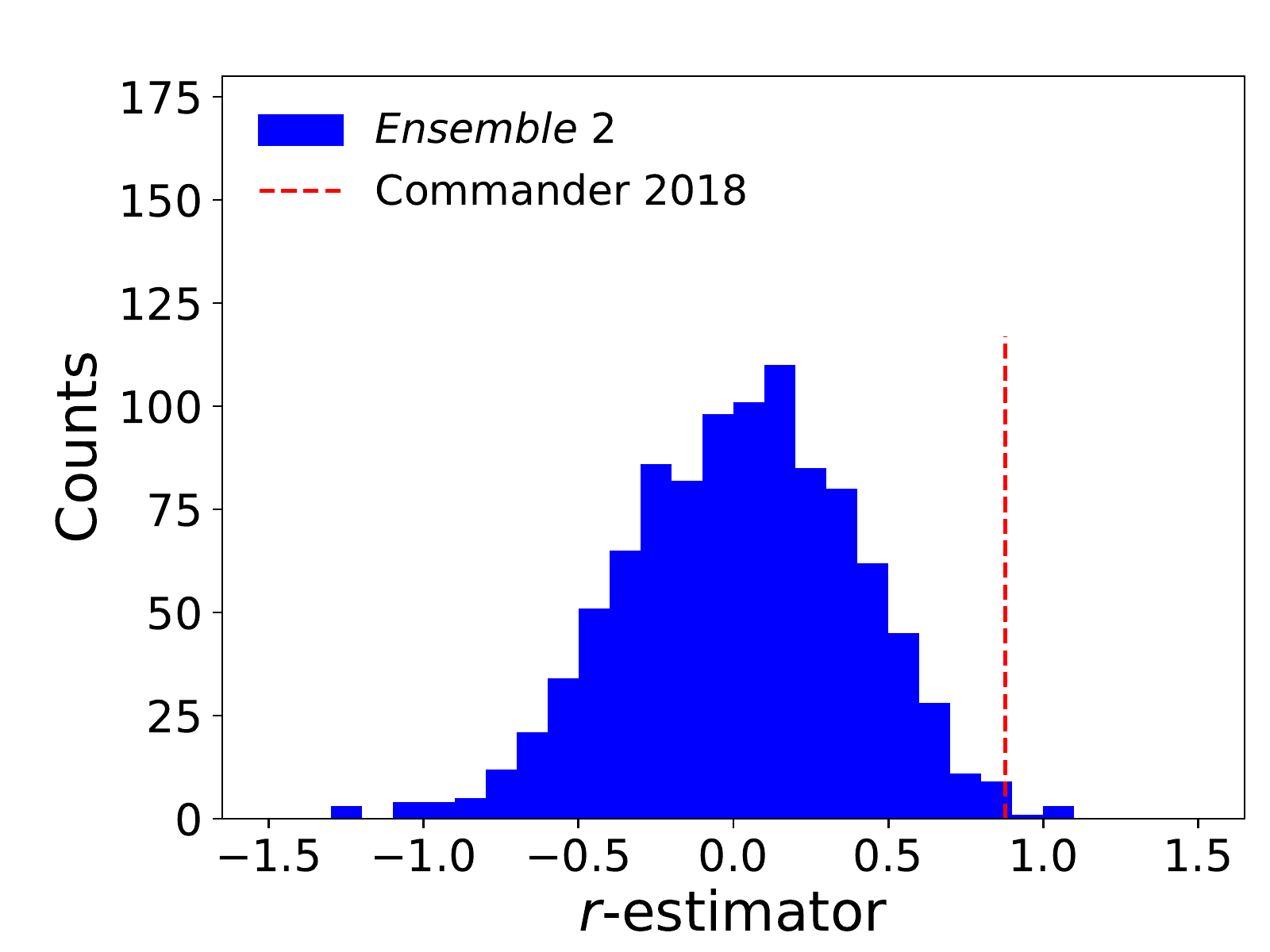}}
	\caption{Left panel: distribution of probability of observing, in a $\Lambda$CDM model with low variance, a lower value with respect to $V_\textrm{c}$ due to random rotations of \ensemble\ 2. Right panel: $r$-estimator computed with Eq. (\ref{eqn:r_estimator}) for the \ensemble\ 2. Both the results have been obtained using the Ext$_{30}$ mask.}\label{fig:rotation_ensemble2}
\end{figure}

\begin{figure}[h]
	\centering
	{\includegraphics[scale=.46]{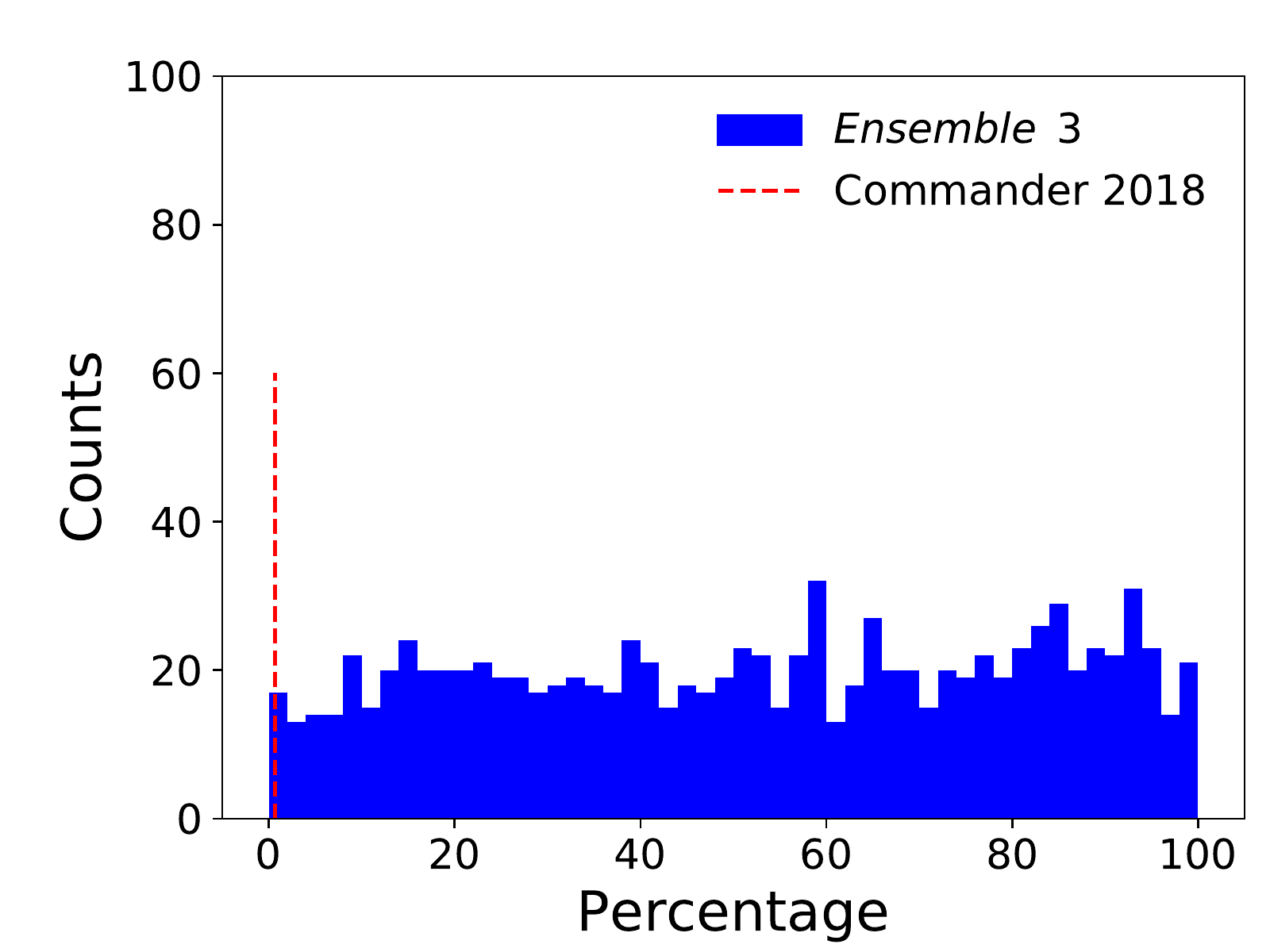}}
	{\includegraphics[scale=.46]{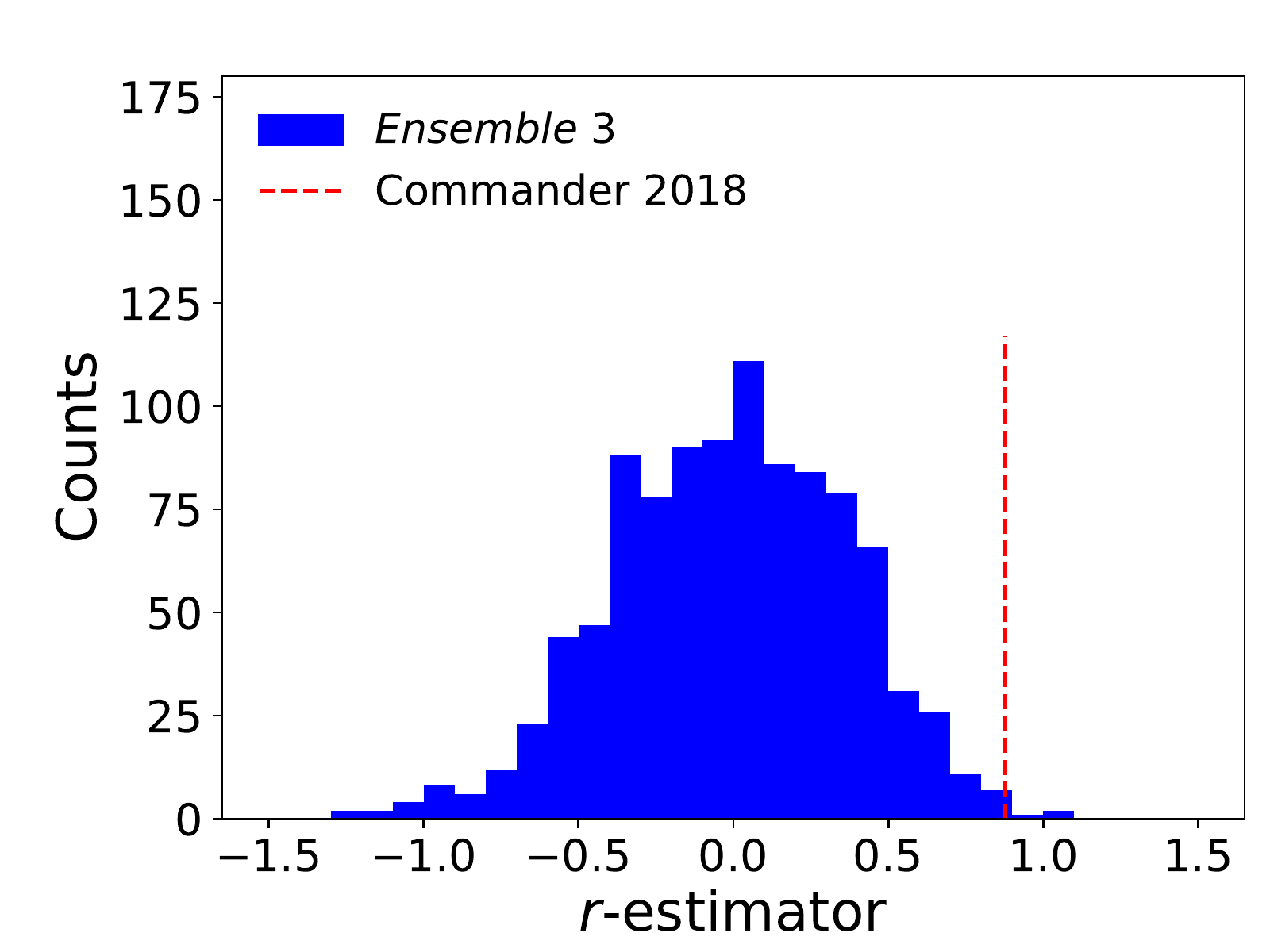}}
	\caption{The same as Fig.\ref{fig:rotation_ensemble2} but for \ensemble\ 3.}\label{fig:rotation_ensemble3}
\end{figure}

As done for the \ensemble\ 1, we can apply random rotations to E2 and E3 and build the LTP-estimator and the $r$-estimator in the Ext$_{30}$ case. 
The distributions of the former are shown in the left panels of Fig.~\ref{fig:rotation_ensemble2} and Fig.~\ref{fig:rotation_ensemble3} for the E2 and E3 case respectively.
The vertical dashed bars stand for the \texttt{Commander} 2018 values of the estimator, see again Table \ref{tab:p_value_comm_SMICA_ruotati}. 
The LTP of the LTP-estimator, turn out to be $0.3 \%$ and $0.2 \%$ for E2  and E3 respectively.
In the right panels of Fig.~\ref{fig:rotation_ensemble2} and Fig.~\ref{fig:rotation_ensemble3} we show the $r$-estimator for the E2 and E3. 
We find UTP$_{E2}(r_\textrm{c}<r_i)$=0.4$\%$ and UTP$_{E3}(r_\textrm{c}<r_i)$=0.5$\%$ for E2 and E3 respectively. 
We conclude that our results are stable with respect to the choice of the threshold which defines the set of constrained realisations.



%

\end{document}